\RequirePackage[-l2tabu,orthodox]{nag}
\documentclass
[11pt,letterpaper]
{article} 

\usepackage[notes=true,later=false,camera=false]{dtrt}

\usepackage[utf8]{inputenc}
\usepackage{ stmaryrd }
\usepackage{xspace,enumerate}
\usepackage[T1]{fontenc}
\usepackage[full]{textcomp}
\usepackage[american]{babel}
\usepackage{mathtools}

\usepackage{tikz}
\usetikzlibrary{arrows.meta,positioning}

\usepackage{amsthm}
\usepackage{thmtools}

\usepackage{tikz}
\usetikzlibrary{arrows.meta,calc,positioning,fit,backgrounds}
\usepackage{pdflscape}
\usepackage{xcolor}
\usepackage{array}
\usepackage{hyperref}

\usepackage[capitalise,nameinlink]{cleveref}

\usepackage{empheq}

\definecolor{citeblue}{HTML}{0055cc}
\hypersetup{
colorlinks=true,
urlcolor=citeblue,
linkcolor=NavyBlue,
citecolor=PineGreen,
linktocpage=true,
}
\renewcommand*\backref[1]{\ifx#1\relax \else (pg. #1) \fi}

\renewcommand{\tilde}{\widetilde}

\usepackage{paralist}
\usepackage{turnstile}
\usepackage[framemethod=TikZ]{mdframed}
\mdfsetup{frametitlealignment=\center}
\usepackage{tikz}
\usepackage{caption}
\DeclareCaptionType{Algorithm}
\usepackage{newfloat}
\usepackage{adjustbox}

\newtheorem{theorem}{Theorem}[section]
\newtheorem{lemma}[theorem]{Lemma}
\newtheorem*{lemma*}{Lemma}
\newtheorem{claim}[theorem]{Claim}
\newtheorem{proposition}[theorem]{Proposition}
\newtheorem{fact}[theorem]{Fact}

\theoremstyle{definition}
\newtheorem{definition}[theorem]{Definition}
\newtheorem*{definition*}{Definition}
\newtheorem{remark}[theorem]{Remark}

\makeatletter
\@for\@tempa:=theorem,lemma,claim,proposition,fact,corollary,conjecture,question,answer,hypothesis,exercise,definition,remark,notation,observation,example,examples\do{%
  \expandafter\def\csname theH\@tempa\endcsname{\thesection.\arabic{theorem}}}
\makeatother

\usepackage[linesnumbered,ruled]{algorithm2e}
\crefname{lemma}{Lemma}{Lemmas}
\crefname{fact}{Fact}{Facts}
\crefname{theorem}{Theorem}{Theorems}
\crefname{mtheorem}{Theorem}{Theorems}
\crefname{itheorem}{Theorem}{Theorems}
\crefname{corollary}{Corollary}{Corollaries}
\crefname{claim}{Claim}{Claims}
\crefname{example}{Example}{Examples}
\crefname{algorithm}{Algorithm}{Algorithms}
\crefname{problem}{Problem}{Problems}
\crefname{definition}{Definition}{Definitions}
\crefname{equation}{Eq.}{Eq.}
\crefname{strategy}{Strategy}{Strategies}
\crefname{observation}{Observation}{Observations}

\Crefname{algocf}{Algorithm}{Algorithms}

\usepackage[
letterpaper,
top=1.2in,
bottom=1.2in,
left=1in,
right=1in]{geometry}
\usepackage{newpxtext} 
\usepackage{textcomp} 
\usepackage[scr=rsfso]{mathalfa}
\usepackage{bm} 

\usepackage{microtype}

\usepackage{footnotebackref}

\usepackage{tikz}
\usetikzlibrary{intersections}

\renewcommand{\bar}{\overline}

\newcommand{\conn}{\operatorname{conn}}
\newcommand{\Ext}{\operatorname{Ext}}
\newcommand{\cT}{\mathcal{T}}
\newcommand{\cN}{\mathcal{N}}

\newcommand{\op}{\operatorname{op}}

\newcommand{\AND}{\operatorname{AND}}

\allowdisplaybreaks
\newcommand{\FormatAuthor}[3]{
\begin{tabular}{c}
#1 \\ {\small\texttt{#2}} \\ {\small #3}
\end{tabular}
}

\newcommand{\sfa}{\mathsf a}

\newcommand{\R}{{\mathbb R}}
\newcommand{\N}{{\mathbb N}}

\newcommand{\eps}{\varepsilon}

\newcommand{\cF}{{\mathcal F}}
\newcommand{\cE}{{\mathcal E}}

\newcommand{\1}{\mathbf{1}}

\newcommand{\tr}{\mathrm{tr}}
\newcommand{\Z}{\mathbb Z}

\newcommand{\C}{\mathbb C}

\newcommand{\bra}[1]{\langle #1\rvert}
\newcommand{\ket}[1]{\lvert #1 \rangle}

\newcommand{\Var}{\mathrm{Var}}

\newcommand{\cU}{\mathcal U}

\newcommand{\cS}{\mathcal S}

\newcommand{\Id}{\operatorname{Id}}

\newcommand{\seq}{\subseteq}
\newcommand{\spn}{\operatorname{span}}

\DeclareMathOperator*{\E}{\mathbb{E}}

\newcommand{\cD}{\mathcal D}
\newcommand{\cR}{\mathcal R}

\newcommand{\cM}{\mathcal M}

\newcommand{\cL}{\mathcal L}

\renewcommand{\emptyset}{\varnothing}
\renewcommand{\geq}{\geqslant}
\renewcommand{\ge}{\geqslant}
\renewcommand{\leq}{\leqslant}
\renewcommand{\le}{\leqslant}

\renewcommand{\epsilon}{\varepsilon}

\newcommand{\ignore}[1]{}

\newcommand{\Unif}{\operatorname{Unif}}

\newcommand{\Ree}{\operatorname{Re}}
\newcommand{\Imm}{\operatorname{Im}}
\newcommand{\cP}{\mathcal{P}}

\newcommand{\Cr}{\operatorname{cr}}
\newcommand{\He}{\operatorname{He}}

\renewcommand{\iota}{\mathbf{i}}

\newcommand{\syk}{\operatorname{SYK}}

\AtBeginDocument{%

}

\hypersetup{pdfview=FitH}
\begin{document}

\title{Lee-Yang theorems for $q$-deformed oscillators and doubly-scaled SYK}
\author{
 \FormatAuthor{Arpon Basu}{arpon.basu@princeton.edu}{Princeton University} 
  \FormatAuthor{Siddhant Midha}{siddhantm@princeton.edu}{Princeton University}
  }

\maketitle

\begin{abstract}
    We show that the vacuum partition function of the $q$-deformed oscillator   
    \[
    \mathcal{L}_q(z):= \langle0|e^{zT_q}|0\rangle, \qquad  T_q := \mathsf{a}_q + \mathsf{a}_q^*
    \]
    satisfies the Lee-Yang property, i.e.\ all zeros of $\mathcal{L}_q$ lie on the imaginary line for all $q\in[-1, 1)$, proving the conjectures of~\cite{Okuyama23}. Using the connection between $q$-deformed oscillators and the Sachdev-Ye-Kitaev (SYK) Hamiltonian in the doubly-scaled regime, we show that the annealed SYK partition function satisfies the Lee-Yang property in the thermodynamic limit, that is, all of its Fisher zeros are on the imaginary axis. As a result, we show that the quenched SYK partition function admits a zero-free strip of constant width with high probability, rigorously complementing the large$-k$ replica calculations of~\cite{zlokapaSYKThermalExpectations2026}.

    We show the imaginary-rootedness of $\cL_q$ by expressing $\varphi_q(z):= \mathcal{L}_q(\mathbf{i} z)$ as a locally uniform limit of ``$q$-deformed'' matching polynomials and then showing that those $q$-deformed polynomials are real-rooted. This establishes that $\varphi_q$ is real-rooted, and in fact belongs to the \emph{Laguerre-P\'olya} class. The real-rootedness of the $q$-deformed polynomials yields novel log-concavity relations on the \emph{$q$-deformed matching moments}
    \[M_\ell(q):= \sum_{\pi\in\mathcal{P}_2(2\ell)}q^{\operatorname{cr}(\pi)}.\]
    We show the real-rootedness of the $q$-deformed polynomials by using the Touchard–Riordan formula to derive a partial differential equation relating their $q$-derivatives to their $z$-derivatives, and then analyzing the resulting motion of their roots.
\end{abstract}
\pagenumbering{arabic}
\newpage
\tableofcontents
\newpage 
\section{Introduction}\label{sec:intro}
Consider the Hilbert space $\ell_2(\Z_{\geq 0})$, which is the topological closure of the complex vector space $\cF:= \spn\{\ket{0}, \ket{1}, \ket{2}, \ket{3}, \ldots\}$, where $\{\ket{n}:n\in\Z_{\geq 0}\}$ are orthonormal basis vectors, i.e. $\bra{i}j\rangle:= \delta_{ij}$ for all $i, j\in\Z_{\geq 0}$. For any $q\in[-1, 1]$, define the \emph{$q$-deformed annihilation operator} $\sfa_q$ acting on $\cF$ as 
\[\sfa_q\ket{0}:= 0,\qquad \sfa_q\ket{n}:= \sqrt{[n]_q}\cdot\ket{n - 1}\text{ for all }n\geq 1,\]
where $[n]_q:= 1 + q + \cdots + q^{n - 1}$ is a \emph{$q$-integer}. One may write $\sfa_q$ as the following matrix: 
\[\sfa_q :=
\left(
\begin{array}{cccccc}
0 & \sqrt{[1]_q} & 0 & 0 & 0 & \cdots\\
0 & 0 & \sqrt{[2]_q} & 0 & 0 & \cdots\\
0 & 0 & 0 & \sqrt{[3]_q} & 0 & \cdots\\
0 & 0 & 0 & 0 & \sqrt{[4]_q} & \cdots\\
\vdots & \vdots & \vdots & \vdots & \ddots & \ddots
\end{array}
\right),
\]
where the rows and columns of the matrix above are indexed by $\Z_{\geq 0}$. The annihilation operator $\sfa_q$ satisfies the \emph{$q$-deformed commutation relation}~\cite{ArikC76,bozejko1991example} with the associated creation operator $\sfa_q^*$
\[\sfa_q\sfa_q^* - q\sfa_q^*\sfa_q = \Id,\]
on $\cF$, which recovers the bosonic commutation relation for $q = 1$, and the fermionic anticommutation relation for $q = -1$.\footnote{For $q  = -1$, $\sfa_{-1}, \sfa_{-1}^*$ are block diagonal matrices with $2\times 2$ blocks. Herein, $\sfa_{-1}$ acts on the block spanned by $\{\ket{2n},\ket{2n + 1}\}$ for all $n\geq 0$, as $\sfa_{-1}\ket{2n+1}=\ket{2n}$ and $\sfa_{-1}\ket{2n} = 0$ recovering the standard fermionic algebra.}
We call $\ket{0}$ the \emph{vacuum state}, and we define the transfer operator $T_q:= \sfa_q + \sfa_q^*$. We study the \emph{vacuum partition function} $\cL_q:\C\to\C$ of the $q$-deformed oscillator in this paper, defined as
\[\cL_q(z):= \bra{0}e^{zT_q}\ket{0} = \sum_{k = 0}^\infty\bra{0}T_q^k\ket{0}\frac{z^k}{k!}.\]

\begin{figure}[t]
    \centering 
    \begin{tikzpicture}[line width=0.9pt]
  \draw[name path=e1] (1,0) arc (180:0:1.5);
  \draw[name path=e2] (2,0) arc (180:0:1.5);
  \draw[name path=e3] (3,0) arc (180:0:1.5);

  \path[name intersections={of=e1 and e2,by=c12}];
  \path[name intersections={of=e1 and e3,by=c13}];
  \path[name intersections={of=e2 and e3,by=c23}];
  \foreach \c in {c12,c13,c23}
  \fill[red!80!black] (\c) circle[radius=2pt];

  \foreach \v in {1,...,6} {
    \fill (\v,0) circle[radius=1.8pt];
    \node[below=4pt] at (\v,0) {$\v$};
  };
\end{tikzpicture}
\caption{A perfect matching $\pi\in\cP_2(6)$ with $\Cr(\pi) = 3$}\label{fig:matchingfig}
\end{figure}
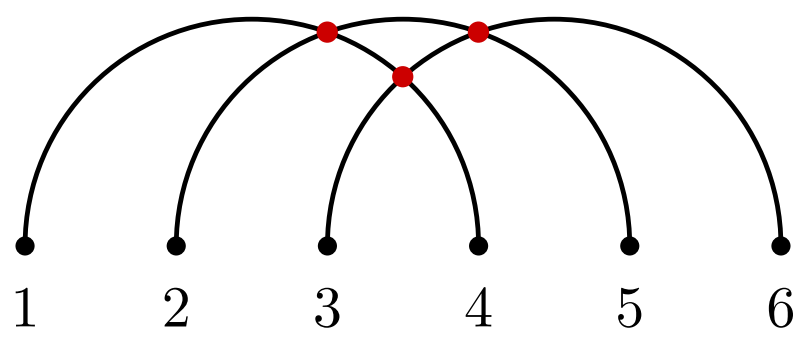

\paragraph{Vacuum Moments and Crossing Numbers} The vacuum moments of the transfer operator admit a combinatorial description in terms of crossing numbers of perfect matchings or \emph{chord diagrams}. For any $\ell\geq 0$, a perfect matching on $[2\ell]:= \{1, 2, \ldots, 2\ell\}$ is a partition $\pi = \{e_1, \ldots, e_\ell\}$ of $[2\ell]$ with 
\[\bigcup_{i = 1}^\ell e_i = [2\ell],\qquad e_i\cap e_j = \emptyset\text{ for all }1\leq i < j\leq\ell,\qquad |e_i| = 2\text{ for all }1\leq i\leq\ell.\]
The sets $e_1, \ldots, e_\ell$ are called \emph{edges/chords} of $\pi$. For any perfect matching $\pi$ containing edges $e = \{a < b\}$ and $e' = \{a' < b'\}$, we say $e, e'$ \emph{cross} each other if $a < a' < b < b'$ or $a' < a < b' < b$. Further, let 
\[\Cr(\pi):= \sum_{1\leq i < j\leq\ell}\1(e_i, e_j\text{ cross each other})\]
be the number of crossings in $\pi = \{e_1, \ldots, e_\ell\}$, see \cref{fig:matchingfig} for an illustration. Finally, let $\cP_2(2\ell)$ be the set of all perfect matchings $\pi$ on $[2\ell]$. Then the \emph{$q$-Wick formula}~\cite{IsmailSV87,bozejko1991example} states that for any $\ell\geq 0$, the even and odd moments satisfy:

\[\bra{0}T_q^{2\ell}\ket{0} = M_\ell(q):= \sum_{\pi\in\cP_2(2\ell)}q^{\Cr(\pi)},\qquad\bra{0}T_q^{2\ell + 1}\ket{0} = 0,\]

where $M_\ell(q)$ are the \emph{$q$-matching moments}\footnote{$M_\ell(q)$ is the $(2\ell)^{\mathrm{th}}$ moment of the $q$-Gaussian distribution $\nu_q$, hence the name ``moment''. We elaborate on this shortly}, with the lower order moments reading $M_0(q) = 1, M_1(q) = 1, M_2(q) = 2 + q$ and so on. The partition function $\cL_q$ can be expressed in terms of the matching moments as
\[\cL_q(z) = \sum_{\ell = 0}^\infty M_\ell(q)\cdot\frac{z^{2\ell}}{(2\ell)!}.\]
For all $q\in[-1, 1]$, one has $|M_{\ell}(q)| \leq M_\ell(1) = |\cP_2(2\ell)| = (2\ell-1)!!$, where $(2\ell-1)!! := (2\ell)!/(2^\ell\ell!)$. As a result $\cL_q$ is an entire function in $z$. Furthermore, 
\[\cL_1(z) = \sum_{\ell = 0}^\infty \frac{z^{2\ell}}{2^\ell\ell!} = e^{z^2/2}\] 
has no zeros in $\C$.

Our first main result is that for $q\in[-1, 1)$, $\cL_q$ has infinitely many zeros which are purely imaginary, i.e. $\cL_q$ has the \emph{Lee-Yang property}. This form of the partition function $\cL_q$ was first conjectured by~\cite{Okuyama23}, where they numerically found that the radius of convergence of the high-temperature expansion of $\log\cL_q$ is set by the first root on the imaginary axis. We resolve these predictions rigorously:

\begin{restatable}[A Lee-Yang property of the $q-$oscillator]{theorem}{momentfunctionimaginaryrooted}\label{thm:moment-function-imaginary-rooted}
    For any real $q\in[-1, 1)$, $\cL_q$ is entire and has infinitely many zeros, all of which lie on the imaginary line $\iota\R$. Furthermore, the zeros are simple, and symmetric about the origin, i.e. the roots of $\cL_q$ can be written as $\{\pm\iota t_k(q)\}_{k\geq 1}$ for some $0 < t_1(q) < t_2(q) < t_3(q) < \cdots$. Moreover, we may write
    \[\cL_q(z) = \prod_{k = 1}^\infty\left(1 + \frac{z^2}{t_k(q)^2}\right).\]
\end{restatable}
\begin{remark}
    For any complex $q$ with $|q| > 1$, the radius of convergence of the series $\sum_{\ell = 0}^\infty M_\ell(q)z^{2\ell}/(2\ell)!$ is $0$, i.e. it is unclear how to even define $\cL_q$ meaningfully for $|q| > 1$.
\end{remark}

The Lee-Yang property originates in the famous circle theorem of Lee and Yang~\cite{LY52}, which showed that for any finite temperature,  the partition function of a finite ferromagnetic Ising model has only purely imaginary zeros in the external field variable. Lee-Yang theorems have subsequently been established for quantum spin models~\cite{Asano70,SuzukiF71}, monomer-dimer systems~\cite{HeilmannL72}, general spin measures~\cite{Newman74,LiebS81}, and certain higher-order interactions~\cite{SuzukiF71,LiuSS19}.

Not only do we obtain a product formula for $\cL_q$; we also obtain precise control on the location of $\{t_k(q)\}_{k\in\N}$, especially in the limit $q\nearrow 1$, when we have $t_k(q)\to\infty$ for all $k\geq 1$. To that end, let $s^*\approx 1.1997$ be the unique positive solution to the equation $s\tanh(s) = 1$, and write $\zeta_*:= 2s^*/\cosh(s^*)\approx 1.3255$. 

\begin{restatable}[Location of the nearest zero]{theorem}{rootbehavior}\label{thm:nearest-root}
 The map $[-1, 1)\ni q\mapsto \sqrt{1 - q}\cdot t_1(q)$ is non-increasing, with $\lim_{q\nearrow 1}\sqrt{1 - q}\cdot t_1(q) = \zeta_*$. In particular the root closest to the real axis obeys: 
\[
t_1(q) > \frac{\zeta_*}{\sqrt{1 - q}}
\] 
 for all $q\in[-1, 1)$. 
\end{restatable}

\paragraph{SYK Hamiltonian} A central motivation to study the $q$-deformed oscillator is the fact that it serves as a tractable approximation for the well-known \emph{Sachdev-Ye-Kitaev (SYK)} Hamiltonian in the \emph{doubly-scaled} regime. The SYK Hamiltonian is a Hamiltonian on $n \geq 2$ Majorana modes\footnote{Formally, for even $n$, the Majorana anticommutation relations admit, up to unitary equivalence, a unique irreducible representation by self-adjoint operators on $\mathbb C^N$, where $N=2^{n/2}$. Furthermore, $N$ is the smallest possible dimension for which such a self-adjoint representation exists.} $\{\gamma_i\}_{i=1}^n$ satisfying the anti-commutation relations $\{\gamma_i, \gamma_j\} = 2\delta_{ij}\cdot\Id$. Then the $k$-wise SYK Hamiltonian, for $k\geq 2$ being an even integer, reads
\[
H_{\syk} := \sqrt{\frac{(k-1)!}{2k n^{k-1}}}\cdot\sum_{S\in\binom{[n]}{k}} g_S\Gamma_S,
\]
where $\{g_S\}_{S\in\binom{[n]}{k}}$ are i.i.d. $\cN(0, 1)$ Gaussians, and for any set $S = \{i_1 < i_2 < \dots < i_k\} \in \binom{[n]}{k}$, $\Gamma_S$ is defined as $\Gamma_S := \iota^{k(k-1)/2}\gamma_{i_1}\gamma_{i_2}\dots \gamma_{i_k}$. This ensures that $\Gamma_S$ is a Hermitian operator, satisfying the commutation relations $\Gamma_S\Gamma_T = \eps_{S,T}\cdot\Gamma_T\Gamma_S$, where $\eps_{S,T}:=(-1)^{|S\cap T|}\in\{\pm 1\}$.

The SYK model has attracted significant attention in high-energy physics,~\cite{Kitaev15,MaldacenaS16,maldacena2016bound,polchinski2016spectrum,cotler2017black,GarciaVerbaaschotPRD,rosenhaus2019introduction,goel2023semiclassicalgeometrydoublescaledsyk}, condensed matter~\cite{SachdevY93,SYK_review,Sachdev24}, as well as quantum information~\cite{hastings2022optimizing,basso2024optimizing,anschuetzStronglyInteractingFermions2025}, and thus understanding the thermodynamic (in the limit $n\to\infty$) properties of the SYK model is of great importance. In this direction,~\cite{SYK_Berkooz2} took an important step by making the following approximation: In the \emph{double-scaling limit}, i.e. in the limit where $n,k\to\infty$\footnote{with even $k$} such that $k^2/n\to \alpha\in(0,\infty)$, we can replace the commutation signs $\eps_{S, T}$ with their average 
\[q:=\lim_{n\to\infty}\E_{S,T\sim\binom{[n]}{k}}\eps_{S,T}=e^{-2\alpha}.\] 
Consequently, they consider the \emph{$q$-deformed oscillator} $\sfa_q$ satisfying the \emph{$q$-deformed commutation relation} $\sfa_q\sfa_q^* - q\sfa_q^*\sfa_q = \1$, and replace the disorder-averaged partition function $\E_{\bm g}\tr\exp(\beta H_{\syk})$\footnote{Here $\tr$ refers to normalized trace, which satisfies $\tr(\Id)=1$} with the vacuum partition function of the $q-$oscillator. ~\cite{SYK_Berkooz2} use this approximation to compute the $4$-point correlation function for the Doubly-Scaled SYK model.

Recently~\cite{BasuKM26} made a mathematically rigorous connection between the disorder averaged SYK model and a deterministic system of oscillators for \emph{all} $n\geq k\geq 2$ with quantitative error bounds. The collective mode on this system of oscillators ``tends to'' the $q-$deformed oscillator when the doubly-scaled limit is taken. Developing on this, we show a Lee-Yang theorem for the (doubly-scaled) SYK model by comparing the SYK partition function with that of the $q-$deformed oscillator $\cL_q$. Towards that end, write $\Omega:=\C\setminus\pm\iota[\zeta_*, \infty)$.

\begin{restatable}[Annealed SYK Partition Function is Asymptotically Lee-Yang]{theorem}{annealedleeyang}
\label{thm:annealed-syk-lee-yang}
    Let $n > k\geq 2$ be even integers with $k=k(n)$ and $k^2/n \to \alpha\in(0,\infty)$ as $n\to\infty$. Then, $Z_n(z)$ converges locally uniformly on $\C$ to 
    \[
    Z_\infty(z) := \cL_q\left(\frac{e^{-\alpha/4}}{\sqrt{2\alpha}}z\right).
    \]
    with $q=e^{-2\alpha}$. As a result, all roots of $Z_\infty$ are on the imaginary axis.
\end{restatable}
Since the complex variable $z$ in the partition function $Z_n(z)$ represents inverse temperature, its zeros are conventionally called \emph{Fisher zeros}~\cite{fisher1967theory}. We use the term \emph{Lee-Yang property} in the sense that all zeros of a moment-generating function lie on the imaginary axis. \cref{thm:annealed-syk-lee-yang} is thus a Lee-Yang result for the SYK partition function in the thermodynamic limit, or, equivalently, stating that all of its Fisher zeros are purely imaginary.

\cref{thm:annealed-syk-lee-yang} may be viewed as a considerable generalization of~\cite{feng2019spectrum}, which shows that the empirical spectral distribution of $H_{\syk}$ in the doubly-scaled limit weakly converges to a $q$-Gaussian distribution, which is the \emph{vacuum spectral measure} of the $q$-deformed oscillator (see~\nameref{para:qgaussian-mgf} for further discussion). Also see~\cite{feng2021spectrum,feng2020spectrum} for a detailed treatment of the empirical spectral distribution of the Doubly-Scaled SYK Hamiltonian, and~\cite{SYK_Berkooz3} for a physics-based introduction to the connections between the SYK model and the $q$-deformed oscillator. 

\begin{remark}\label{remark:non-imaginary-roots}
    Importantly, the localization of zeros on the imaginary axis holds only in the limit $n\to\infty$. One can check that for $n = 8, k = 4$, the annealed partition function $\E_{\bm g}\tr\exp(zH_{\syk})$ is \textbf{not} imaginary-rooted, i.e. it has roots with non-zero real parts (see~\href{https://github.com/siddhant-midha/syk-roots}{\texttt{github.com/siddhant-midha/syk-roots}} for numerical results). This should be contrasted with other seminal Lee-Yang results, such as the $2$-local ferromagnetic Ising model~\cite{LY52}, or the monomer-dimer model~\cite{HeilmannL72}, which exhibit imaginary-rootedness of the partition function for \emph{all} system sizes $n\in\N$. Of course, these models are deterministic, while the SYK model is disordered, and it would be interesting to understand the role of randomness in the asymptotic Lee-Yang phenomenon.
\end{remark}
Given the Lee-Yang behavior of the annealed partition function, one can further establish \emph{zero-freeness} (see the discussion~\nameref{par:zerofree} for consequences and algorithmic implications) for the \emph{quenched} partition function with high-probability over the disorder configurations. Towards that end, define the following closed strip extending along the real axis,
\[
\cR_n := \{z \in \C : |\Ree z| \leq n^{0.08}, \quad |\Imm z| \leq \zeta_*\}\subset\C.
\]
We show that the doubly-scaled SYK model is zero-free in the strip $\cR_n$ with high probability.

\begin{restatable}[Zero-Freeness of the Quenched SYK Partition Function]{theorem}{quenchedstrip}
\label{thm:sykstrip}
    For even integers $n > k\geq 2$ with $k \geq \omega(\sqrt{n/\log(n)})$, the SYK partition function $Z_n(z;\bm g)$ is zero free on the strip $\cR_n$ with probability at least $1 - e^{-\Omega(\sqrt{n \log n})}$ over the disorder $\bm g$.
\end{restatable}

The agreement between our strip half-width constant $\zeta_*\approx1.32$ and that obtained in~\cite{zlokapaSYKThermalExpectations2026} reflects a direct connection between the two computations: the partition function $Z_\infty(z)$, is a rescaling of the moment generating function of the $q-$Gaussian distribution (see~\nameref{para:qgaussian-mgf}). Upon taking the $\alpha\to 0$ limit, $2\alpha \log Z_{\infty}(z)$ coincides with the large-$k$ saddle-point action of~\cite{zlokapaSYKThermalExpectations2026} under a replica symmetric ansatz. Our~\cref{thm:sykstrip} establishes this predicted strip width $\zeta_*$ rigorously in the doubly-scaled regime.

\subsection{Related Work and Discussion}
\paragraph{Algorithmic tractability of the SYK model} 
The SYK model has been considered as a possible candidate benchmark for a quantum advantage in simulation in recent years~\cite{Garcia_Qsim,Babbush_Qsim,luo_qsim,jafferis_qsim}. Evidence supporting both quantum and classical tractability of the SYK model has since emerged.~\cite{hastings2022optimizing} gave an efficient quantum algorithm that prepares a state whose energy is within a constant factor of the ground state energy of the $k=4$ SYK model. At the same time, they showed that Gaussian (free-fermionic) states, a natural efficient classical ansatz, can only attain a vanishing fraction of the optimum.~\cite{basso2024optimizing} subsequently obtained a constant-factor approximation to the ground-state energy~\cite{BasuKM26} through a simple dissipative quantum algorithm, in a framework covering random local Hamiltonians beyond SYK. More recently,~\cite{CaiZZ26} construct a quantum Gibbs sampler for the $k=4$ model which mixes in polynomial time at sufficiently high temperatures.

On the classical side, it was very recently shown that any local thermal expectation value of an SYK Gibbs state can be estimated in polynomial time~\cite{SchmidhuberZ26}. These algorithmic results are complemented by an increasingly complete structural picture. The SYK model was shown not to be glassy, in the sense that its annealed and quenched free energies coincide~\cite{anschuetzStronglyInteractingFermions2025}, thereby ruling out the type of landscape obstruction that typically underlies algorithmic hardness in random optimization problems~\cite{Gamarnik_2021}. Furthermore,~\cite{gamarnik2026free,he2026spectraledgequarticsyk,SchmidhuberZ26} characterize the limiting free energy of SYK at any real inverse-temperature $\beta > 0$, as a consequence precluding a first-order phase transition for all $\beta > 0$.

Taken together, the SYK model appears far more tractable than its reputation as chaotic and entangled system would suggest. The mathematical techniques behind these results are, however, strikingly heterogeneous, each tailored to the specific algorithmic task at hand and often relying on particular properties of that task. This raises a natural question: is there a single unifying mathematical principle from which these algorithmic successes follow in a more principled manner?

\paragraph{Zero-Freeness as a unifying principle}\label{par:zerofree} In light of these advances, it is pertinent to revisit the Lee-Yang theory of partition function zeros~\cite{LY52,LY52_0,fisher1967theory} as a candidate for the algorithmic tractability for quantum states at equilibrium. To see why, let $H= H_n$ be a $k-$local Hamiltonian on $n$ particles and let $Z(z)$ be its partition function. Suppose that for some $\beta > 0$ there is a $r\geq \Omega_n(1)$ such that 
\[
  Z(z)\neq 0 \qquad\text{for all } z\in\C \text{ in the $r$-neighborhood }
  \operatorname{dist}\bigl(z,[0,\beta]\bigr)\le r.
\]
This has consequences of three kinds:
\begin{enumerate}[(1)]
    \item \textbf{Thermodynamics}. The free-energy density, 
    \[
    f(z) = \frac{1}{n}\log Z_n(z)
    \]
    is analytic and uniformly bounded on the $r-$neighborhood of $[0,\beta]$. As a result, there is no phase transition in this temperature range. 
    \item \textbf{Classical Algorithms}. Since computing the partition function at $\beta = 0$ is trivial, Barvinok's interpolation method~\cite{Barvinok16,PatelR22,HarrowMS20} applies and a controlled analytic continuation of $\log Z$ of degree $m=O_{\beta/r}(\log(n/\eps))$ approximates $\log Z(\beta)$ to additive error $\eps$. As a result, the free energy is computed deterministically in time $n^{O_{\beta/r}(k\log(n/\eps))}$. If the same zero-freeness holds for the perturbed Hamiltonians $H+\lambda \mathfrak{O}$, where $\mathfrak{O}$ is a local observable and $|\lambda|$ is small, differentiating in $\lambda$ gives the thermal expectation $\langle \mathfrak{O}\rangle_\beta$ to additive error $\eps$ in the same time~\cite{HarrowMS20}; zero-freeness also implies exponential decay of correlations in the Gibbs state~\cite{HarrowMS20} for geometrically local systems.
    \item \textbf{Sampling.} For classical spin systems, suitable zero-freeness conditions imply optimal $O(n\log n)$ mixing of Glauber dynamics on bounded-degree graphs~\cite{chen2023optimalmixingglauberdynamics,Chen_2024}. For quantum Hamiltonians we are not aware of a direct analogue of this route. The regimes in which quantum Gibbs samplers~\cite{chen2025efficientexactnoncommutativequantum} are proven to mix rapidly, namely sufficiently high temperature on lattices~\cite{Rouz__2026} and, very recently, for all-to-all $k$-local Hamiltonians~\cite{bergamaschi2026fastmixingalltoallquantum}, are nonetheless high-temperature regimes of the kind in which cluster expansions converge, and a convergent cluster expansion is the standard source of a zero-free disc~\cite{HarrowMS20,Mann_2021}. Whether a zero-free \emph{strip} can be converted into rapid mixing of a quantum Gibbs sampler at every constant $\beta$, as it can be classically, is an intriguing open direction.
\end{enumerate}

It is against this backdrop that we view \cref{thm:annealed-syk-lee-yang,thm:sykstrip}. These results show that the doubly scaled SYK has no phase transition at any finite temperature, and through Barvinok's method yield algorithms for computing (e.g.) free energy that are exponentially faster than exact diagonalization.\footnote{Owing to the doubly scaled limit, the problem description itself is not polynomial in $n$} \cref{thm:sykstrip} can also potentially lead towards efficient samplers for the Gibbs state: Indeed, the Gibbs samplers of \cite{CaiZZ26} at sufficiently high temperatures were preceded by a result \cite{zlokapaRigorousQuasipolynomialtimeClassical2026} showing that the SYK partition function was zero-free on the disc $\{\beta\in\C:|\beta|\leq O_k(1)\}$. In the double-scaled regime, \cref{thm:sykstrip} is a strong complement of the zero-free disc of~\cite{zlokapaRigorousQuasipolynomialtimeClassical2026} for finite $k$, extending the disc of constant radius to a strip of constant width around the real axis for every constant $\beta > 0$.

Thus, we hope that investigating the zeros of the $k$-SYK partition function starting from $k\sim\sqrt n$ (what we currently handle) all the way down to $k = 4$ will clarify our understanding of algorithmic tractability of SYK. More broadly, our techniques suggest a general approach to establishing Lee-Yang type theorems for disordered systems beyond SYK. Indeed, the twisted-bosonic construction of \cite{BasuKM26} expresses disorder-averaged moments through a deterministic operator derived from the \emph{sign-matrix} of the Hamiltonian operators~\footnote{For $H= \sum_{i=1}^m g_i A_i$, with $A_i A_j = \varepsilon_{ij}A_jA_i$, $g_i \sim \cN(0,1)$ i.i.d., the sign matrix is $\cE := m^{-1}\cdot (\varepsilon_{ij})_{i,j}$.}. When this sign matrix has an approximately rank-one structure, its dominant eigenvector gives rise to an effective $q$-deformed oscillator, with deviations from this collective mode controlled by the rest of the spectrum. Combined with our quantitative root-location result (\cref{thm:nearest-root}) and partition-function comparison (\cref{lem:partfntransfer}), this suggests a powerful recipe for proving zero-freeness in many regimes of disordered systems. An interesting direction is therefore to characterize which other disordered models admit such a sign-matrix representation and an approximately $q$-deformed description.

\paragraph{Connections between $q$-deformed oscillators and the $q$-Gaussian Distribution}\label{para:qgaussian-mgf}
\cref{thm:moment-function-imaginary-rooted} also has a purely mathematical interpretation in terms of \emph{Moment Generating Functions of the $q$-Gaussian distributions}. \cite{bozejko1991example,bozejko1997q} showed that the $q$-matching moment $M_\ell(q)$ is the $(2\ell)^{\mathrm{th}}$ moment of the \emph{$q$-Gaussian distribution $\nu_q$}, i.e. $M_\ell(q) = \int_{\R}x^{2\ell}\nu_q(\operatorname{d}\!x)$ for all $\ell\geq 0$. Consequently 
\[\cL_q(z) = \sum_{\ell = 0}^\infty M_\ell(q)\frac{z^{2\ell}}{(2\ell)!} = \int_{\R}e^{zx}\nu_q(\operatorname{d}\!x)\]
can also be viewed as the \emph{Moment Generating Function (MGF)} of the distribution $\nu_q$, and thus \cref{thm:moment-function-imaginary-rooted} can be interpreted as saying that 
\[\text{All roots of the MGF of }\nu_q\text{ are purely imaginary}.\]
For $q = 1$, $\nu_1$ is the usual $\cN(0, 1)$ standard Gaussian, and thus $\cL_1(z) = e^{z^2/2}$. For $q = -1$, $\nu_q = \frac{1}{2}(\delta_{-1} + \delta_{1})$ is the \emph{Rademacher distribution}, and thus $\cL_{-1}(z) = \cosh(z)$, and consequently \cref{thm:moment-function-imaginary-rooted} can be verified directly for $q = -1$.

For $q\in(-1, 1)$, $\nu_q$ is a distribution supported on $[-\frac{2}{\sqrt{1-q}}, \frac{2}{\sqrt{1-q}}]$ whose density function is given as~\cite[Theorem~1.10]{bozejko1997q}
\[\nu_q(\operatorname{d}\!x):= \frac{\sqrt{1 - q}}{\pi}\sin(\theta)\cdot\left(\prod_{k = 1}^\infty(1-  q^k)|1-q^ke^{2\iota\theta}|^2\right)\operatorname{d}\!x,\]
where $x$ is parametrized as $x = \frac{2}{\sqrt{1-q}}\cos(\theta)$ for $\theta\in[0, \pi]$, where the parametrization ensures that $\Var(\nu_q) = 1$ for all $q\in[-1, 1]$.\footnote{Note that this implies that $M_\ell(q)\geq 1$ for all $q\in[-1, 1],\ell\geq 0$: Indeed, $M_\ell(q) = \int_\R x^{2\ell}\nu_q(\operatorname{d}\!x)\overset{\text{Jensen}}{\geq}\left(\int_\R x^{2}\nu_q(\operatorname{d}\!x)\right)^\ell = \Var(\nu_q)^\ell = 1$} $\nu_q$ is particularly notable for $q = 0$, where it becomes the \emph{semi-circular distribution} 
\[\frac{1}{2\pi}\sqrt{4-x^2}\cdot\1_{[-2, 2]}(x)\operatorname{d}\!x,\]
which plays the role of the Gaussian distribution in \emph{free probability} (see~\cite{NicaS06} for a very nice introduction to free probability). The $q$-Gaussian distribution $\nu_q$ thus interpolates between three very important and well-studied distributions in probability theory: The Rademacher distribution for $q = -1$, the semi-circular distribution for $q = 0$, and the Gaussian distribution for $q = 1$.

Furthermore, as a consequence of the theory of free probability, $\cL_0$ admits a description in terms of the \emph{modified Bessel function} $I_1(\cdot)$, i.e. 
\[\cL_0(z) = \frac{I_1(2z)}{z}\]
for all $z\in\C$. One may then appeal to the theory of Bessel functions~\cite{Watson62,NIST:DLMF} to directly verify \cref{thm:moment-function-imaginary-rooted} at $q=0$.


\subsection{Technical Overview}\label{sec:techov}

\paragraph{$q$-deformed Matching Polynomials} We now describe the ideas which go into proving \cref{thm:moment-function-imaginary-rooted}. It suffices to show that all roots of the function $\varphi_q(z):= \cL_q(\iota z)$ are real. We in fact show something stronger, which is that $\varphi_q$ is the (locally uniform) limit of real-rooted polynomials, which by Hurwitz's theorem from complex analysis (see \cref{item:hurwitz} of \cref{fact:complex-analysis}) implies $\varphi_q$ is also real-rooted. The class of entire functions which are the limit of real-rooted polynomials is known as the \emph{Laguerre-P\'olya class}~\cite{Levin80}. By studying the \emph{order of growth} of the function $z\mapsto\varphi_q(z)$,\footnote{For an entire function $f$, we define its order of growth $\rho(f)$ as $\limsup_{r\to\infty}(\log\log M_f(r))/\log r$, where $M_f(r):= \sup_{|z| = r}|f(z)|$. We have $\rho(\varphi_q) = 1$ for $q\in[-1, 1)$, and $\rho(\varphi_1) = \rho(e^{-z^2/2}) = 2$.} and the \emph{Hadamard product formula} (see \cref{item:hadamard} of \cref{fact:complex-analysis}), we further show that 
\[\varphi_q(z) = \prod_{k = 1}^\infty\left(1 - \frac{z^2}{t_k(q)^2}\right),\]
where $\pm t_1(q), \pm t_2(q), \ldots$ are the roots of $\varphi_q$. Applying the rotation $z\mapsto\iota z$ then yields that $\cL_q(z) = \prod_{k = 1}^\infty\left(1 + \frac{z^2}{t_k(q)^2}\right)$. 

Thus consider the \emph{$q$-deformed matching polynomial} defined as 
\begin{equation}\label{eq:mudef}
    \mu_{n, q}(z):= \sum_{\ell = 0}^{\lfloor n/2\rfloor}(-1)^\ell\binom{n}{2\ell} M_\ell(q)z^{n-2\ell}.
\end{equation}
Using some basic analysis it is not difficult to see that
\[\varphi_q(z) = \lim_{n\to\infty}\left(\frac{z}{n}\right)^n\cdot\mu_{n, q}\left(\frac{n}{z}\right)\]
locally uniformly on $\C$. We now claim that $\mu_{n, q}$ is real-rooted\footnote{A polynomial $p$ with real coefficients is defined to be \emph{real-rooted} if all of its roots are real, or if it is a constant polynomial.} for all $n\geq 0, q\in[-1, 1]$. 
\begin{restatable}[$\mu_{n, q}$ is real-rooted]{theorem}{maintheorem}\label{thm:main-thm}
    For any real $q\in[-1, 1]$, and any integer $n\geq 0$, the polynomial $\mu_{n, q}$ is real-rooted. Furthermore, for $n\geq 2$, all roots of $\mu_{n, q}$ are distinct. 
\end{restatable}
\begin{remark}\label{remark:qgeq1notrr}
We also prove that real-rootedness is strictly constrained to $q$ with $|q| \leq 1$. Concretely, for any real $q$ with $|q| > 1$, there exists an integer $n = n(q)$ for which $\mu_{n, q}$ is not real-rooted.
\end{remark}
For $q = 1$, $\mu_{n, 1} = \He_n$ equals the (probabilist's) \emph{Hermite polynomial}, all of whose roots are real~\cite{szeg1939orthogonal}. Alternatively, $\mu_{n, 1}$ can also be viewed as the \emph{Heilmann-Lieb matching polynomial} of the complete graph $K_n$~\cite{HeilmannL72}, and thus for $q = 1$, the real-rootedness of $\mu_{n, 1}$ can be viewed as a special case of the \emph{Heilmann-Lieb matching theorem}, which says that the matching polynomial of \emph{any} graph is real-rooted. As with the imaginary-rootedness of $\cL_q$ in our paper, the real-rootedness of the Heilmann-Lieb matching polynomials was used by Heilmann and Lieb to show that the partition function of the \emph{monomer-dimer model} on any graph was imaginary rooted (in the variable representing monomer-activity). See \cref{thm:main-thm-vertex-weighted,prop:edge-weighted-failure} for further generalizations of \cref{thm:main-thm} along the lines of the Heilmann-Lieb theorem.

The real-rootedness of $\mu_{n, q}$ also has some purely combinatorial consequences. For any $q\in[-1, 1]$, write 
\[c_\ell(q):= \E_{\pi\sim\cP_2(2\ell)}q^{\Cr(\pi)} = \frac{M_\ell(q)}{(2\ell - 1)!!}. \]
Then by \emph{Newton's inequality} (see \cref{fact:newton}) applied to $\mu_{n, q}$, we obtain that:
\begin{restatable}{theorem}{logconcavity}\label{thm:log-concave}
    For any $q\in[-1, 1]$, the sequence $(c_0(q), c_1(q), c_2(q), \ldots)$ is \textbf{log-concave}, i.e. for any $\ell\geq 1$, we have $c_\ell(q)^2\geq c_{\ell + 1}(q)\cdot c_{\ell - 1}(q)$. 
\end{restatable}
\begin{remark}
    We can in fact obtain an entire family of inequalities amongst the coefficients $c_\ell(q)$ using the theory of \emph{P\'olya frequency sequences}; see \cref{thm:polya-frequency}.
\end{remark}

\paragraph{Real-Rootedness and Log-Concavity}
There is a deep and fascinating interplay between inequalities in enumerative combinatorics and the geometry of polynomials, beginning from Newton~\cite{Newton1707,Branden14}, who observed that if all roots of a polynomial $p(x) = \sum_{t = 0}^n (-1)^{n - t} a_t x^t$ are non-negative real numbers, then the coefficients $(a_0, a_1, \ldots, a_n)$ are \emph{ultra-log-concave}, i.e. for all $1\leq t < n$, we have
\[\left(\frac{a_t}{\binom{n}{t}}\right)^2\geq\frac{a_{t - 1}}{\binom{n}{t - 1}}\cdot\frac{a_{t + 1}}{\binom{n}{t + 1}}.\]
One can often associate polynomials to combinatorial objects; consequently real-rootedness for such polynomials immediately translates into combinatorial inequalities~\cite{Stanley89,Brenti92,Wagner09,Branden14}. It should be noted though that log-concavity is a more general phenomenon than real-rootedness: For instance, Newton's inequalities are \emph{not} a sufficient condition for real-rootedness. See~\cite{Huh23,Vinzant26} for a survey of the connections between log-concavity and the geometry of polynomials. 


We now move on to explaining our proof of \cref{thm:main-thm}.

\paragraph{A Differential Equation for $\mu_{n, q}$} We now discuss the main technical ingredient in proving \cref{thm:main-thm}, which is a differential equation involving $\mu_{n, q}$:
\begin{restatable}[Differential Equation Governing $\mu_{n, q}$]{theorem}{munqpde}\label{thm:munq-pde}
    Write $b_n(q):= 4 + (2n - 3)(1 - q), c_n(q):= n(4 + (n - 2)(1 - q))$. Then, as a polynomial identity in $(q, z)$, we have
    \begin{equation}\label{eq:munq-pde}
        8q(1 - q)\cdot \partial_q\mu_{n, q} = 4\partial^2_z\mu_{n, q} + \left((1 - q)z^2\cdot\partial^2_z - b_n(q)z\cdot\partial_z + c_n(q)\right)\mu_{n, q}.
    \end{equation}
\end{restatable}
One may also obtain a similar differential equation for $\cL_q$ itself. Write $\mathscr{L}_q(z):= \cL_q(\sqrt{1 - q}\cdot z)$. 
\begin{restatable}{corollary}{scrlpde}\label{cor:scrl-pde}
    For $q\in[-1, 1)$, we have
    \begin{equation}\label{eq:lq-pde}
        8q\partial_q \mathscr{L}_q = \left(z^2\partial_z^2 + 3z\partial_z - 4z^2\right)\mathscr{L}_q.
    \end{equation}
\end{restatable}
\cref{cor:scrl-pde} is also proved in a different way in~\cite[Section~6.3]{AlfinitoB26}, using the theory of modular forms from number theory. 

Once we have \cref{thm:munq-pde}, the real-rootedness of $\mu_{n, q}$ follows by tracking the evolution of the roots of $\mu_{n, q}$ as $q$ varies. See~\cite{Tao17} for an excellent introduction into the general technique of studying the roots of a polynomial by analyzing the differential equation it obeys.

More precisely, for $q = 1$, by properties of Hermite polynomials (see~\cref{fact:hermite-simple-real}), we know that $\mu_{n, 1}$ has distinct real roots, say $\lambda_1(1) < \cdots < \lambda_n(1)$. If we denote the roots of $\mu_{n, q}$ as $\{\lambda_i(q)\}_{i\in[n]}$, then simply by the distinctness of the real roots of $\mu_{n, 1}$, we obtain that $\{\lambda_i(q)\}_{i\in[n]}$ remain distinct and real for $q$ in some open neighborhood of $1$. Now, since $\mu_{n, q}$ has real coefficients, all its (hypothetical) non-real roots must occur in conjugate pairs. Furthermore, since $\lambda_i(q)$ vary smoothly with $q$ in a neighborhood of $1$, using some complex analysis it can be shown that the only way we can have non-real roots is if a \emph{collision} occurs at some $q^* < 1$, i.e. we have $\lambda_i(q^*) = \lambda_{i + 1}(q^*)$ for some $i\in[n - 1]$, and then the roots split off into the complex plane, i.e. $\lambda_i(q^* - \eps) = \bar{\lambda_{i + 1}(q^* - \eps)}\notin\R$ for some small enough $\eps > 0$. 

We claim that no such collision can occur: Indeed, using \cref{eq:munq-pde}, we control the evolution of $\{\lambda_i(q)\}_{i\in[n]}$ and show that the \emph{repulsion} between the roots \emph{stays bounded} as $q$ decreases from $1$ (see \cref{lem:root-evolution}), where for any $i\in[n]$, we define the repulsion experienced by $\lambda_i(q)$ to be $\sum_{j\neq i}(\lambda_i(q) - \lambda_j(q))^{-1}$. Note that bounded repulsion implies that the roots remain distinct, and hence real, as desired. 

\paragraph{Real-Stability Preserving Operators} There is a more abstract way to prove the real-rootedness of $\mu_{n, q}$, using the general theory of \emph{real-stability preserving operators}~\cite{BorceaB09a,BorceaB09b}. More precisely,~\cite{BorceaB09a} characterizes differential operators which preserve real-rootedness. Using \cref{thm:munq-pde}, we can express $\mu_{n,q}$ as $E_{n,q}\mu_{n,1}$ for some differential operator $E_{n,q}$, at which point we apply the criteria of~\cite{BorceaB09a} to $E_{n,q}$, thus concluding the real-rootedness of $\mu_{n,q}$. See~\cref{subsec:alternate-proof} for further details.

There is a rich history of establishing real-rootedness of a polynomial by expressing it as the output of a real-rootedness preserving operator applied to a polynomial whose real-rootedness is apparent. Perhaps the most celebrated example of this technique is Marcus, Spielman and Srivastava's proof that the expected characteristic polynomial of a random $2$-lift of a graph is real-rooted~\cite{MarcusSS15a}. 

Furthermore, just as we are able to exploit \cref{thm:munq-pde} to control the position of the roots $\pm\iota t_1(q)$ of $\cL_q$ (see \cref{thm:nearest-root}), real-rootedness preserving operators can also be used to bound the roots themselves; for instance, in the resolution of the \emph{Kadison-Singer conjecture} by~\cite{MarcusSS15}, they use a multivariate barrier function to track the effect of the (real-stability preserving) operator $\Id-\partial_{z_i}$ on the largest root of a certain ``mixed characteristic polynomial''. Even before this,~\cite{BatsonSS14} studied the evolution of the polynomial $z^n$ under the application of the (real-rootedness preserving) \emph{Laguerre operator} $\Id-\alpha\partial_z$ to obtain ``twice-Ramanujan'' sparsifiers for any graph. This is analogous to our proof in \cref{subsec:alternate-proof} where we obtain $\mu_{n,q}$, up to rescaling, by applying the exponential of a second-order differential operator $D_n$ to $z^n$.\footnote{More precisely, we have the operator $e^{-tD_n}$ being applied to $z^n$, where $D_n:=(4+z^2)\partial_z^2-(2n+1)z\partial_z+n(n+2)$, and $t:= \frac{1}{8}\log\frac{1}{q}$ for $q\in(0,1)$. We then write $e^{-tD_n} = \lim_{m\to\infty}(\Id - tD_n/m)^m$, and thus our $\mu_{n,q}$ can be viewed as the iterated application of $\Id - hD_n$ on $z^n$ for small enough $h> 0$.}

There is also a direct relation between the ``root repulsion'' $\sum_{j\neq i}(\lambda_i -\lambda_j)^{-1}$ (see~\cref{lem:root-collision,lem:root-evolution}) we consider, the ``barrier function methods'' employed by~\cite{BatsonSS14,MarcusSS15}: For any polynomial $p$ with roots $\lambda_1<\cdots<\lambda_n$, the \emph{logarithmic derivative} $B(z):= \frac{\operatorname{d}}{\operatorname{d}\!z}\log p(z) = \sum_i(z-\lambda_i)^{-1}$ is a very important quantity associated to the polynomial. \cite{BatsonSS14} uses $B(z)$ as its barrier function to analyze a sparsification algorithm; \cite{MarcusSS15} uses a multivariate analog of $B(z)$ to resolve the Kadison-Singer conjecture, and $B(z)$ also plays a very important role in the theory of finite free convolutions introduced in~\cite{MarcusSS22}. In our own argument, we have $\sum_{j\neq i}(\lambda_i- \lambda_j)^{-1} = \lim_{z\to\lambda_i}(B(z) - (z-\lambda_i)^{-1})$, i.e. the ``repulsion'' we consider is also $B(z)$, with its pole at $z=\lambda_i$ removed. 

\paragraph{A Differential Recurrence for $M_\ell(q)$ using the Touchard--Riordan formula} Thus establishing \cref{thm:munq-pde} will complete our proof outline for \cref{thm:moment-function-imaginary-rooted}. The key technical ingredient in the proof of \cref{thm:munq-pde} is the \emph{Touchard-Riordan formula} (see \cref{fact:touchard-riordan}), which is a combinatorial description of the moments $M_\ell(q)$. Differentiating the Touchard Riordan formula w.r.t. $q$ yields the following differential recurrence for $M_\ell(q)$ (see \cref{lem:mlq-recurrence})
\[2q(1 - q)M'_\ell(q) = \ell(2 + (\ell - 1)(1 - q))M_\ell(q) - 2\ell(2\ell - 1)M_{\ell - 1}(q),\]
and manipulating this recurrence then yields \cref{thm:munq-pde}.

We now expound on the mathematical techniques that go into proving \cref{thm:nearest-root,thm:annealed-syk-lee-yang,thm:sykstrip}:
\paragraph{Tutte polynomials and \cref{thm:nearest-root}} The multivariate Tutte polynomial~\cite{Sokal_2005} of a weighted graph $G = G(V, E, \bm w)$,
\begin{equation}\label{eq:graph-polynomial-intro}
 P_{G}(t;\bm w):=\sum_{F\subseteq E}t^{\conn(F)}
                   \prod_{e\in F}w_e, 
\end{equation}
plays a key role as a common graph-theoretic framework for showing the nearest-root result of \cref{thm:nearest-root} as well as the SYK results of \cref{thm:annealed-syk-lee-yang,thm:sykstrip}. Here $\conn(F)$ counts the connected components of the spanning subgraph $(V,F)$ including isolated vertices. Its connected coefficient, 
\begin{equation}
    C_G(\bm w) := [t] P_G(t;\bm w) = \sum_{\substack{F\subseteq E \\ (V,F) \text{ connected}} } \prod_{e\in F}w_e,
\end{equation}
which sums over connected spanning subgraphs. For connected $G$, fixing every edge's weight to $w_e := q-1$ and denoting $C_G(\bm w) = C_G(q-1)$ gives, 
\[
C_G(q-1) = (q-1)^{|V|-1}T_G(1,q), \quad T_G(1,q) := \sum_{\substack{F\subseteq E \\ (V,F) \text{ connected}}}(q-1)^{|F|-|V|+1}.
\]
The connection to the oscillator comes through \emph{cumulants} (of the $q$-Gaussian distribution). Indeed, write
\[
\log\cL_q(z) = \sum_{m\geq 1}\kappa_{2m}(q) \frac{z^{2m}}{(2m)!}.
\]
Then the cumulant formula of Josuat-Verg\`{e}s~\cite{VergesCumulants} reads,
\begin{equation}\label{eq:JVcumulant}
    \frac{\kappa_{2m}}{(q-1)^{m-1}} = \sum_{\substack{\pi\in\cP_2(2m)\\ {G_\pi\text{ connected}}} } T_{G_{\pi}}(1,q).
\end{equation}
For any $m \geq 1$ define,
\[
b_m(q) := \frac{m}{(2m)!} \sum_{\substack{\pi \in \cP_2(2m) \\ G_\pi \text{ connected}}} T_{G_\pi}(1,q)
\]
where $G_\pi$ is the crossing graph of the matching $\pi\in\cP_2(2m)$.\footnote{For any $\pi = \{e_1, \ldots, e_m\}\in\cP_2(2m)$, the crossing graph $G_\pi$ has vertices $e_1,\ldots,e_m$, with $e_i,e_j$ being connected if they cross each other}
Combining the cumulant formula \cref{eq:JVcumulant} with our \cref{thm:moment-function-imaginary-rooted}, we get,  
\[
(1-q)^{m-1}b_m(q) = \sum_{j\geq 1}\frac{1}{t_j(q)^{2m}}.
\]
Thus locating the nearest root becomes a question about the growth of $b_m$ as $m\to\infty$. A careful analysis of $b_m$ then results in the proof of \cref{thm:nearest-root}.

\paragraph{Controlling the moments of SYK} The Tutte polynomials of \cref{eq:graph-polynomial-intro} also aid in a quantitative comparison between the SYK and oscillator moments. For a finite simple graph $G=(V,E)$, define,
\[
W_G(n,k) := \E_{S_v \sim \Unif\binom{[n]}{k}} \prod_{\{u,v\}\in E(G)} (-1)^{|S_u\cap S_v|}
\]
where the sets are chosen independently. Using Isserlis-Wick formula~\cite{Isserlis18,Wick50}, the moments of the SYK Hamiltonian can be expressed in terms of $W_G(n,k)$ for any $\ell \geq 1$:
\[
\E \tr H_{\syk}^{2\ell} = V_n^{\ell} \sum_{\pi\in\cP_2(2\ell)} W_{G_\pi}(n,k),
\]
where $G_\pi$ is the crossing graph of $\pi \in \cP_2(2\ell)$ and $V_n := \E_{\bm g} \tr H_{\syk}^2$ is the variance of the SYK Hamiltonian. Crucially, the oscillator moments have the same pairing expansion with $W_{G_\pi}$ replaced by $q_n^{\Cr(\pi)} = q_n^{|E(G_\pi)|}$, recall, 
\[
\bra{0}T_q^{2\ell}\ket{0} = M_\ell(q):= \sum_{\pi\in\cP_2(2\ell)}q^{\Cr(\pi)}
\]
To complete this comparison, we express $W_G$ as a ratio of multivariate Tutte polynomials on an auxilliary graph, and show that for sufficiently large $n$,
\[
\log W_G(n,k) = |E(G)| \log q_n + O\left(\frac{\ell^3k^3}{n^2}\right).
\]

\paragraph{Mapping SYK to $q-$oscillators}  We compare the annealed SYK partition function $Z_n(z)$ to a rescaled $q-$deformed function defined as:
\[
Q_n(z) := \cL_{q_n}(\sqrt{V_n} z)
\]
The zeroes of $Q_n$ are at $\pm \iota t_j(q_n)/\sqrt{V_n}$. An elementary bound on the variance $V_n$ gives $V_n(1-q_n) < 1$, so \cref{thm:nearest-root} implies
\[
\frac{t_j(q_n)}{\sqrt{V_n}} > \frac{\zeta_*}{\sqrt{V_n(1-q_n)}} > \zeta_*.
\]
Thus $Q_n$ has no zeros in the closed strip $|\Imm z| \leq \zeta_*$. We then bound the error between the annealed SYK $Z_n(z)$ and $q-$deformed partition functions $Q_n(z)$ in \cref{lem:partfntransfer}. In the doubly-scaled limit, $k^2/n \to \alpha\in(0,\infty)$, a careful analysis of \cref{lem:partfntransfer} suffices to show the annealed Lee-Yang result of \cref{thm:annealed-syk-lee-yang}. Upon accounting for the fluctuations of the quenched partition function $Z_n(z;\bm g)$ around $Z_n(z)$ (in~\cref{lem:sykmomentbound}) and again using \cref{lem:partfntransfer}, we then prove \cref{thm:sykstrip}.

\subsection*{Acknowledgements}

\paragraph{Human} A.B. thanks Joao Basso, Jake Hofgard, Kuikui Liu, Nikhil Srivastava, and Daniel Spielman for useful discussions. S.M. thanks Herman Verlinde for helpful discussions. A.B. and S.M. thank Pravesh Kothari for useful discussions and collaboration on related work. A.B. and S.M. also thank David Gamarnik for helpful comments, and pointing out interesting research directions.

A.B. thanks Joao Basso and Jake Hofgard for pointing out that the annealed $4$-SYK partition function has non-imaginary zeros (see~\cref{remark:non-imaginary-roots}). A.B. also thanks Joao Basso for pointing out numerous connections to Lee-Yang theory, the theory of real stability preserving operators, and pointing out that one may relate $q$-deformed oscillators with SYK using the Duhamel formula (see \cref{fact:duhamel}). A.B. thanks Nikhil Srivastava for pointing~\cite{Gribinski25} out to us. A.B. is very grateful to Joao Basso, Nikhil Srivastava and Daniel Spielman for their encouragement in writing the paper.

Part of this work was done while A.B. was attending the Summer School in High-Dimensional Probability and its Applications at Northwestern University in August 2026, and the Local to Global Principles Workshop at Simons Institute, Berkeley, in September 2026, and he would like to thank Northwestern University, the Simons Institute, and the workshop organizers for creating an intellectually stimulating environment. 

\paragraph{AI} We adopt the recommendation of the Simons Working Group for the AI Methodology disclosure.
\begin{flushleft}
\renewcommand{\arraystretch}{1.3}
\begin{tabular}{@{}l@{\qquad}c@{}}
\textbf{Activity} & \textbf{AI used} \\
Asking the research question & No \\
Coming up with the approach & No \\
Proof development & Yes \\
Writing and exposition & Yes \\
Checking for bugs & Yes \\
Other supporting tasks & Yes \\
\end{tabular}
\end{flushleft}

\begin{sloppypar}
\paragraph{Details}
The key technical statement which underlies most of the results in this paper is \cref{thm:main-thm}, which was proven in multiple sequential interactions with ChatGPT-5.6-Sol-Plus with significant technical input from the authors. The proof of \cref{thm:main-thm} was subsequently cleaned up and broken up into the intermediate statements~\cref{lem:mlq-recurrence,thm:munq-pde}. Once \cref{thm:main-thm} was established, \cref{thm:moment-function-imaginary-rooted,thm:main-thm-vertex-weighted,thm:log-concave,thm:polya-frequency} follow using standard (but non-trivial) connections in the literature; and the authors established those results themselves. Multiple sessions of ChatGPT-5.6-Sol and 6-Astra were then used to look for further connections and applications to our theorem; GPT pointed us out to \cite{Okuyama23}, which motivated the authors to formally state and prove \cref{thm:nearest-root}. Although~\cref{thm:nearest-root} admits a direct proof in terms of the cumulants of the $q$-Gaussian, a few conversations with GPT-6-Astra resulted in a more illuminating proof involving the theory of Tutte polynomials, which also included GPT-6-Astra pointing out~\cite{VergesCumulants} to us. Following up with subsequent interactions, the authors then realized that this machinery could also be generalized to prove statements about the doubly-scaled SYK Model itself, and a few more conversations with GPT-6-Astra (in which the authors prompted the broad proof strategy) yielded the proofs of \cref{thm:annealed-syk-lee-yang,thm:sykstrip}.

AI helped in writing the proofs of \cref{claim:gnq-local-conv,lem:bmq-facts,lem:graph-log,cl:sykproductbound,cl:sykcouplingdecay}. AI was also used to find references and typos in the manuscript. AI also helped generate the \emph{Leitfaden} for this paper.

Claude Opus 5.5 was used to organize the \LaTeX{} code, check for errors/typos, and prepare the \href{https://github.com/siddhant-midha/syk-roots}{supplementary numerics}.

\end{sloppypar}

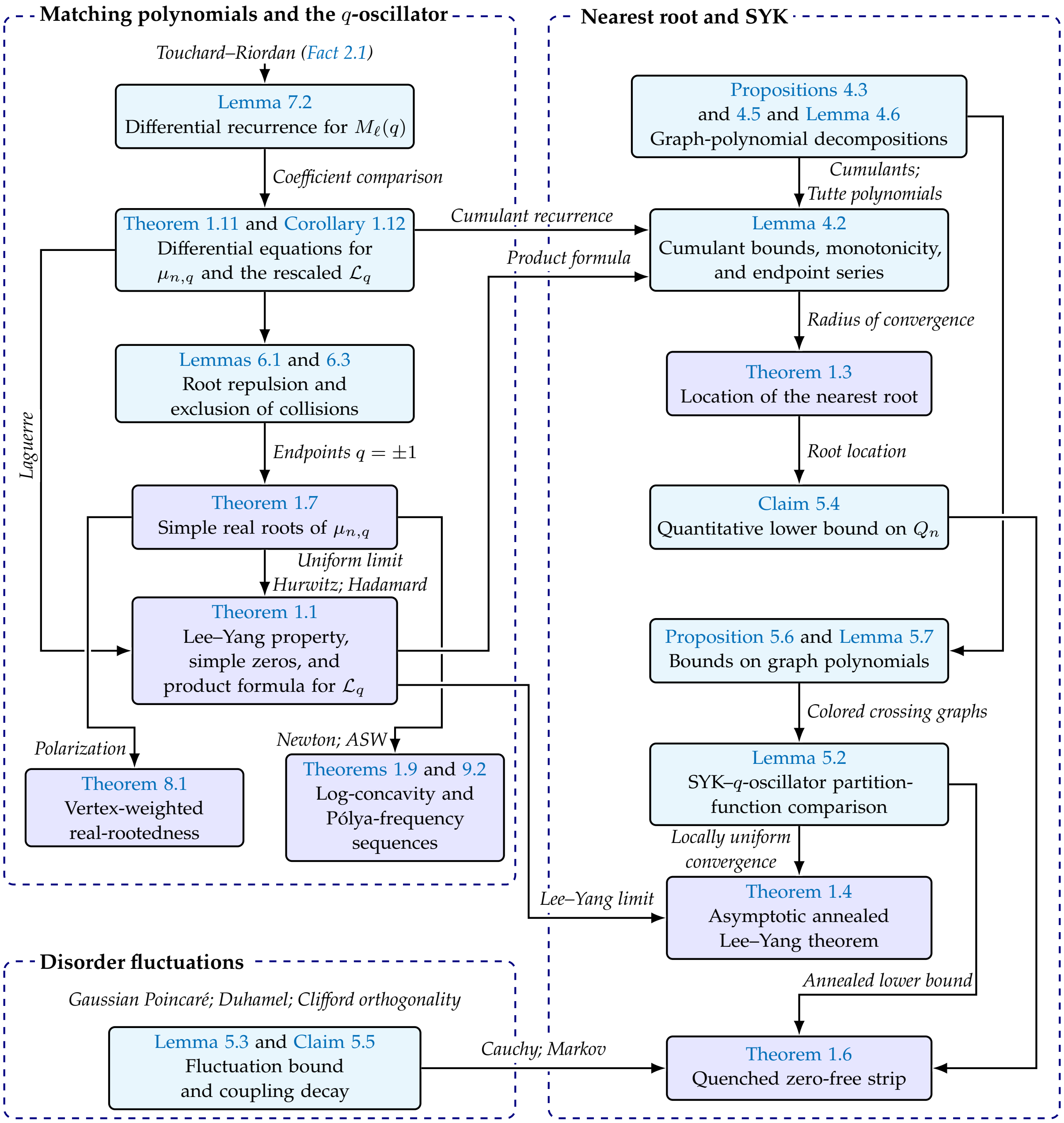
\begin{figure}[p]

\section*{A \emph{Leitfaden} (Guide) of our Main Results}

\centering
\scriptsize

\begin{adjustbox}{
  max totalsize={\textwidth}{0.80\textheight},
  center
}
\begin{tikzpicture}[x=1cm,y=1cm]

\tikzset{
  lemmanode/.style={
    draw,
    rounded corners=2.5pt,
    thick,
    align=center,
    fill=cyan!8,
    text width=4.25cm,
    minimum height=0.95cm,
    inner sep=3pt
  },
  resultnode/.style={
    draw,
    rounded corners=2.5pt,
    thick,
    align=center,
    fill=blue!10,
    text width=3.75cm,
    minimum height=0.95cm,
    inner sep=3pt
  },
  techlabel/.style={
    font=\scriptsize\itshape,
    align=center,
    fill=white,
    inner sep=1pt
  },
  dep/.style={
    -{Latex[length=2.2mm]},
    thick,
    preaction={draw=white,line width=2.8pt,-}
  },
  branchbox/.style={
    draw=blue!50!black,
    rounded corners=4pt,
    dashed,
    thick
  },
  branchlabel/.style={
    font=\bfseries\footnotesize,
    fill=white,
    inner sep=2pt
  }
}


\draw[branchbox] (-7.9,1.5) rectangle (-0.25,-11.5);
\node[branchlabel,anchor=west] at (-7.45,1.5)
  {Matching polynomials and the $q$-oscillator};

\draw[branchbox] (0.25,1.5) rectangle (7.9,-15.0);
\node[branchlabel,anchor=west] at (0.65,1.5)
  {Nearest root and SYK};

\draw[branchbox] (-7.9,-12.65) rectangle (-0.25,-15.0);
\node[branchlabel,anchor=west] at (-7.45,-12.65)
  {Disorder fluctuations};


\node[lemmanode] (recurrence) at (-4.0,0.0)
  {\cref{lem:mlq-recurrence}\\
   Differential recurrence for $M_\ell(q)$};

\node[lemmanode] (pde) at (-4.0,-2.0)
  {\cref{thm:munq-pde,cor:scrl-pde}\\
   Differential equations for\\$\mu_{n,q}$ and the rescaled $\mathcal L_q$};

\node[lemmanode] (repulsion) at (-4.0,-4.0)
  {\cref{lem:root-evolution,lem:root-collision}\\
   Root repulsion and\\exclusion of collisions};

\node[resultnode] (realroots) at (-4.0,-6.0)
  {\cref{thm:main-thm}\\
   Simple real roots of $\mu_{n,q}$};

\node[resultnode] (oscillator) at (-4.0,-8.0)
  {\cref{thm:moment-function-imaginary-rooted}\\
   Lee--Yang property,\\simple zeros, and\\
   product formula for $\mathcal L_q$};

\node[resultnode,text width=3.05cm] (weighted) at (-5.95,-10.35)
  {\cref{thm:main-thm-vertex-weighted}\\
   Vertex-weighted\\real-rootedness};

\node[resultnode,text width=3.05cm] (combinatorial) at (-2.05,-10.35)
  {\cref{thm:log-concave,thm:polya-frequency}\\
   Log-concavity and\\P\'olya-frequency sequences};

\draw[dep] (recurrence.south) --
  node[techlabel,right=2pt,midway] {Coefficient comparison}
  (pde.north);

\node[techlabel] (touchard) at (-4.0,0.95)
  {Touchard--Riordan (\cref{fact:touchard-riordan})};
\draw[dep] (touchard.south) -- (recurrence.north);

\draw[dep] (pde.south) -- (repulsion.north);

\draw[dep] (repulsion.south) --
  node[techlabel,right=2pt,midway] {Endpoints $q=\pm1$}
  (realroots.north);

\draw[dep] (realroots.south) --
  node[techlabel,right=2pt,midway] {Uniform limit\\Hurwitz; Hadamard}
  (oscillator.north);

\draw[dep] (pde.west) -- (-7.35,-2.0) --
  node[techlabel,rotate=90,above=2pt,midway] {Laguerre}
  (-7.35,-8.0) -- (oscillator.west);

\draw[dep] (realroots.west) -- (-6.65,-6.0) -- (-6.65,-9.15) --
  (-5.95,-9.15) -- node[techlabel,left=2pt,midway] {Polarization}
  (weighted.north);

\draw[dep] (realroots.east) -- (-1.35,-6.0) -- (-1.35,-9.15) --
  (-2.05,-9.15) -- node[techlabel,left=2pt,midway] {Newton; ASW}
  (combinatorial.north);


\node[lemmanode,text width=4.8cm] (graphs) at (4.0,0.0)
  {\cref{prop:pottspoly,prop:familydecomposition,prop:spanningtreedecomp}\\
   Graph-polynomial decompositions};

\node[lemmanode] (cumulants) at (4.0,-2.0)
  {\cref{lem:bmq-facts}\\
   Cumulant bounds, monotonicity,\\and endpoint series};

\node[resultnode] (nearest) at (4.0,-4.0)
  {\cref{thm:nearest-root}\\
   Location of the nearest root};

\node[lemmanode] (product) at (4.0,-6.0)
  {\cref{cl:sykproductbound}\\
   Quantitative lower bound on $Q_n$};

\node[lemmanode] (graphbounds) at (4.0,-8.0)
  {\cref{prop:connpolybound,lem:graph-log}\\
   Bounds on graph polynomials};

\node[lemmanode] (transfer) at (4.0,-10.0)
  {\cref{lem:partfntransfer}\\
   SYK--$q$-oscillator partition-function comparison};

\node[resultnode] (annealed) at (4.0,-12.0)
  {\cref{thm:annealed-syk-lee-yang}\\
   Asymptotic annealed Lee--Yang theorem};

\node[resultnode] (quenched) at (4.0,-14.25)
  {\cref{thm:sykstrip}\\
   Quenched zero-free strip};

\draw[dep] (graphs.south) --
  node[techlabel,right=2pt,midway] {Cumulants;\\Tutte polynomials}
  (cumulants.north);

\draw[dep] (cumulants.south) --
  node[techlabel,right=2pt,midway] {Radius of convergence}
  (nearest.north);

\draw[dep] (nearest.south) --
  node[techlabel,right=2pt,midway] {Root location}
  (product.north);


\draw[dep] (graphs.east) -- (7.05,0.0) -- (7.05,-8.0)
  -- (graphbounds.east);

\draw[dep] (graphbounds.south) --
  node[techlabel,right=2pt,midway] {Colored crossing graphs}
  (transfer.north);

\draw[dep] (transfer.south) --
  node[techlabel,left=2pt,midway] {Locally uniform\\convergence}
  (annealed.north);

\draw[dep] (product.east) -- (7.55,-6.0) -- (7.55,-14.25)
  -- (quenched.east);

\draw[dep] (transfer.east) -- (6.65,-10.0) -- (6.65,-13.15)
  -- node[techlabel,above=2pt,midway] {Annealed lower bound}
  (4.0,-13.15) -- (quenched.north);


\draw[dep]
  (oscillator.east) -- (-0.65,-8.0) -- (-0.65,-2.4)
  -- node[techlabel,above=2pt,midway] {Product formula}
  ([yshift=-4mm]cumulants.west);

\draw[dep]
  ([yshift=3mm]pde.east)
  -- node[techlabel,above=2pt,midway] {Cumulant recurrence}
  ([yshift=3mm]cumulants.west);
  
\draw[dep]
  ([yshift=3mm]oscillator.south east) -| (-0.05,-12.0)
  -- node[techlabel,above=2pt,midway] {Lee--Yang limit}
  (annealed.west);


\node[lemmanode,text width=4.45cm] (fluctuations) at (-4.0,-14.25)
  {\cref{lem:sykmomentbound,cl:sykcouplingdecay}\\
   Fluctuation bound and coupling decay};

\node[techlabel] at (-4.0,-13.25)
  {Gaussian Poincar\'e; Duhamel; Clifford orthogonality};

\draw[dep] (fluctuations.east) --
  node[techlabel,above=2pt,midway] {Cauchy; Markov}
  (quenched.west);

\end{tikzpicture}
\end{adjustbox}

\caption{Arrows indicate the principal proof
dependencies}
\label{fig:leitfaden}
\end{figure}
\newpage
\section{Preliminaries}

Throughout the paper we adopt the convention that for any integer $k\geq 0$, $\binom{k}{t} := 0$ if $t < 0$ or $t > k$. Recall that, for $\ell\geq 1$ one has:
\[M_\ell(q):= \sum_{\pi\in\cP_2(2\ell)}q^{\Cr(\pi)},\] 
The Touchard-Riordan formula relates the $M_\ell$ polynomials to the so-called \emph{ballot numbers} $B_{\ell ,r}$.
\begin{fact}[Touchard-Riordan Formula (\cite{touchard1952probleme,riordan1975distribution})]\label{fact:touchard-riordan}
    For any $\ell\geq 0$ and $q\in\R$, we have 
    \[(1 - q)^\ell M_\ell(q) = \sum_{r = 0}^\ell(-1)^r B_{\ell, r}\cdot q^{r(r + 1)/2},\]
    where 
    \[B_{\ell, r}:= \binom{2\ell}{\ell - r} - \binom{2\ell}{\ell - r - 1} = \frac{2r + 1}{\ell + r + 1}\binom{2\ell}{\ell - r}.\]
\end{fact}
\begin{remark}
    There is a very intriguing restatement of \cref{fact:touchard-riordan} in terms of Hilbert spaces (see also \cite{Szabłowski09,SYK_Berkooz2}). For $q\in(-1, 1)$, write $X_q:= \sqrt{1 - q}\cdot T_q$. Also define $\ket{\Omega_q}:= \sum_{r\geq 0}(-1)^rq^{r(r + 1)/2}\ket{2r}\in\ell_2(\Z_{\geq 0})$. Then for any $\ell\geq 0$ we have
    \[\bra{0}X_q^{2\ell}\ket{0} = \bra{0}X_0^{2\ell}\ket{\Omega_q},\]
    i.e. the entire $q$-deformation in $T_q$ can be transferred from the operator $T_q$ to a particular vector $\ket{\Omega_q}$. 
\end{remark}

\parhead{Polynomials} We say that a (univariate) polynomial $p$ is \emph{simple} if all of its roots are distinct. Also recall that a polynomial $p$ is \emph{real-rooted} if all of its roots are real, and we also declare constant polynomials to be real-rooted. In particular, the zero polynomial is real-rooted.
\begin{fact}\label{fact:hermite-simple-real}
    The (probabilist's) Hermite polynomials $\He_n(x) := \sum_{\ell = 0}^{\lfloor n/2\rfloor}(-1)^\ell(2\ell - 1)!!\binom{n}{2\ell}x^{n - 2\ell} = \mu_{n, 1}(x)$ are simple real-rooted polynomials for all $n\geq 1$.
\end{fact}

$\He_n$ are degree$-n$ orthogonal polynomials under the standard Gaussian measure on $\R$, and hence have $n$ real distinct roots by the standard zero theorem for orthogonal polynomials~\cite{szeg1939orthogonal}.

We can also explicitly analyze the polynomial $\mu_{n, -1}$ for $q=-1$:
\begin{lemma}\label{lem:mu-1-real-rooted}
    For any $n\geq 1$, we have that 
    \[\mu_{n, -1}(z) = \sum_{\ell = 0}^{\lfloor n/2\rfloor}(-1)^\ell\binom{n}{2\ell}z^{n - 2\ell} = \prod_{k = 1}^n\left(z - \cot\left(\frac{(2k - 1)\pi}{2n}\right)\right).\]
    In particular $\mu_{n, -1}$ is simple and real-rooted.
\end{lemma}
\begin{proof}
    Firstly, note that since $M_\ell(-1)$ is the $(2\ell)^{\mathrm{th}}$ moment of the Rademacher distribution, we have $M_\ell(-1) = 1$ for all $\ell\geq 0$. Consequently, $\mu_{n, -1}(z) = \sum_{\ell = 0}^{\lfloor n/2\rfloor}(-1)^\ell\binom{n}{2\ell}z^{n - 2\ell}$. Note that we have 
    \[\sum_{\ell = 0}^{\lfloor n/2\rfloor}(-1)^\ell\binom{n}{2\ell}z^{n - 2\ell} = \frac{(z + \iota)^n + (z - \iota)^n}{2},\]
    and then it is easy to determine that the roots are $\cot\left(\frac{(2k - 1)\pi}{2n}\right)$ for $k\in[n]$. 
\end{proof}

\newcommand{\Real}{\operatorname{RR}}
\paragraph{Analytic Preliminaries} Let $\cP_n\subset\R[X]$ be the set of monic polynomials of degree $n$,\footnote{There is an unfortunate clash of notation between $\cP_n$ (the set of monic degree $n$ polynomials) and $\cP_2(2\ell)$ (the set of perfect matchings on $2\ell$ elements): the meaning of the symbol $\cP$ will be clear from context.} i.e. every $f\in\cP_n$ can be written as $f(X) = X^n + \sum_{i < n}a_iX^i$. For any $f, g\in\cP_n$, where $f(X) = X^n + \sum_{i < n}a_iX^i, g(X) = X^n + \sum_{i < n}b_iX^i$, define the \emph{coefficient metric} $\|f - g\|:= \max_{i < n}|a_i - b_i|$. The coefficient metric $\|\cdot\|$ induces the \emph{coefficient topology} on $\cP_n$. Let $\Real_n\subset\cP_n$ be the set of real-rooted polynomials, and let $\cS_n\subset\cP_n$ be the set of simple polynomials.\footnote{Recall that a function is called simple if all of its zeros have multiplicity equal to $1$.} For any set $X\seq\cP_n$ let $\operatorname{int}(X)$ denote the interior of $X$, and let $\partial X$ denote the boundary of $X$. 
\begin{fact}[Standard consequences of Theorem~1.3.1,~\cite{RahmanS02}]\label{fact:real-rooted-topology}
    For any $n\geq 1$ we have:
    \begin{enumerate}[(1)]
        \item $\Real_n$ is closed and $\cS_n$ is open.
        \item $\operatorname{int}(\Real_n) = \Real_n\cap\cS_n$. In words, every polynomial in the interior of $\Real_n$ is simple, and every polynomial in the boundary of $\Real_n$ is \textbf{not} simple.
        \item For any $f\in\Real_n$ with roots $\lambda_1(f)\leq\cdots\leq\lambda_n(f)$, the map $\Real_n\ni f\mapsto(\lambda_1(f), \ldots, \lambda_n(f))\in\R^n$ is continuous in the coefficient topology on $\Real_n$. 
    \end{enumerate}
\end{fact}

\begin{fact}[Implicit function theorem; Theorem~9.28,~\cite{Rudin76}]\label{fact:differentiable-roots}
    Fix any $n\geq 1$, and let $T\seq\R$ be a non-empty open set. Let $p_t\in\Real_n\cap\cS_n\subset\cP_n$ be a parametrized family of \textbf{real-rooted simple} polynomials with roots $\lambda_1(t) < \cdots < \lambda_n(t)$. Suppose the map $\R\times T\ni (z, t)\mapsto p_t(z)\in\R$ is $C^1$. Then for any $i\in[n]$ the map $T\ni t\mapsto\lambda_i(t)\in\R$ is $C^1$ with 
        \begin{equation}\label{eq:eigenvalue-derivative}
            \frac{\operatorname{d}\!\lambda_i(t)}{\operatorname{d}\!t} = -\frac{\partial_t p_t(\lambda_i(t))}{\partial_z p_t(\lambda_i(t))}.
        \end{equation}
\end{fact}
We will eventually apply \cref{fact:differentiable-roots} to the polynomial $\mu_{n, q}$, and it is readily verified that $\R^2\ni(z, q)\mapsto\mu_{n, q}(z)\in\R$ is a $C^1$ map for all $n\geq 0$.

We collect some standard facts from complex analysis. 

Fix $\mathbb{K}:= \R$ or $\C$, and let $\Omega\seq\mathbb{K}^n$ be open, for some $n\geq 1$. Let $\{f_\ell\}_{\ell\geq 0}$ be a sequence of functions $\Omega\to\mathbb{K}$. We say $\{f_\ell\}_{\ell\geq 0}$ converges \emph{locally uniformly} to some $f:\Omega\to\mathbb{K}$ if for all compact sets $K\subset\Omega$ we have $\lim_{\ell\to\infty}\sup_{z\in K}|f_\ell(z) - f(z)| = 0$.

For an open set $\Omega\seq\C^n$, we say $f:\Omega\to\C$ is \emph{holomorphic} if for all $(w_{1}, \ldots, w_{n})\in\Omega$, we can write 
\[f(z_1, \ldots, z_n) = \sum_{\bm\alpha:=(\alpha_1, \ldots, \alpha_n)\in\Z_{\geq 0}^n}a_{\alpha_1, \ldots, \alpha_n}(z_1 - w_{1})^{\alpha_1}\cdots(z_n - w_{n})^{\alpha_n}\] 
for all $\bm z:= (z_1, \ldots, z_n)$ in some polydisc $\{\bm z\in\Omega:\max_{i\in[n]}|z_i-w_i| < r\}$, for some $r > 0$. Also for any multi-index $\bm\alpha\in\Z_{\geq 0}^n$ denote by $\partial^{|\bm\alpha|}f/\partial \bm z^{\bm\alpha}$ the partial derivative $\partial^{\alpha_1}_{z_1}\cdots\partial^{\alpha_n}_{z_n}f$. For holomorphic functions the order of taking partial derivatives does not matter.

For a function $f:\Omega\to\C$ let $Z(f):= \{z\in\Omega: f(z) = 0\}$ be the zero set of $f$. A holomorphic function $f:\C\to\C$ is called \emph{entire}.

A standard reference for the facts below are~\cite{SteinS03}. The content of \cref{item:multivar-weierstrass} can be found in \cite[Corollary~2.2.4]{Hormander90}.
\begin{fact}\label{fact:complex-analysis}
We will use the following standard facts from complex analysis:
    \begin{enumerate}[(1)]
        \item\label{item:cauchy-hadamard-pringsheim}\textbf{Cauchy-Hadamard and Pringsheim's Theorem} Consider the power series $f(z):= \sum_{\ell\geq 0}a_\ell z^\ell$, and write $R:= (\limsup_{\ell\to\infty}|a_\ell|^{1/\ell})^{-1}$. Suppose $0 < R < \infty$. Then $f$ is holomorphic on $\{z\in\C:|z| < R\}$. Furthermore, if $a_\ell\geq 0$ for all $\ell\geq 0$, then for every $\eps > 0$, there does \emph{not} exist a holomorphic function $g:\Omega\to\C$, where $\Omega:= \{z:|z - R| < \eps\}$, which satisfies $g(z) = f(z)$ for all $z\in\Omega$ for which $|z| < R$.
        \item\label{item:weierstrass} \textbf{Weierstrass's Theorem} Let $\Omega\seq\C$ be an open set, and let $\{g_\ell\}_{\ell\geq 0}$ be a sequence of holomorphic functions $\Omega\to\C$ such that for any compact $K\subset\Omega$ we have $\sum_{\ell\geq 0}\sup_{z\in K}|g_\ell(z)| < \infty$. Then $\sum_{\ell\geq 0}g_\ell(z)$ is a holomorphic function $\Omega\to\C$, and furthermore $\sum_{n\geq\ell\geq 0}g_\ell$ converges locally uniformly to $\sum_{\ell\geq 0}g_\ell$.
        \item\label{item:multivar-weierstrass} Let $\Omega\seq\C^n$ be open, and let $f_\ell:\Omega\to\C$ be holomorphic for all $\ell\geq 0$. Suppose $\{f_\ell\}_{\ell\geq 0}$ converges locally uniformly to $f:\Omega\to\C$. Then $f$ is holomorphic, and furthermore for all multi-indices $\bm\alpha\in\Z_{\geq 0}^n$, $\{\partial^{|\bm\alpha|}f_\ell/\partial \bm z^{\bm\alpha}\}_{\ell\geq 0}$ converges locally uniformly to $\partial^{|\bm\alpha|}f/\partial \bm z^{\bm\alpha}$.
        \item\label{item:hurwitz} \textbf{Hurwitz's Theorem} Let $\Omega\seq\C$ be open and connected, and let $\{f_\ell\}_{\ell\geq 0}$ be a sequence of holomorphic functions $\Omega\to\C$ converging locally uniformly to $f:\Omega\to\C$. Suppose $f\not\equiv 0$, and let $z_0$ be a root of $f$ of multiplicity $m$. Then there exists $r>0$ such that, for all sufficiently large $\ell$, the function $f_\ell$ has exactly $m$ zeros $z_{\ell, 1}, \ldots, z_{\ell, m}\in B(z_0; r):= \{z:0\leq|z - z_0| < r\}$. Furthermore, $\lim_{\ell\to\infty}\sup_{j\in[m]}|z_{\ell, j} - z_0| = 0$. Finally, the zeros of $f$ account for every zero of $f_\ell$ as $\ell\to\infty$, i.e. for every compact set $K\subseteq\Omega$ and any open set $Z\supset Z(f)$, we have $Z(f_\ell)\cap K\seq Z$ for large enough $\ell$. Consequently, if $f$ has no zeros in $\Omega$, then for every compact set $K\subseteq\Omega$, the function $f_\ell$ has no zeros in $K$ for all sufficiently large $\ell$.      
        \item\label{item:hadamard} \textbf{Hadamard Factorization} Let $f$ be an entire function with $f(0)\neq 0$. Suppose there exist constants $C_1, C_2 > 0$ and $\alpha\in(0, 1)$ such that $|f(z)| < C_1\exp(C_2|z|^\alpha)$ for all $z\in\C$. Let $\{t_1, t_2, \ldots\}$ be an enumeration of the zeros of $f$ (with multiplicity). Then we have 
        \[f(z) = f(0)\prod\left(1 - \frac{z}{t_i}\right),\]
        with locally uniform convergence, i.e. for every $R > 0$ we have 
        \[\lim_{N\to\infty}\sup_{|z|\leq R}\left|f(z) - f(0)\prod_{i\leq N}\left(1 - \frac{z}{t_i}\right)\right| = 0.\]
    \end{enumerate}
\end{fact}
We shall also need the following complex Gaussian-Poincar\'e inequality, which is a straightforward consequence of the (more usual) real Poincar\'e inequality.
\begin{fact}[Complex Gaussian-Poincar\'{e} Inequality] \label{fact:gaussianpoincare}
    Let $\bm g = (g_1, \dots g_m) \sim \cN(0, \Id_m)$ be a vector of independent standard real Gaussian variables. Let $f:\R^m \to \C$ be continuously differentiable, and suppose that, 
    \[
    \E_{\bm g}|f(\bm g)|^2 < \infty, \quad \sum_{j=1}^m \E_{\bm g}\left| \frac{\partial f}{\partial g_j}(\bm g)\right|^2 < \infty.
    \]
Then
\[
\E_{\bm g}\left|f(\bm g) - \E_{\bm g} f(\bm g)\right|^2 \leq \sum_{j=1}^m \E_{\bm g}  \left| \frac{\partial f}{\partial g_j} (\bm g)\right|^2.
\]
\end{fact}
\begin{proof}
    Let $f=u +\iota v$, where $u, v : \R^m \to \R$ are real valued functions. Then
    \[\E_{\bm g}|f(\bm g) - \E_{\bm g} f(\bm g)|^2 = \Var(u(\bm g)) + \Var(v(\bm g))\leq \sum_{j=1}^m \E_{\bm g}\left[ \left(\frac{\partial u}{\partial g_j} (\bm g)\right)^2 + \left(\frac{\partial v}{\partial g_j} (\bm g)\right)^2 \right] = \sum_{j=1}^m \E_{\bm g}  \left| \frac{\partial f}{\partial g_j} (\bm g)\right|^2,\]
    where the real valued Gaussian-Poincar\'{e} inequality~\cite{boucheron2013non} is applied to $u,v$ separately. 
\end{proof}

We also recall the standard \emph{Duhamel's formula}:
\begin{fact}[Duhamel's Formula~\cite{higham2008functions}]\label{fact:duhamel}
    Let $I \subseteq \R$ be an open interval, and let $A:I \to \C^{N\times N}$ be continuously differentiable. Then, 
\[
\frac{\operatorname{d}}{\operatorname{d}\!t}e^{A(t)} = \int_{0}^1 e^{sA(t)} A'(t) e^{(1-s)A(t)}\operatorname{d}\!s.
\]
\end{fact}

\paragraph{Real Stability} For variables $z_1, \ldots, z_n$ and any $0\leq k\leq n$, define the elementary symmetric polynomials 
\[e_k(z_1, \ldots, z_n):= \sum_{S\in\binom{[n]}{k}}\prod_{i\in S}z_i.\]
\begin{definition}[Polarization]\label{def:polarization}
    For any non-zero univariate polynomial $p(z) = \sum_{k = 0}^na_kz^k$ of degree $\leq n$, we define its \emph{polarization} on $n$ variables to be 
\[\operatorname{Pol}_n(p)(z_1, \ldots, z_n) = \sum_{k = 0}^n\frac{a_k}{\binom{n}{k}}e_k(z_1, \ldots, z_n).\]
Note that $\operatorname{Pol}_n(p)(z, \ldots, z) = p(z)$. 
\end{definition}

Now, we say that a polynomial $P(z_1, \ldots, z_n)\in\R[z_1, \ldots, z_n]$ is \emph{real stable} if for all $z_1, \ldots, z_n\in\C$ with $\Imm(z_i) > 0$ for all $i\in[n]$, we have $P(z_1, \ldots, z_n)\neq 0$. 

\begin{fact}[Grace-Walsh-Szeg\H{o} Theorem (see~\cite{RahmanS02})]\label{fact:Grace-Walsh-Szego}
    For any non-zero univariate polynomial $p(z)\in\R[z]$ of degree $\leq n$, $\operatorname{Pol}_n(p)$ is real-stable if and only if $p$ is real-rooted.
\end{fact}

We will also need the following fact about \emph{real-rootedness preserving operators}. Let $\R[z]_{\leq n}$ denote the space of univariate polynomials with real coefficients of degree $\leq n$, and let $D:\R[z]_{\leq n}\to\R[z]_{\leq n}$ be a real-rootedness preserving linear operator, i.e. for any real-rooted polynomial $p\in\R[z]_{\leq n}$, $D[p]$ is also real-rooted. Then we have: 
\begin{fact}[Real-Rootedness Preserving Operator (see Theorem 2.2(b) in~\cite{BorceaB09a})]\label{fact:rr-preserve}
    $D$ is real-rootedness preserving if $D[(z+w)^n]\in\R[z,w]$ is real-stable.
\end{fact}
In the expression $D[(z+w)^n]$, $D$ treats the variable $w$ as a constant, i.e. for any $g(z,w) = \sum_{k=  0}^np_k(z)w^k\in\R[z,w]$, we have $D[g] := \sum_{k=  0}^nD[p_k(z)]\cdot w^k\in\R[z,w]$.

We shall also need the following inequalities satisfied by real-rooted polynomials.

\begin{fact}[Newton's Inequalities (see Lemma 1.1 in~\cite{Branden14})]\label{fact:newton}
    Let $\lambda_1, \ldots, \lambda_m\geq 0$ be real numbers, and write $e_\ell:= e_\ell(\lambda_1, \ldots, \lambda_m)$ for all $0\leq\ell\leq m$. Then for all $1\leq\ell\leq m - 1$ we have 
    \[\left(\frac{e_\ell}{\binom{m}{\ell}}\right)^2\geq\frac{e_{\ell + 1}}{\binom{m}{\ell + 1}}\cdot\frac{e_{\ell - 1}}{\binom{m}{\ell - 1}}\implies e_\ell^2\geq \frac{(\ell + 1)(m - \ell + 1)}{\ell(m - \ell)}e_{\ell + 1}e_{\ell - 1}.\]
\end{fact}

Along similar lines as above, we shall need the following \emph{Laguerre's inequalities} for functions from the \emph{Laguerre-P\'olya ($\mathcal{LP}$) class}. Recall that an entire function $f:\C\to\C$ is said to belong to the $\mathcal{LP}$ class if it is the locally uniform limit of real-rooted polynomials with real coefficients. Functions in $\mathcal{LP}$ satisfy the so-called \emph{Laguerre inequality}:

\begin{fact}[Laguerre's Inequality, see Eq.(3.7) in~\cite{CravenC94}]\label{fact:laguerre}
    For any function $\varphi\in\mathcal{LP}$, and any integer $k\geq 1$, we have, for all $z_0\in\R$,
    \[\varphi^{(k)}(z_0)^2\geq\varphi^{(k - 1)}(z_0)\cdot\varphi^{(k + 1)}(z_0),\]
    where $\varphi^{(k)}:= \operatorname{d}^k\!\varphi/\operatorname{d}\!z^k$.
\end{fact}
\section{Imaginary-Rootedness of $\cL_q$: Proof of \cref{thm:moment-function-imaginary-rooted}}
We are now ready to prove \cref{thm:moment-function-imaginary-rooted} assuming \cref{thm:main-thm}. We first show that (a suitable rescaling) of the polynomials $\mu_{n, q}$ locally uniformly converges to $\varphi_q(z):= \cL_q(\iota z)$, thus establishing the real-rootedness of $\varphi_q$ (and thus imaginary-rootedness of $\cL_q$). We then study the order of growth of the function $\cL_q$ to prove the product formula $\cL_q(z) = \prod_{k = 1}^\infty(1 + z^2/t_k(q)^2)$. Finally, we show that the roots of $\cL_q$ are simple by applying \emph{Laguerre's inequality} (see \cref{fact:laguerre}) for functions in the Laguerre-P\'olya ($\mathcal{LP}$) class, where we use the fact that $\varphi_q$, being the locally uniform limit of real-rooted polynomials, is in the $\mathcal{LP}$ class.
\momentfunctionimaginaryrooted*
\begin{proof}[Proof of \cref{thm:moment-function-imaginary-rooted} assuming \cref{thm:main-thm}]
Fix any $q\in[-1, 1)$ throughout the proof, and write $\varphi_q(z):= \cL_q(\iota z)$. For any $n\geq 1$ consider the polynomial 
\[g_{n, q}(z):= \sum_{\ell = 0}^{\lfloor n/2\rfloor}(-1)^\ell\frac{\binom{n}{2\ell}}{n^{2\ell}}M_\ell(q)z^{2\ell} = \left(\frac{z}{n}\right)^n\mu_{n, q}\left(\frac{n}{z}\right).\]
By \cref{thm:main-thm} we have that $g_{n, q}$ is real-rooted. We now claim that:\footnote{$g_{n,q}$ are also known as the (rescaled) \emph{Jensen polynomials} associated to $\varphi_q$. The locally uniform convergence of Jensen polynomials is also the content of \cite[Lemma~2.2]{CravenC89}. We nevertheless include a short proof for the sake of completeness}
\begin{claim}\label{claim:gnq-local-conv}
    $g_{n, s}(z)\to\varphi_s(z)$ locally uniformly for $(s, z)\in\{w\in\C:|w| \leq 1\}\times\C$, i.e. for any $R > 0$, we have
    \[\lim_{n\to\infty}\sup_{\substack{|s|\leq 1\\ |z|\leq R}}|g_{n, s}(z) - \varphi_s(z)| = 0,\]
    where we treat both $s, z$ as complex variables.\footnote{Note that for complex $s\in\C$, $M_\ell(s)$ is defined exactly the same way for all $\ell\geq 0$, and we can then define $\varphi_s, \cL_s$ through their power series description in terms of $\{M_\ell(s)\}_{\ell\geq 0}$. The proof of $\cL_s$ being entire is exactly the same for all $|s| < 1$.}
\end{claim}
Now assume for the sake of contradiction that $\varphi_q(z)$ has a non-real root $z_0\in\C\setminus\R$. Choose $\eps > 0$ such that $B(z_0;\eps)\subset\C\setminus\R$. By \cref{claim:gnq-local-conv}, $g_{n, q}$ converge locally uniformly to $\varphi_q(z)$, and thus by \cref{item:hurwitz} of \cref{fact:complex-analysis} we have that $g_{n, q}$ has a root in $B(z_0;\eps)\subset\C\setminus\R$ for sufficiently large $n$, contradicting the real-rootedness of $g_{n, q}$.

Thus $\cL_q$ is imaginary-rooted. Now note that if we define $f_q(z):= \sum_{\ell\geq 0}M_\ell(q)z^\ell/(2\ell)!$, then $\cL_q(z) = f_q(z^2)$, and thus $\cL_q$ is an even function. It remains to demonstrate the product formula for $\cL_q$, and show that it has infinitely many distinct roots. To do this, we need to bound the growth of $\cL_q$, for which we need the following claim:

\begin{claim}\label{claim:tq-operator-norm}
    $\|T_q\|\leq\sqrt{8/(1 - q)}$.
\end{claim}
Now note that $M_\ell(q) = \bra{0}T_q^{2\ell}\ket{0}\leq\|T_q\|^{2\ell} \leq (8/(1 - q))^\ell$, and consequently, for any $R > 0$, 
\begin{equation}\label{eq:lq-growth}
    \sum_{\ell\geq 0}\sup_{|z|\leq R}\left|M_\ell(q)\frac{z^{2\ell}}{(2\ell)!}\right|\leq\sum_{\ell\geq 0}\left(\sqrt{\frac{8}{1 - q}}\right)^{2\ell}\frac{R^{2\ell}}{(2\ell)!}\leq\exp\left(\sqrt{\frac{8}{1 - q}}\cdot R\right) < \infty.
\end{equation}
Consequently by \cref{item:weierstrass} of \cref{fact:complex-analysis} we have that $\cL_q$ is entire. Furthermore by \cref{eq:lq-growth} we have that $|f_q(z)|\leq\exp(\sqrt{8|z|/(1 - q)})$ for all $z\in\C$, and thus the order of growth of $f_q$ is $\leq 1/2$. Since $f_q(0) = 1$, by \cref{item:hadamard} of \cref{fact:complex-analysis} we have 
\[f_q(z) = \prod\left(1 - \frac{z}{-t_k(q)^2}\right)\implies\cL_q(z) = \prod\left(1 + \frac{z^2}{t_k(q)^2}\right),\]
where $\{-t_k(q)^2\}_{k\in\N}$ are the roots of $f_q$. Further note that $f_q$ (and thus $\cL_q$) has to have infinitely many roots, otherwise $f_q$ will equal a polynomial, which contradicts the fact that $M_\ell(q)\geq 1$ for all $\ell\geq 0$ (and the Taylor series for $f_q$ has $M_\ell(q)/(2\ell)!$ as the coefficient of $z^\ell$). 

It remains to show that the roots $t_1(q), t_2(q), \ldots$ are distinct. WLOG assume $q\in(-1, 1)$, since for $q = -1$, $\cL_{-1}(z) = \cosh(z)$, all of whose roots are distinct. To do so, we adapt the argument in~\cite[Theorem~3.10]{CravenC94}. 

Define the function $H:B(1)\times\C\to\C$ as $H(s, z):= \varphi_s(\sqrt{1 - s}\cdot z)$, where $B(1):= \{w\in\C:|w| < 1\}$, and we choose the analytic branch of $\sqrt{1 - s}$ on $B(1)$. By \cref{item:multivar-weierstrass} of \cref{fact:complex-analysis} and \cref{claim:gnq-local-conv}, we have that $H$ is holomorphic. Furthermore, by~\cref{claim:gnq-local-conv,thm:main-thm}, $H(s,\cdot)$ is in the Laguerre-P\'olya class for all $s\in[-1, 1)$.

By~\cref{cor:scrl-pde} we obtain
\begin{equation}\label{eq:Hpde}
    8s\partial_sH=(z^2\partial_z^2+3z\partial_z+4z^2)H.
\end{equation}
Assume for the sake of contradiction $H(q_0,\cdot)$ has a root $z_0\in\R$ of multiplicity $m\geq2$ for some $q_0\in(-1,1)$.\footnote{Note that $m$ is finite since $H$ is non-zero and holomorphic} Since $H(q_0,0)=1$, $z_0\neq0$.
Write $h(s,z):=\partial_z^{m-2}H(s,z), A:=\partial_z^mH(q_0,z_0)$ and observe that $h(q_0,z_0)=\partial_zh(q_0,z_0)=0$ and $\partial_z^2h(q_0,z_0)=A\neq 0$. Also define the function $E:(-1,1)\to\R_{\geq 0}$ given as 
\[E(s):= h_z(s,z_0)^2 -h(s,z_0)h_{zz}(s,z_0),\]
where $h_z$ stands for $\partial_zh$, $h_{zz}$ stands for $\partial_z^2h$, and so on. 

Intuitively, the proof of simplicity is the following: We can show that \cref{eq:Hpde} is a \emph{repulsive} differential equation (see \cref{sec:q-deformed-real-rooted} for a detailed investigation of this phenomena), i.e. it tends to push the roots of $H$ away from each other. One way we can quantify this ``repulsion'' is by saying that $E'(q_0)$ is non-zero; indeed, since $h(q_0,\cdot)$ has a root at $z_0$, for $s$ away from $q_0$, the roots will also be repelled from $z_0$. But on the other hand, $E(q_0) = 0$, and thus $E'(q_0)$ is forced to be $0$, since $E(s)\geq 0$ for all $s\in(-1,1)$, yielding a contradiction.

By \cref{fact:laguerre}, $E(s)\geq 0$ for all $s\in(-1, 1)$. Since $E:(-1, 1)\to\R_{\geq 0}$ is differentiable, and since $E(q_0) = 0$, we have $E'(q_0) = 0$. However, 
\[E'(q_0)=\left[2h_zh_{sz}-h_sh_{zz}-h_{szz}h\right]_{(s,z)=(q_0,z_0)}=-A\cdot h_s(q_0,z_0),
\]

Now apply $\partial_z^{m - 2}$ to both sides of \cref{eq:Hpde} and evaluate at $(s, z) = (q_0, z_0)$. Every term involving a $z$-derivative of $H$ of order $< m$ vanishes, yielding $8q_0\,h_s(q_0,z_0)=z_0^2A$. Consequently, $0 = 8q_0E'(q_0) = -z_0^2A^2 < 0$, yielding a contradiction. 

Thus all roots of $H(q, \cdot)$, and hence $\cL_q$, are simple.

We finish off with proofs of \cref{claim:tq-operator-norm,claim:gnq-local-conv}.
\begin{proof}[Proof of \cref{claim:tq-operator-norm}]
    Note that $[n]_q:= 1 + q + \cdots + q^{n - 1} = (1 - q^n)/(1 - q)\leq 2/(1 - q)$. Thus if we have some finitely supported $\psi\in\spn\{\ket{n}:n\geq 0\}$ with $\psi = \sum_{n\geq 0}\psi_n\ket{n}$, then (setting $\psi_{-1}:= 0, \ket{-1}:= 0$) we have
\[T_q\psi = \sum_{n\geq 0}\psi_n\left(\sqrt{[n]_q}\ket{n - 1} + \sqrt{[n + 1]_q}\ket{n + 1}\right) = \sum_{n\geq 0}\left(\sqrt{[n + 1]_q}\psi_{n + 1} + \sqrt{[n]_q}\psi_{n - 1}\right)\ket{n}\]
\[\implies\|T_q\psi\|_2^2 = \sum_{n\geq 0}\left|\sqrt{[n + 1]_q}\psi_{n + 1} + \sqrt{[n]_q}\psi_{n - 1}\right|^2\leq 2\sum_{n\geq 0}\left([n + 1]_q|\psi_{n + 1}|^2 + [n]_q|\psi_{n - 1}|^2\right)\]
\[\leq\frac{4}{1 - q}\sum_{n\geq 0}\left(|\psi_{n + 1}|^2 + |\psi_{n - 1}|^2\right) \leq \frac{8}{1 - q}\|\psi\|^2\implies\|T_q\psi\|_2\leq\sqrt{\frac{8}{1 - q}}\cdot\|\psi\|_2.\]
Since finitely supported vectors are dense in $\ell_2(\Z_{\geq 0})$ we have $\|T_q\|_2\leq\sqrt{8/(1 - q)}$. 
\end{proof}
\begin{proof}[Proof of \cref{claim:gnq-local-conv}]
    Fix any $R \geq 1$ and assume $|z|\leq R$. Also write $c_{n, \ell}:= \begin{cases}
        \binom{n}{2\ell}/n^{2\ell} & \text{if }2\ell\leq n,\\
        0 & \text{if }2\ell > n.
    \end{cases}$. Then 
\[|g_{n, s}(z) - \varphi_s(z)|\leq\sum_{\ell\geq 0}\left|c_{n, \ell} - \frac{1}{(2\ell)!}\right|\cdot|M_\ell(s)|R^{2\ell}.\]
Note that $0\leq c_{n, \ell}\leq 1/(2\ell)!$ for all $\ell, n\geq 0$. Also note that for any $s\in\C$ with $|s|\leq 1$, we have $|M_\ell(s)|\leq M_\ell(1) = (2\ell - 1)!!$. We thus have
\[\left|c_{n, \ell} - \frac{1}{(2\ell)!}\right|\cdot |M_\ell(s)|\cdot R^{2\ell}\leq \frac{1}{(2\ell)!}\cdot (2\ell - 1)!!\cdot R^{2\ell} = \frac{R^{2\ell}}{2^\ell\ell!}.\]
Note that $\sum_{\ell\geq 0}\frac{R^{2\ell}}{2^\ell\ell!} = e^{R^2/2} < \infty$, and thus for any $\eps > 0$ there exists $L = L(\eps, R)\geq 1$ such that $\sum_{\ell\geq L}\frac{R^{2\ell}}{2^\ell\ell!}\leq\eps$. But also note that for any \emph{fixed} $\ell$, we have that $\lim_{n\to\infty}c_{n, \ell} = 1/(2\ell)!$. Thus choose $n_0$ large enough w.r.t. $\eps, R$ such that for all $n\geq n_0$ and $0\leq\ell\leq L$ we have 
\[\left|c_{n, \ell} - \frac{1}{(2\ell)!}\right|\leq\frac{\eps}{L\cdot (2L - 1)!!\cdot R^{2L}}.\]
Then we have for all $n\geq n_0 = n_0(\eps, R)$ that
\[|g_{n, s}(z) - \varphi_s(z)|\leq\sum_{\ell\geq L}\left|c_{n, \ell} - \frac{1}{(2\ell)!}\right|\cdot M_\ell(1)R^{2\ell} + \sum_{\ell < L}\left|c_{n, \ell} - \frac{1}{(2\ell)!}\right|\cdot M_\ell(1)R^{2\ell}\] 
\[\leq\sum_{\ell\geq L}\frac{(2\ell - 1)!!\cdot R^{2\ell}}{(2\ell)!} + \sum_{\ell < L}\frac{\eps}{L\cdot (2L - 1)!!\cdot R^{2L}}\cdot (2\ell - 1)!!\cdot R^{2\ell}\leq 2\eps.\]
Since $\eps > 0$ was arbitrary, we're done.
\end{proof}
\end{proof}
\begin{remark}
A few remarks are due:
\begin{enumerate}[(1)]
    \item Note that for any $R > 0$ we have $\cL_q(R)\geq\cosh(R)$, since $M_\ell(q)\geq 1$ for all $q\in[-1, 1], \ell\geq 0$, and that shows that the order of growth of $\cL_q$ (for $q\in[-1, 1)$) is exactly $1$.
    \item Note that even though $g_{n, q}$ are simple polynomials (i.e. all their roots have multiplicity $1$), we still have to give a separate argument for the simplicity of $\varphi_q$ because \emph{a priori}, two distinct roots of $g_{n, q}$ could be separated by $o_n(1)$ and coalesce into a single root with multiplicity $\geq 2$ in the limit $n\to\infty$. This is why an additional argument was needed to show the simplicity of $\varphi_q$.
    \item $\cL_q$ is \textbf{not} imaginary-rooted for non-real $q$ with $|q| < 1$. Indeed, assume for the sake of contradiction that there exists $q\in\C\setminus\R$ with $|q| < 1$ for which $\cL_q$ is imaginary-rooted. Note that $|\cL_q(z)|\leq\exp(|z|\cdot \sqrt{8/(1 - |q|)})$, and thus \cref{item:hadamard} of \cref{fact:complex-analysis} gives us that $\cL_q(z) = \prod_{k = 1}^\infty(1 + z^2/t_k(q)^2)$, which would imply that all Taylor coefficients of $\cL_q$ are real. On the other hand, $\cL_q(z) = 1 + \frac{1}{2}z^2 + \frac{2 + q}{24}z^4 + O(z^6)$, and thus the coefficient of $z^4$ in $\cL_q$ is not real, if $q$ is not real, leading to a contradiction.
    \item For $q\in[0,1)$, the product representation in \cref{thm:moment-function-imaginary-rooted} was anticipated in~\cite[Eq. (3.10)]{Okuyama23}.
\end{enumerate}
\end{remark}
Using the differential equation governing $\cL_q$, one can obtain far stronger control on the roots $t_1(q) < t_2(q) < \cdots$ than merely their distinctness.

\section{Location of the nearest root: Proof of \cref{thm:nearest-root}}
We shall now prove \cref{thm:nearest-root}:
\rootbehavior*
Proving \cref{thm:nearest-root} will require us to reason about $\log\cL_q(z)$. Since $\cL_q$ is the moment generating function of the $q$-Gaussian distribution $\nu_q$, the Taylor series of $\log\cL_q$ defines the \emph{cumulants} of $\nu_q$:
\[
\log \cL_q(z) = \sum_{m \geq 1} \kappa_{2m}\frac{z^{2m}}{(2m)!}.
\]
The cumulants of the $q$-Gaussian distribution admit some beautiful connections to \emph{Tutte polynomials} thanks to~\cite{VergesCumulants}. Thus define the Tutte polynomial for a connected graph $G$, as in~\cite{tutte1954contribution}, to be
\[T_G(1,q) := \sum_{\substack{F\subseteq E \\ (V,F) \text{ connected}}}(q-1)^{|F|-m+1}.\]
For any $m \geq 1$ also define the polynomial
\[
b_m(q) := \frac{m}{(2m)!} \sum_{\substack{\pi \in \cP_2(2m) \\ G_\pi \text{ connected}}} T_{G_\pi}(1,q),
\]
where $G_\pi$ is the crossing graph of the matching $\pi\in\cP_2(2m)$. The result of Josuat-Verg\`{e}s gives us the following relation between the cumulants $\kappa_{2m}$ and the polynomials $b_m$.

\begin{lemma}[Josuat-Verg\`{e}s~\cite{VergesCumulants}]\label{lem:JVcumulant}
    For every $m \geq 1$ one has the polynomial identity: 
    \[
    \kappa_{2m}(q) = (q-1)^{m-1}\sum_{\substack{\pi\in\cP_2(2m)\\ G_{\pi}\text{ connected}}} T_{G_\pi}(1,q) = (q-1)^{m-1}\frac{(2m)!}{m}b_m(q)
    \]
\end{lemma}

The proof of \cref{thm:nearest-root} requires the following technical facts about the polynomial $b_m(q)$. Recall from \cref{sec:intro} the following definitions: Let $s_*\approx 1.1997$ be the unique solution of $s_*\tanh(s_*) = 1$ in $\R_{\geq0}$, and let $\zeta_* := 2s_*/\cosh(s_*)\approx 1.3255, R_*:= \zeta_*^2\approx 1.7570$.
\begin{lemma}\label{lem:bmq-facts}
    The polynomial $b_m(q)$ satisfies the following properties for all $m\geq 1$:
    \begin{enumerate}[(1)]
        \item (\cref{lem:bmrootsrelation}) Each $b_m$ has nonnegative coefficients. Moreover, for $-1\leq q < 1$, $(1-q)^{m-1} b_m(q) = \sum_{j \geq 1}t_j(q)^{-2m}$, where $\{t_j(q)\}_{j\geq 1}$ are as in \cref{thm:moment-function-imaginary-rooted}. In particular, $\sum_{j \geq 1}1/t_j(q)^2 = 1/2$.
        \item (\cref{prop:bm-monotonicity}) The map $[-1,1]\ni q\mapsto b_m(q)$ is non-decreasing.
        \item (\cref{lem:endpointseries}) The power series $B(u) := \sum_{m \geq 1} b_m(1) u^m$ has radius of convergence $R_*$, with $B(R_*)=2$.
    \end{enumerate}
\end{lemma}

We are now ready to show the location of the nearest root assuming \cref{lem:bmq-facts}:
\begin{proof}[Proof of \cref{thm:nearest-root} assuming \cref{lem:bmq-facts}]
    Let $r_q := (1-q)t_1(q)^2$. By \cref{lem:bmrootsrelation} we have, 
    \begin{align*}
            b_m(q) &:= (1-q)^{1-m} \sum_{j\geq 1}\frac{1}{t_j(q)^{2m}} =  (1-q)^{1-m} \frac{t_1(q)^{2}}{t_1(q)^{2m}}\sum_{j\geq 1}\left(\frac{t_1(q)}{t_j(q)}\right)^{2m} \frac{1}{t_1(q)^2} \\ 
        &\leq  (1-q)^{1-m} \frac{t_1(q)^{2}}{t_1(q)^{2m}}\sum_{j\geq 1}\left(\frac{t_1(q)}{t_j(q)}\right)^{2} \frac{1}{t_1(q)^2} \leq  (1-q)^{1-m} \frac{t_1(q)^{2}}{t_1(q)^{2m}}\frac{1}{2},
    \end{align*}
    where we used the fact that $t_1(q) \leq t_j(q)$ and $\sum_{j} t_j(q)^{-2} = 1/2$. Combining this with $b_m(q) \geq (1-q)^{1-m}t_1(q)^{-2m}$ we get: 
    \begin{equation}\label{eq:bmrqsandwich}
            \frac{1-q}{r_q^m} \leq b_m(q) \leq \frac{1}{2}r_q^{1-m}.
    \end{equation}
    As a result, $\lim_{m\to\infty} b_m(q)^{1/m} = r_q^{-1}$. Since $[-1,1]\ni q\mapsto b_m(q)$ is non-decreasing by \cref{prop:bm-monotonicity}, it follows that $[-1,1]\ni q\mapsto r_q$ is non-increasing. Now, invoking~\cref{lem:endpointseries}, 
    \[
    \sum_{m\geq 1}b_m(q)R_*^m \leq \sum_{m\geq 1}b_m(1)R_*^m = B(R_*)=2
    \]
    where we used coefficient monotonicity $b_m(q) \leq b_m(1)$ from \cref{prop:bm-monotonicity}. We can now show that $r_q > R_*$ for $q\in[-1,1)$. Assume otherwise, i.e., $r_q \leq R_*$, then \cref{eq:bmrqsandwich} gives 
    \[
    b_m(q) R_*^m \geq (1-q) \left(\frac{R_*}{r_q}\right)^m \geq 1-q > 0.
    \]
    for every $m$, which is a contradiction to the convergence of the sum $\sum_{m\geq 1}b_m(q)R_*^m$. Thus $r_q > R_*$ for $-1 \leq q < 1$. Since $r_q$ is nonincreasing and bounded below, the limit $r := \lim_{q\nearrow 1} r_q \geq R_*$ exists. In fact, we claim that $r = R_*$. Indeed, note that~\cref{lem:bmrootsrelation} yields
    \begin{align*}
        r_q b_{m+1}(q) &= \frac{1}{(1-q)^{m-1}}\sum_{j \geq 1} \frac{t_1(q)^2}{t_j(q)^{2m+2}} \leq  \frac{1}{(1-q)^{m-1}}\sum_{j \geq 1} \frac{1}{t_j(q)^{2m}} \leq b_m(q).
    \end{align*}
    Since each $b_m$ is a polynomial, letting $q\nearrow 1$ yields $r b_{m+1}(1) \leq b_m(1)$. Starting from $b_1(1)=1/2$, we obtain $b_m(1) \leq \frac{1}{2}r^{1-m}$. Thus $B(u) = \sum_{m \geq 1}b_m(1)u^m$ converges for $|u|<r$. Then,~\cref{lem:endpointseries} gives $r\leq R_*$. Combining the lower and upper bounds, $r=R_*$. Now, recall that $R_* = \zeta_*^2$, $r_q := (1-q)t_1(q)^2$, and taking square roots gives us $t_1(q) > \frac{\zeta_*}{\sqrt{1-q}}$ and 
    \[
    \lim_{q\nearrow 1} \sqrt{1-q}\cdot t_1(q) = \zeta_*,
    \]
    where the monotonicity of $r_q$ gives that of $\sqrt{1-q}\cdot t_1(q)$.
\end{proof}

Thus it only remains to show \cref{lem:bmq-facts}. To do so, we shall develop some preliminaries about Tutte polynomials, which will be necessary in the proof of \cref{lem:bmq-facts}, and also later on, in the proofs of \cref{thm:annealed-syk-lee-yang,thm:sykstrip}.

\subsection{Graph polynomials and partition functions}
Define the multivariate Tutte polynomial~\cite{Sokal_2005} of a weighted graph $G = G(V, E, \bm w)$ as:
\begin{equation}\label{eq:graph-polynomial}
 P_{G}(t;\bm w):=\sum_{F\subseteq E}t^{\conn(F)}
                   \prod_{e\in F}w_e.
\end{equation}
If $w_e = w$ for all $e\in E(G)$, we write $P_{G}(t;\bm w)$ simply as $P_{G}(t;w)$. 
Consider a simple undirected graph $G = (V,E)$ equipped with edge weights $\bm w\in \C^{E}$. For $F \subseteq E$, let $\conn(F)$ denote the number of connected components of the subgraph $(V,F)$ including isolated vertices. In particular, $\conn(\varnothing)=|V|$. 
It is easy to see that $P_{G}$ is a monic polynomial in $t$ of degree $|V|$. Further define the connected graph polynomial in $\bm w$:
\begin{equation}
    C_G(\bm w) := [t] P_G(t;\bm w) = \sum_{\substack{F\subseteq E \\ (V,F) \text{ connected}} } \prod_{e\in F}w_e.
\end{equation}

Thus, $C_G$ vanishes for a disconnected graph and $C_G=1$ for a graph with a single vertex. We first show that, $P_G(N;\bm w)$ for $N\in\N$ has a natural interpretation in terms of functions from vertices to $[N]$, which can be thought of as assigning one of $N$ colors to each vertex.
\begin{proposition}[Coloring Decomposition]\label{prop:pottspoly}
    For a positive integer $N$, 
    \[
    P_G(N;\bm w) = \sum_{f: V\to [N]} \prod_{e=\{u,v\}\in E}\left(1 + w_e\cdot \1_{f(u) = f(v)}\right).
    \]    

\end{proposition}
\begin{proof}
    Expand the product:
    \[
     \prod_{e=\{u,v\}\in E}\left(1 + w_e\cdot \1_{f(u) = f(v)}\right) = \sum_{F\subseteq E} \left[\prod_{e\in F}w_e\right] \left[\prod_{e=\{u,v\}\in F}\1_{f(u) = f(v)}\right]
    \]
    For a subset of edges $F\subseteq E$, the sum is nonzero iff $f$ is constant on every connected component of the spanning subgraph $(V,F)$. Given an $F$, there are $N^{\conn(F)}$ such maps. Reorganizing the sum, we get, 
    \[
    P_G(N;\bm w) = \sum_{F\subseteq E} N^{\conn(F)}\left[\prod_{e\in F}w_e\right]  
    \]
    precisely the form of \cref{eq:graph-polynomial}.
\end{proof}

\begin{remark}
    \cref{prop:pottspoly} can be understood as the familiar $N-$state Potts model partition function~\cite{fortuin1972random}. To see this, consider the $N-$state Potts model defined by the Hamiltonian expressed using spins $\bm \sigma: V\to [N]$ as
    \[
    H_G(\bm \sigma) := -\sum_{\{u,v\}\in E} h_{\{u,v\}} \1_{\sigma_u = \sigma_v}.
    \]
    Then, if we define $w_e := e^{\beta h_e}-1$, see that, 
    \[
    P_G(N;\bm w) = \sum_{\bm\sigma: V\to [N]} e^{-\beta H_G(\bm \sigma)} \equiv Z_{\beta,\operatorname{Potts}}.
    \]
\end{remark}

For a finite vertex set $V$, define the set of unordered families of pairwise disjoint subsets:
\[
\cD(V):= \{ \{S_1,\dots , S_r\}~:~ S_i \subseteq V, |S_i|\geq 2, S_i \cap S_j = \emptyset\text{ for all }1\leq i < j\leq r\}.
\]
These subsets need not cover $V$. We now show that $P_G$ admits another decomposition in terms of connected polynomials of families of the vertex set.

\begin{proposition}[Family Decomposition]\label{prop:familydecomposition}
    For $t\neq 0$, we have 
    \[
    \frac{P_G(t;\bm w)}{t^{|V|}} = \sum_{\cF \in \cD(V)} \prod_{S\in \cF} \frac{C_{G[S]}(\bm w)}{t^{|S|-1}},
    \]
    where $G[S]$ is the subgraph induced by $S$ with inherited weights. 
\end{proposition}
\begin{proof}
    Given a subset of edges $F\subseteq E$, we map it to a family $\Phi(F) \in \cD(V)$ by defining the $S_i \in \cF$ to be the sets of vertices that form connected components of size at least two in the subgraph induced by $F$. Vertices outside the family are precisely the isolated vertices in the spanning subgraph. Thus,
    \[ \conn(F) = \left|V \setminus \bigcup_{S\in \Phi(F)} S\right| + |\cF| = |V| - \sum_{S\in \Phi(F)}|S| + |\cF| =|V| - \sum_{S\in \Phi(F)}(|S|-1).\]
    Conversely, given a family $\cF\in \cD(V)$, one can obtain edge sets $F$ with $\Phi(F)=\cF$ by independently choosing a connected spanning edge set $F_S \subseteq E(G[S])$ for each $S\in\cF$ and taking $F$ to be the union of all the $F_S$.
    \begin{align*}
        \frac{P_G(t;\bm w)}{t^{|V|}} &=\sum_{F\subseteq E} t^{\conn(F)-|V|} \prod_{e\in F} w_e = \sum_{\cF \in \cD(V)} t^{-\sum_{S\in \Phi(F)}(|S|-1)}\sum_{\substack{F\subseteq E \\ \Phi(F) = \cF}} \prod_{e\in F}w_e \\ 
        &=  \sum_{\cF \in \cD(V)}\prod_{S\in \cF}\left(t^{1-|S|}\sum_{\substack{F_S \subseteq E(G[S])\\ (S,F_S) \text{ connected }}} \prod_{e\in F_S} w_e\right)=  \sum_{\cF \in \cD(V)}\prod_{S\in \cF} \frac{C_{G[S]}(\bm w)}{t^{|S|-1}}.\qedhere
    \end{align*}
\end{proof}

Let $G=(V,E)$ be a connected graph on $m\geq 1$ vertices, and let $\cT(G)$ be the set of its spanning trees. By fixing a total order on the set of edges $E$, define for every $T\in\cT(G)$, 
\[
\Ext(T) := \{e\in E\setminus T~:~e\text{ is larger than every edge on the }T\text{ path joining its endpoints}\}.
\]

Now note that $T_G(1,q) = (q-1)^{1-m}C_G(q-1)$, where $C_G(q-1)$ denotes the connected graph polynomial evaluated at every edge weight equal to $q-1$.

We now show that, the connected graph polynomial $C_G$ can be decomposed into a sum over spanning trees of the graph.
\begin{restatable}[Spanning Tree Decomposition]{lemma}{spannindecomp}\label{prop:spanningtreedecomp}
    \[
    C_G(\bm w) = \sum_{T\in \cT(G)} \left(\prod_{e\in T}w_e\right) \left(\prod_{e\in \Ext(T)}(1+w_e)\right).
    \]
    Furthermore as polynomial identities in $q$,
    \[
    T_G(1,q) = \sum_{T\in \cT(G)}q^{|\Ext(T)|}.
    \]
    Consequently, $T_G(1,q)$ has nonnegative integer coefficients.
\end{restatable}

\begin{proof}
    Given a connected spanning edge set $F\subseteq E$, obtain a spanning tree in the following manner: scan the edges in $F$ in increasing order, retaining an edge only if it joins two different components of the edges already retained. This is precisely Kruskal's spanning tree construction~\cite{kruskal1956shortest}.

    For a fixed spanning tree $T$, this procedure returns $T$ exactly for the edges satisfying $ T \subseteq F\subseteq T \cup \Ext(T)$. To see this, note that every discarded edge in the scan had its endpoints joined by a path of previously retained edges, which are smaller than the discarded edge. Hence, every edge of $F \setminus T$ belongs to $\Ext(T)$ by construction. Conversely, suppose $T \subseteq F\subseteq T \cup \Ext(T)$. Induct through the scanning procedure: the retained edges are exactly the tree edges already encountered. An edge of $T$ is retained because the other tree edges cannot connect its endpoint vertices. Any edge in $F\setminus T$ is discarded as they have order higher than the smaller tree edges already retained.

    As a result, the connected spanning edge sets can be partitioned into the intervals organized by spanning trees:
    \[
    \{T \cup A ~:~ A \subseteq \Ext(T)\}
    \]
    for $T\in \cT(G)$. Reorganizing the summation of the connected polynomial, we get, 
    \begin{align*}
        C_G(\bm w) &= \sum_{T\in\cT(G)} \sum_{A\subseteq \Ext(T)} \left(\prod_{e\in T}w_e\right)\left(\prod_{f\in A}w_f\right) = \sum_{T\in\cT(G)}\left(\prod_{e\in T}w_e\right) \left[\prod_{f\in \Ext(T)}(1+w_f)\right].
    \end{align*}
    This proves the first part. Now, consider every edge weight equal to $q-1$, and use the fact that $C_G(q-1) = (q-1)^{m-1}T_G(1,q) $ to conclude 
    \[
    T_G(1,q) = \sum_{T\in \cT(G)}q^{|\Ext(T)|},
    \]
    as desired.
\end{proof}

Now we are finally ready to prove \cref{lem:bmq-facts}.
\subsection{Proof of \cref{lem:bmq-facts}}
We have for $q\in [-1,1)$, 
\[
\cL_q(z) = \prod_{j \geq 1}\left(1 + \frac{z^2}{t_j(q)^2}\right), \qquad 0 < t_1(q) \leq t_2(q) \leq \dots 
\]
Furthermore, by \cref{cor:scrl-pde}, $\varphi_q(z) := \cL_q(\iota z)$ satisfies
\[
8q(1-q)\partial_q\varphi_q = (1-q)z^2 \partial_z^2\varphi_q + (3+q)z\partial_z\varphi_q + 4z^2 \varphi_q.
\]
See~\cite{SYK_Jia} for a physics perspective on these computations.
\begin{lemma}\label{lem:bmrootsrelation}
    Each $b_m$ has nonnegative coefficients. Moreover, for $-1\leq q < 1$, 
    \[
    (1-q)^{m-1} b_m(q) = \sum_{j \geq 1}t_j(q)^{-2m}.
    \]
    In particular, $\sum_{j \geq 1}1/t_j(q)^2 = 1/2$.
\end{lemma}
\begin{proof}
    By~\cref{lem:JVcumulant}, we have 
    \[
    \kappa_{2m}(q) = (q-1)^{m-1}\frac{(2m)!}{m}b_m(q).
    \]
    By \cref{prop:spanningtreedecomp}, every $T_{G_\pi}(1,q)$ has nonnegative coefficients, and so does $b_m(q)$. 
    
    We now obtain the expression for log of $\varphi_q(z)$. Substituting the cumulant formula yields
   \begin{align*}
        \log \varphi_q(z) &= \sum_{m\geq 1}\frac{\kappa_{2m}(q)}{(2m)!} (\iota z)^{2m} = \sum_{m \geq 1}\frac{(-1)^m(q-1)^{m -1 }b_{m}(q)}{m} z^{2m} = -\sum_{m \geq 1}\frac{(1-q)^{m -1 }b_{m}(q)}{m} z^{2m}.
   \end{align*}
   Thus, 
   \[
   -\log\varphi_q(z) = \sum_{m \geq 1}\frac{(1-q)^{m -1 }b_{m}(q)}{m} z^{2m}.
   \]
    Recall the product formula from \cref{thm:moment-function-imaginary-rooted}:
    \[
    \varphi_q(z) = \prod_{j\geq 1}\left(1- \frac{z^2}{t_j(q)^2}\right)
    \]
    Since $\varphi_q(z) = 1-z^2/2 + O(z^4)$, comparison of the $z^2$ coefficient in the locally uniformly convergent product formula gives: 
    \[
    \sum_{j \geq 1}\frac{1}{t_j(q)^2} = \frac{1}{2}.
    \]
    Thus for $|z| < t_1(q)$, we take the log of the product and expand each factor: 

    \begin{align*}
        -\log\varphi_q(z) &= -\sum_{j \geq 1}\log\left(1-\frac{z^2}{t_j(q)^2}\right) = \sum_{j \geq 1} \sum_{m\geq 1}\frac{z^{2m}}{m t_j(q)^{2m}} = \sum_{m\geq 1} \frac{z^{2m}}{m}\left(\sum_{j\geq 1} \frac{1}{t_j(q)^{2m}}\right)
    \end{align*}
    where the interchanging sums is justified by absolutely and locally uniform convergence. 

    As a result, 
    \[
    (1-q)^{m-1}b_m(q) = \sum_{j\geq 1}\frac{1}{t_j(q)^{2m}}.\qedhere
    \]
\end{proof}
Now, we will show that $b_m(q)$ is nondecreasing in $q$, which will later be used for \cref{lem:endpointseries}.
\begin{proposition}\label{prop:bm-monotonicity}
    For every $m\geq 1$, the map $[-1,1]\ni q\mapsto b_m(q)$ is non-decreasing.
\end{proposition}

\begin{proof}
    By \cref{prop:spanningtreedecomp}, each $b_m$ has nonnegative
    coefficients, so it is nondecreasing on $[0,1]$.
    It remains to prove monotonicity on $[-1,0]$. From~\cref{lem:bmrootsrelation} and~\cref{lem:JVcumulant} we have:
    \[
        \log\varphi_q(z)
        =-\sum_{m\geq 1}
        \frac{(1-q)^{m-1}b_m(q)}{m}z^{2m}.
    \]
    Write $\ell_q(z):=\log\varphi_q(z)$. Dividing the differential
    equation for $\ell_q$ by $\varphi_q$ gives
    \[
        8q(1-q)\partial_q\ell_q
        =(1-q)z^2\bigl(\partial_z^2\ell_q
        +(\partial_z\ell_q)^2\bigr)
        +(3+q)z\partial_z\ell_q+4z^2.
    \]
    Substituting the logarithmic expansion and comparing coefficients
    of $z^{2m}$ yields, for every $m\geq 2$,
    \begin{equation}\label{eq:bm-recurrence}
        2q(1-q)b_m'(q)
        =
        \bigl[m(m+1)-(m-1)(m-2)q\bigr]b_m(q)
        -m\sum_{j=1}^{m-1}b_j(q)b_{m-j}(q).
    \end{equation}
    This is a polynomial identity in $q$.

    We first determine $b_m'(-1)$. Since
    $\varphi_{-1}(z)=\cos z$, evaluating the differential
    equation at $q=-1$ gives
    \[
        \left.\partial_q\ell_q(z)\right|_{q=-1}
        =\frac{z\tan z-z^2}{8}.
    \]
    Put $p_m(q):=(1-q)^{m-1}b_m(q)$. Then
    \[
        \ell_q(z)=-\sum_{m\geq1}\frac{p_m(q)}m z^{2m},
        \qquad
        z\tan z=-z\ell_{-1}'(z)
        =2\sum_{m\geq1}p_m(-1)z^{2m}.
    \]
    Comparing coefficients of $z^{2m}$, for $m\geq2$, gives
    \[
        p_m'(-1)=-\frac m4p_m(-1).
    \]
    Differentiating $p_m(q)=(1-q)^{m-1}b_m(q)$ therefore yields
    \begin{equation}\label{eq:bm-derivative-minus-one}
        b_m'(-1)=\frac{m-2}{4}b_m(-1),
        \qquad m\geq2.
    \end{equation}

    We now prove $b_m'(q)\geq0$ on $[-1,0]$ by induction on $m$.
    The cases $m=1,2$ are immediate, since
    \[
        b_1(q)=\frac12,
        \qquad b_2(q)=\frac1{12}.
    \]
    Fix $m\geq3$ and suppose the assertion holds for all $j<m$.
    Define
    \[
        H_m(q):=\sum_{j=1}^{m-1}b_j(q)b_{m-j}(q).
    \]
    By~\cref{lem:bmrootsrelation}, $b_j(q)>0$ for $-1\leq q<1$.
    The induction hypothesis therefore implies
    \[
        H_m'(q)\geq0
        \qquad (-1\leq q\leq0).
    \]
    Moreover, \cref{eq:bm-derivative-minus-one} gives
    $b_m'(-1)>0$.

    Suppose, for a contradiction, that $b_m'$ vanishes at some
    $q_0\in(-1,0)$. Choose $q_0$ to be its first zero after $-1$.
    Then $b_m'(q)>0$ for $-1\leq q<q_0$, and hence
    $b_m''(q_0)\leq0$.

    Differentiating \cref{eq:bm-recurrence} and evaluating at $q_0$,
    where $b_m'(q_0)=0$, gives
    \[
        2q_0(1-q_0)b_m''(q_0)
        =
        -(m-1)(m-2)b_m(q_0)-mH_m'(q_0)<0.
    \]
    Since $q_0(1-q_0)<0$, this implies $b_m''(q_0)>0$,
    a contradiction.

    Thus $b_m'(q)>0$ on $[-1,0)$ for $m\geq3$, and continuity
    gives $b_m'(0)\geq0$. This completes the induction and,
    together with coefficient nonnegativity on $[0,1]$,
    proves the proposition.
\end{proof}
Now, we have the final technical result to prove \cref{lem:bmq-facts}, which is the radius of convergence of the power series $B(u) = \sum_{m\geq 1}b_m(1) u^m$.
\begin{lemma}\label{lem:endpointseries}
    The power series $B(u)$ has radius of convergence $R_*$ and $B(R_*)=2$.
\end{lemma}

\begin{proof}
    Evaluating \cref{eq:bm-recurrence} at $q=1$ gives
    \[
        b_1(1)=\frac12,
        \qquad
        (4m-2)b_m(1)
        =m\sum_{j=1}^{m-1}b_j(1)b_{m-j}(1),
        \qquad m\geq2.
    \]
    Multiplying by $u^m$ and summing over $m\geq2$, we obtain
    \begin{equation}\label{eq:B-ode}
        2u(2-B(u))B'(u)=2B(u)+u.
    \end{equation}

    We claim that the solution with $B(0)=0$ admits the
    parametrization
    \begin{equation}\label{eq:B-param}
        u=\frac{4s^2}{\cosh^2s},
        \qquad
        B(u)=2s\tanh s.
    \end{equation}
    Indeed, both expressions are even analytic functions of $s$
    near the origin, with
    \[
        u=4s^2+O(s^4),
        \qquad
        B(u)=2s^2+O(s^4).
    \]
    Applying the analytic inverse function theorem in the variable
    $s^2$ therefore defines an analytic function $B$ near $u=0$.
    Direct differentiation verifies \cref{eq:B-ode}, and the
    preceding recurrence uniquely determines its Taylor coefficients.
    Thus \cref{eq:B-param} parametrizes the required power series.

    Recall that $s_*>0$ is the unique solution of
    $s_*\tanh s_*=1$. Differentiating the first equation in
    \cref{eq:B-param} gives
    \[
        \frac{du}{ds}
        =\frac{8s}{\cosh^2s}(1-s\tanh s)>0,
        \qquad 0<s<s_*.
    \]
    Hence $u(s)$ increases from $0$ to
    \[
        u(s_*)=\frac{4s_*^2}{\cosh^2s_*}
        =\zeta_*^2=R_*.
    \]
    Since $du/ds\neq0$ for $0<s<s_*$, the parametrization
    analytically continues $B$ along the entire interval $(0,R_*)$.

    By~\cref{lem:bmrootsrelation}, $b_m(1)\geq0$ for every $m$.
    \cref{item:cauchy-hadamard-pringsheim} of \cref{fact:complex-analysis} therefore implies that the radius of
    convergence of $B$ cannot be smaller than $R_*$, since
    otherwise $B$ would have a singularity at an earlier positive
    point.

    On the other hand, as $u\nearrow R_*$, the parametrization gives
    \[
        B(u)\longrightarrow 2s_*\tanh s_*=2.
    \]
    Consequently, \cref{eq:B-ode} implies
    \[
        B'(u)
        =\frac{2B(u)+u}{2u(2-B(u))}
        \longrightarrow+\infty.
    \]
    Thus $B$ cannot be analytic at $R_*$, and its radius of
    convergence is exactly $R_*$.

    Finally, since all coefficients of $B$ are nonnegative,
    monotone convergence yields
    \[
        \sum_{m\geq1}b_m(1)R_*^m
        =\lim_{u\nearrow R_*}B(u)
        =2,
    \]
    completing the proof.
\end{proof}

\section{Asymptotic Lee-Yang property of the SYK model: Proofs of \cref{thm:annealed-syk-lee-yang,thm:sykstrip}}\label{sec:syk}
We now prove \cref{thm:annealed-syk-lee-yang,thm:sykstrip}. Throughout this section $2\leq k < n$ are even integers, and $\tr$ denotes normalized trace, i.e. $\tr(\Id)=1$. We have the Hamiltonian,
\[
H_{\syk}(\bm g) = \sigma_{n,k}\sum_{|S| = k}g_S\Gamma_S, \qquad \sigma_{n,k}^2:= \frac{(k-1)!}{2kn^{k-1}}.
\]
Recall, 
\[
\alpha_n := \frac{k^2}{n}, \quad q_n:=e^{-2\alpha_n}, \text{ and } V_n := \E_{\bm g}\tr H_{\syk}^2.
\]
We will frequently use the following elementary bound on the variance: 

\begin{proposition}[Bounds on variance]\label{prop:sykvariancebound}
    For every $n \geq k \geq 2$, we have 
    \[
    V_n \leq \frac{e^{-\alpha_n/4}}{2\alpha_n}\leq\frac{1}{2\alpha_n}.
    \]
    In particular, if $\alpha_n \log n \to \infty$ as $n\to\infty$, $V_n = o(\log n)$.
    \end{proposition}
\begin{proof}
    We have, 
    \[
    V_n = \sigma_{n,k}^2 \binom{n}{k}  = \frac{{(n)_k}}{n^k (2\alpha_n)} = \frac{1}{2\alpha_n}\cdot\prod_{j=0}^{k-1}\left(1-\frac{j}{n}\right) \leq \frac{1}{2\alpha_n}\cdot\exp{\left(-\frac{k(k-1)}{2n}\right)} \leq\frac{e^{-\alpha_n/4}}{2\alpha_n}.\qedhere
   \]
\end{proof}

Now, we have the quenched and annealed partition functions as,  
\[
Z_n(z; \bm g) := \tr e^{z H_{\syk}(\bm g)}, \qquad Z_n(z) := \E_{\bm g}Z_n(z;\bm g).
\]
Recall also that $\Omega := \C \setminus \{\iota y: y \in \R, |y| \geq \zeta_*\}$.
\annealedleeyang*
Now, recall the closed strip along the real axis:
\[
\cR_n := \{z \in \C : |\Ree z| \leq n^{0.08}, \quad |\Imm z| \leq \zeta_*\}\subset\C.
\]

\quenchedstrip*

We will prove \cref{thm:annealed-syk-lee-yang,thm:sykstrip} by comparing the moments of the SYK Hamiltonian to the $q-$oscillator, and using \cref{thm:moment-function-imaginary-rooted,thm:nearest-root} for the zeroes of the $q-$oscillator. To do so, we define a rescaled partition function for the $q-$oscillator as: 
\[
Q_n(z) := \cL_q(\sqrt V_n z)
\]
The zeroes of $Q_n$ are at $\pm \iota t_j(q) /\sqrt{V_n}$. Using \cref{prop:sykvariancebound}, since 
\[
V_n(1-q_n) \leq \frac{1-e^{-2\alpha_n}}{2\alpha_n} < 1
\]
the zeroes of $Q_n$ are strictly outside the closed strip of height $\zeta_*$ by \cref{thm:nearest-root}.
We will first show a moment comparison bound between $Z_n(z)$ and $Q_n(z)$ in the following.
\begin{restatable}{lemma}{partfntransfer}\label{lem:partfntransfer}
    Fix $r\geq 1$ and $D > 0$ and choose $A_D \geq \max \{e,D/\log 2\}$. Set,
    \[
        L:=\left\lceil A_D(Vr^2+\log n)\right\rceil \quad d:=\frac{2BL^3k^3}{n^2}
    \]
    Then, whenever $n\geq ALk$, we have for all
    $z\in\mathbb{C}$ with $|z|\leq r$,
    \[
        |Z_n(z)-Q_n(z)|
        \leq (e^d-1)Q_n(|z|)+2n^{-D},
    \]
where $A,B$ are absolute constants.
\end{restatable}

\cref{lem:partfntransfer} will suffice to show the annealed Lee-Yang result of \cref{thm:annealed-syk-lee-yang}. 

\begin{proof}[Proof of \cref{thm:annealed-syk-lee-yang} using \cref{lem:partfntransfer}]
    Suppose $\alpha_n \to \alpha \in(0,\infty).$ The variance formula in \cref{prop:sykvariancebound} gives:
    \[
    V_n = \frac{1}{2\alpha_n} \prod_{j=0}^{k-1}\left(1-\frac{j}{n}\right)
    \]
    Now, 
    \[
    \log \prod_{j=0}^{k-1}\left(1-\frac{j}{n}\right) = -\sum_{j=0}^{k-1}\left(\frac{j}{n} + O\left(\frac{j}{n}\right)^2\right) =  -\frac{k(k-1)}{2n} + O\left(\frac{k^3}{n^2}\right)
    \]
    Thus as $n,k\to\infty$,
    \[
    V_n \to v_\alpha := \frac{e^{-\alpha/2}}{2\alpha}, \quad q_n=e^{-2\alpha_n} \to q_\alpha =e^{-2\alpha}.
    \]
    For any $r\geq 1$ and fixed $D=10$, \cref{lem:partfntransfer} one has:
    \[
    L = O_r(\log n), \quad d = O\left(\frac{(\log n)^3}{\sqrt n}\right) = o(1)
    \]
    and thus $n \geq ALk$ eventually. Since $Q_n(r) \leq e^{Vr^2/2} = O_r(1)$, it follows that, 
    \[
   \sup_{z:|z| \leq r} |Z_n(z) - Q_n(z)| \to 0.
    \]
    Joint continuity from \cref{claim:gnq-local-conv} gives,
    \[
    Q_n(z) = \cL_{q_n}(\sqrt{V_n}z) \to \cL_{q_\alpha}(\sqrt{v_\alpha}z) = Z_\infty(z),
    \]
    locally uniformly.
\end{proof}

To show \cref{thm:sykstrip} for the quenched partition function, we will need to bound the pointwise fluctuations of $Z_n(z;\bm g)$ around $Z_n(g)$, which we do as the following \cref{lem:sykmomentbound}.
\begin{restatable}{lemma}{sykmomentbound}\label{lem:sykmomentbound}
    For every $z= x + \iota y\in\C$, we have that 
    \[
    \E_{\bm g} \left|Z_n(z;\bm g) - Z_n(z)\right|^2 \leq \sigma_{n,k}^2 |z|^2 e^{2V_n x^2}
    \]
    where 
    \[
    \sigma_{n,k}^2 :=\frac{(k-1)!}{2k n^{k-1}}.
    \]
\end{restatable}
We are now ready to prove \cref{thm:annealed-syk-lee-yang,thm:sykstrip}. 
\begin{proof}[Proof of \cref{thm:sykstrip} assuming \cref{lem:partfntransfer,lem:sykmomentbound}]
Fix a sequence of even integers $k=k(n)$, with $n$ also even, such that $2\leq k<n$, $k(n)/\sqrt{n/\log n}=\omega(1)$. Let $K$ be a compact set satisfying
\[
    K\subset\Omega
    :=\C\setminus\{\iota y:y\in\R,\ |y|\geq\zeta_*\}.
\]
For any constant $\gamma\in(0,1/12)$ define the closed strip,
\[
    \cR_n
    :=\{z\in\C:|\Ree z|\leq n^\gamma,\quad |\Imm z|\leq \zeta_*\}.
\]
We will show annealed zero freeness on $K \cup \cR_n$, and quenched zero freeness on $\cR_n$. All sufficiently large $n$ thresholds are along the given sequence of $k=k(n)$.

Set $R:=n^\gamma$, and $r:=1+\max\{R+\zeta_*,\max_{z\in K}|z|\}$. This ensures that the disk $\cD_r:=\{z\in\C:|z|\leq r\}$ contains both $\cR_n$ and $K$, with $r=O_{K}(n^\gamma)$. The sequence $k=k(n)$ gives
\[
    \alpha_n\log n=\frac{k^2\log n}{n}\longrightarrow\infty\overset{\text{\cref{prop:sykvariancebound}}}{\implies}
    \frac{V_n}{\log n}
    \leq\frac{1}{2\alpha_n\log n}\longrightarrow0,
\]
so $V_n=o(\log n)$. 

\begin{claim}\label{cl:sykproductbound}
    Let
    \[
        \cU:=K\cup\{z\in\C:|\Imm z|\leq \zeta_*\}.
    \]
    Then $Q_n$ has no zeros on $\cU$. Furthermore, there is a constant $C=C(K)>0$, independent of $n,k$, such that for every $z\in\cU$,
    \[
        |Q_n(z)|
        \geq e^{-CV_n}Q_n(|z|)
        \geq e^{-CV_n}.
    \]
\end{claim}

Using this claim, we will first show that for all sufficiently large $n$,
\begin{equation}\label{eq:sykannealedlower}
    \inf_{z\in K\cup\cR_n}|Z_n(z)|
    \geq\frac12e^{-CV_n}.
\end{equation}
Let us divide the $k=k(n) = \omega(\sqrt{n/\log n})$ into two categories:
\[
\underbrace{\sqrt{\frac{n}{\log n}} \ll k \leq 4\sqrt{n \log n},}_{\text{case (i)}} \text{ and } \underbrace{k \geq 4\sqrt{n\log n}}_{\text{case (ii)}}. 
\]
For (i), we will need to compare to the $q-$oscillator, while (ii) will constrain the annealed partition function to be unity on the disk of concern.

\textbf{Case (i)} Suppose that $k\leq4\sqrt{n\log n}$. Then, upon choosing
$D=10$ in \cref{lem:partfntransfer}, we have
\begin{align*}
    L&:=\left\lceil A_D(V_n r^2+\log n)\right\rceil
        =O(n^{2\gamma}\log n),\\
    d&:=\frac{2BL^3k^3}{n^2}
        =O\left(n^{6\gamma-1/2}(\log n)^{9/2}\right)
        \leq n^{-\eta+o(1)},\\
    \frac{Lk}{n}
        &=O\left(n^{2\gamma-1/2}(\log n)^{3/2}\right)
        =o(1),
\end{align*}
where $\eta:=1/2-6\gamma>0$. As a result, $n\geq ALk$ for all sufficiently large $n$, and \cref{lem:partfntransfer} applies. For $z\in\cD_r$, we then have
\[
    |Z_n(z)-Q_n(z)|
    \leq(e^d-1)Q_n(|z|)+2n^{-10}.
\]
For $z\in K\cup\cR_n$, dividing by $|Q_n(z)|$ and using \cref{cl:sykproductbound} gives
\[
    \sup_{z\in K\cup\cR_n}
    \left|\frac{Z_n(z)}{Q_n(z)}-1\right|
    \leq(e^d-1+2n^{-10})e^{CV_n}
    =o(1),
\]
since $d\leq n^{-\eta+o(1)}$ and $e^{CV_n}=n^{o(1)}$. As a result, for all sufficiently large $n$, the supremum is less than $1/2$, and hence for $z\in K\cup\cR_n$
\[
    |Z_n(z)|
    \geq\frac12|Q_n(z)|
    \geq\frac12e^{-CV_n}.
\]

\textbf{Case (ii)} Now consider $k>4\sqrt{n\log n}$. Then $\alpha_n>16\log n$, and \cref{prop:sykvariancebound} gives
\[
    V_n\leq\frac{e^{-\alpha_n/4}}{2\alpha_n}
      \leq\frac{n^{-4}}{2\alpha_n}.
\]
Thus $V_nr^2=o(1)$. Using the vanishing of the annealed odd moments and the upper bound on the even moments, we obtain
\[
    \sup_{|z|\leq r}|Z_n(z)-1|
    \leq\sum_{\ell\geq1}
        \frac{(2\ell-1)!!\,V_n^\ell r^{2\ell}}{(2\ell)!}
    =e^{V_nr^2/2}-1
    =o(1).
\]
Thus, for all sufficiently large $n$,
\[
    \inf_{z\in K\cup\cR_n}|Z_n(z)|
    \geq\frac12
    \geq\frac12e^{-CV_n}.
\]
This proves \eqref{eq:sykannealedlower} in both regimes of $k$.

In particular, $Z_n$ is zero-free on $\cR_n\cup K$ for all sufficiently large $n$, in particular establishing annealed zero freeness on the closed strip $\cR_n$.

We now show uniform quenched concentration to obtain a zero-free strip for the quenched partition function. Denote,
\[
    G_n(z;\bm g):=Z_n(z;\bm g)-Z_n(z).
\]
For each $\bm g$, this is an entire function of $z$. Cauchy's integral formula on the circle of radius $2r$ gives, for every $z\in \cD_r$,
\[
    G_n(z;\bm g)
    =\int_0^{2\pi}
       G_n(2re^{\iota\theta};\bm g)
       \frac{2re^{\iota\theta}}{2re^{\iota\theta}-z}
       \frac{d\theta}{2\pi}.
\]
Now, since $|z|\leq r$, we have
\[
    \left|
        \frac{2re^{\iota\theta}}{2re^{\iota\theta}-z}
    \right|
    \leq\frac{2r}{2r-|z|}
    \leq2,
\]
and an application of Cauchy-Schwarz gives
\[
    \sup_{|z|\leq r}|G_n(z;\bm g)|^2
    \leq4\int_0^{2\pi}
          |G_n(2re^{\iota\theta};\bm g)|^2
          \frac{d\theta}{2\pi}.
\]
By \cref{lem:sykmomentbound}, for $z=x+\iota y$ on the
circle of radius $2r$,
\[
    \E_{\bm g}|G_n(z;\bm g)|^2
    \leq\sigma_{n,k}^2|z|^2e^{2V_nx^2}
    \leq4\sigma_{n,k}^2r^2e^{8V_nr^2},
\]
where
\[
    \sigma_{n,k}^2
    :=\frac{(k-1)!}{2k\,n^{k-1}}.
\]
Taking expectations we conclude that
\begin{align*}
    \E_{\bm g}\sup_{|z|\leq r}|G_n(z;\bm g)|^2
    &\leq4\int_0^{2\pi}
          \E_{\bm g}|G_n(2re^{\iota\theta};\bm g)|^2
          \frac{d\theta}{2\pi}\\
    &\leq16\sigma_{n,k}^2r^2e^{8V_nr^2}.
\end{align*}
Recall that $\cR_n\subset\cD_r$. Applying Markov's
inequality gives
\[
    \Pr_{\bm g}\left(
        \sup_{z\in\cR_n}|G_n(z;\bm g)|
        \geq\frac{e^{-CV_n}}2
    \right)
    \leq64\sigma_{n,k}^2r^2e^{8V_nr^2+2CV_n}.
\]
It remains to bound this exceptional probability.

\begin{claim}\label{cl:sykcouplingdecay}
    For all sufficiently large $n$ along the fixed sequence
    $k=k(n)$, we have
    \[
        \sigma_{n,k}
        \leq e^{-\sqrt{n\log n}/4}.
    \]
\end{claim}

Using this claim, for all sufficiently large $n$ we note that, 
\begin{align*}
    \sigma_{n,k}^2
        &\leq e^{-\sqrt{n\log n}/2},\\
    V_nr^2
        &=O(n^{2\gamma}\log n)
          =o(\sqrt{n\log n}),\\
    CV_n+\log r
        &=o(\sqrt{n\log n}).
\end{align*}
Here the second estimate uses $\gamma<1/12$, and the third
uses $V_n=o(\log n)$ and $r=O_{K}(n^\gamma)$.
As a result,
\begin{align*}
    \Pr_{\bm g}\left(
        \sup_{z\in\cR_n}|G_n(z;\bm g)|
        \geq\frac{e^{-CV_n}}2
    \right)
    &\leq
       \exp\left(
          -\frac12\sqrt{n\log n}
          +o(\sqrt{n\log n})
       \right)\\
    &\leq e^{-c\sqrt{n\log n}}
\end{align*}
for an absolute constant $c>0$ and all sufficiently large $n$.
Consequently, except with probability at most
$e^{-c\sqrt{n\log n}}$, we have simultaneously for every
$z\in\cR_n$,
\[
    |G_n(z;\bm g)|
    <\frac{e^{-CV_n}}2
    \leq|Z_n(z)|,
\]
where the last inequality follows from
\eqref{eq:sykannealedlower}. Therefore
\[
    |Z_n(z;\bm g)|
    \geq|Z_n(z)|-|G_n(z;\bm g)|
    >0
    \qquad(z\in\cR_n).
\]
This proves \cref{thm:sykstrip}.
\end{proof}

Now, to prove \cref{lem:partfntransfer,lem:sykmomentbound}, we need some technical preliminaries about graph polynomials, which we now prove. We defer the proofs of \cref{cl:sykcouplingdecay,cl:sykproductbound} to the very end of this section. 

\subsection{Bounds on graph polynomials}
We start with a simple bound on the connected graph polynomial.
\begin{proposition}\label{prop:connpolybound}
    Let $G=(V,E)$ be a simple graph on $m\geq 2$ vertices with real edge weights $w_e \in [-2,0]$. Then, 
    \[
    |C_G(\bm w)| \leq 2^{m-1} m^{m-2}
    \]
\end{proposition}
\begin{proof}
    If $G$ is disconnected, then $C_G(\bm w)=0$ and we are done. Otherwise, apply \cref{prop:spanningtreedecomp}. Note that $|w_e|\leq 2$ and $|1+w_e|\leq 1$ by assumption. We obtain: 
    \[|C_G(\bm w)| = \left|\sum_{T\in\cT(G)} \left(\prod_{e\in T}w_e\right)\left[\prod_{f\in\Ext(T)}(1+w_f)\right]\right|\leq \sum_{T\in\cT(G)} \left(\prod_{e\in T}|w_e|\right)\left[\prod_{f\in\Ext(T)}|1+w_f|\right]\]
    \[\leq \sum_{T\in\cT(G)}2^{m-1}\leq 2^{m-1} |\cT(G)|\leq 2^{m-1}m^{m-2},\]
    where the last inequality $|\cT(G)| \leq m^{m-2}$ follows from Cayley's formula~\cite{Cayley_2009}.
\end{proof}
Now, we will prove a key technical result that will be used in comparing the moments of doubly-scaled SYK to the $q-$oscillator in~\cref{lem:partfntransfer}. The proof of the following will involve an argument akin to the cavity method.
\begin{lemma}\label{lem:graph-log}
    There are absolute constants $A,B > 0$ such that for every simple graph on $M \ge 1$ vertices with real edge weights $w_e \in [-2,0]$ and every real $N \geq AM$ one has $P_G(N;\bm w) > 0$ and 
    \[
    \left|\log\frac{P_G(N;\bm w)}{N^M} - \frac{1}{N}\sum_{e\in E}w_e\right| \leq B\frac{M^3}{N^2}    
    \]
\end{lemma}

\begin{proof}
For every $U\subseteq V$, define
\[
Z(U):=\frac{P_{G[U]}(N;\bm w)}{N^{|U|}}
\]
and let $Z(\emptyset):=1$. By \cref{prop:familydecomposition},
\[
Z(U) = \sum_{\mathcal F\in\mathcal D(U)} \prod_{S\in\mathcal F} \frac{C_{G[S]}(\bm w)}{N^{|S|-1}}.
\]
Fix a vertex $v\in U$. A family $\cF\in \cD(U)$ either has no member containing $v$, or has a unique such member $S$; in the latter case, its remaining members form a family on $U\setminus S$. Thus,
\begin{equation}\label{eq:graph-log-deletion}
Z(U) = Z(U\setminus\{v\})+ \sum_{\substack{S\subseteq U\\v\in S,\ |S|\geq 2}} \frac{C_{G[S]}(\bm w)}{N^{|S|-1}}\,Z(U\setminus S).
\end{equation}
Choose a positive constant $A \ge 16e$, so that
\[
\delta := \sum_{s\geq 2}(4e\varepsilon)^{s-1}= \frac{4e\eps}{1-4e\eps} \leq \frac13<\frac12.
\]
where $\eps:=M/N$. In particular, $\delta=O(\eps)$.

We will show by induction on $|U|$ that $Z(U)>0$ and, for every $v\in U$ the ratio 
\begin{equation}\label{eq:graph-log-ratio}
R_U(v):=\frac{Z(U)}{Z(U\setminus\{v\})}
\qquad\text{satisfies}\qquad
|R_U(v)-1|\leq\delta.
\end{equation}
The assertion is immediate for the empty set and singletons.

Suppose this holds for all proper subsets of $U$. For $S\subseteq U$ containing $v$, remove the vertices of $S\setminus\{v\}$ one at a time. This expresses ${Z(U\setminus S)}/{Z(U\setminus\{v\})}$ as a product of $|S|-1$ inverse deletion ratios on proper subsets of $U$. By the induction hypothesis, each factor is positive and at most $2$. Consequently,
\[
0< \frac{Z(U\setminus S)}{Z(U\setminus\{v\})} \leq 2^{|S|-1}.
\]
Divide \cref{eq:graph-log-deletion} by $Z(U\setminus\{v\})$, which is positive by induction. Applying \cref{prop:connpolybound} gives us,
\[|R_U(v)-1| \leq \sum_{\substack{S\subseteq U\\v\in S,\ |S|\geq 2}}\frac{|C_{G[S]}(\bm w)|}{N^{|S|-1}}\,2^{|S|-1}\leq \sum_{s=2}^{|U|} \binom{|U|-1}{s-1} \frac{4^{s-1}s^{s-2}}{N^{s-1}}\leq \sum_{s\geq 2}(4e\eps)^{s-1} =\delta.\]
Here we used
\[
\binom{|U|-1}{s-1}\leq\frac{M^{s-1}}{(s-1)!}, \qquad
\frac{s^{s-2}}{(s-1)!}\leq e^{s-1}.
\]
Since all quantities are real and $\delta<1/2$, we obtain $R_U(v)>0$,
and hence $Z(U)>0$. This completes the induction. We now separate the terms with $|S|=2$ in \cref{eq:graph-log-deletion}. Write $w_{\{u,v\}}=0$ when $\{u,v\}\notin E$. Since
\[
C_{G[\{u,v\}]}(\bm w)=w_{\{u,v\}}, \qquad \frac{Z(U\setminus\{u,v\})}{Z(U\setminus\{v\})} = \frac{1}{R_{U\setminus\{v\}}(u)},
\]
the divided recursion becomes
\begin{equation}\label{eq:graph-log-pair-terms}
R_U(v)-1=\frac1N \sum_{u\in U\setminus\{v\}} \frac{w_{\{u,v\}}}{R_{U\setminus\{v\}}(u)}+ O(\eps^2).
\end{equation}
Indeed, the same estimate used in the induction bounds the total contribution from $|S|\geq 3$ by
\[
\sum_{s\geq 3}(4e\eps)^{s-1}=
\frac{(4e\eps)^2}{1-4e\eps}=O(\eps^2).
\]

By \cref{eq:graph-log-ratio}, uniformly over all deletion ratios,
\[
\left|\frac1{R_U(v)}-1\right|=\frac{|R_U(v)-1|}{R_U(v)}
\leq 2\delta = O(\eps).
\]
Moreover,
\[
\frac1N \sum_{u\in U\setminus\{v\}}|w_{\{u,v\}}| \leq \frac{2M}{N}
=
2\varepsilon.
\]
Thus replacing the inverse ratios in \cref{eq:graph-log-pair-terms} by $1$ introduces another error of order $\varepsilon^2$. We conclude that
\[
R_U(v)-1
= \frac1N \sum_{u\in U\setminus\{v\}}w_{\{u,v\}}+O(\varepsilon^2).
\]
Because $R_U(v)=1+O(\varepsilon)$ and $|R_U(v)-1|<1/2$, the expansion $\log(1+x)=x+O(x^2)$ gives
\begin{equation}\label{eq:graph-log-local}
\log R_U(v)=
\frac1N \sum_{u\in U\setminus\{v\}}w_{\{u,v\}}+O(\varepsilon^2).
\end{equation}

Finally, order the vertices as $v_1,\dots,v_M$ and set
\[
U_i:=\{v_i,\dots,v_M\}, \qquad U_{M+1}:=\varnothing.
\]
The deletion ratios telescope:
\[
\log Z(V)
= \sum_{i=1}^M
\log\frac{Z(U_i)}{Z(U_{i+1})}= \sum_{i=1}^M\log R_{U_i}(v_i).
\]
Summing \cref{eq:graph-log-local}, each edge appears exactly once, when its first endpoint is deleted. Therefore
\[
\log Z(V) =
\frac1N\sum_{e\in E}w_e + O(M\varepsilon^2) =
\frac1N\sum_{e\in E}w_e + O\!\left(\frac{M^3}{N^2}\right).
\]
Choosing an absolute constant $B$ to bound the final error proves the stated estimate. The positivity established above also gives
\[
P_G(N;\bm w)=N^M Z(V)>0.\qedhere
\]
\end{proof}

We are now ready to show \cref{lem:partfntransfer,lem:sykmomentbound}.

\subsection{Proof of \cref{lem:partfntransfer,lem:sykmomentbound}}
For a finite graph $G=(V,E)$,
\[
W_G(n,k) := \E_{S_v \sim \Unif\binom{[n]}{k}} \prod_{\{u,v\}\in E(G)} (-1)^{|S_u\cap S_v|}
\]
where the sets are chosen independently. Using Isserlis-Wick formula~\cite{Isserlis18,Wick50}, the moments of the SYK Hamiltonian can be expressed in terms of $W_G(n,k)$ for any $\ell \geq 1$:
\[
\E \tr H_{\syk}^{2\ell} = V^{\ell} \sum_{\pi\in\cP_2(2\ell)} W_{G_\pi}(n,k),
\]
where $G_\pi$ is the crossing graph of $\pi \in \cP_2(2\ell)$, and $V$ is the variance: 
\[
V = \E_{\bm g}\tr H_{\syk}^2.
\]
See~\cite{BasuKM26} for a derivation of the same.

\partfntransfer*
\begin{proof}[Proof of \cref{lem:partfntransfer}]
    Given a simple graph $G$ on $[\ell]$, construct a weighted graph $J$ on $[\ell]\times[k]$. Each row $r\in[\ell]$ is a clique with edge weights $-1$: join every two distinct vertices $(r,c)$ and $(r,c')$ (where $c\neq c'$) with an edge of weight $-1$.  If $\{r,r'\}\in E(G)$, join every vertex in row $r$ to every vertex in row $r'$ with an edge of weight $-2$.

    Using \cref{prop:pottspoly}, for the graph polynomial $P_J(n)$, the edges of weight $-1$ force the $k$ colors within each row to be distinct. Thus, the colors in a row $r$ form a set $S_r \in \binom{[n]}{k}$. For $r,r'\in E(G)$, the product of the between-row factors is, 
    \[
     \prod_{c,c'=1}^k\bigl(1-2\cdot\mathbf1[f(r,c)=f(r',c')]\bigr)
 =(-1)^{|S_r\cap S_{r'}|}.
    \]
    Every $k$-subset arises from exactly $k!$ row colorings, so the
resulting sum is $(n)_k^\ell W_G(n,k)$, where
$(n)_k=n(n-1)\cdots(n-k+1)$.  Therefore
\begin{equation}\label{eq:potts-id}
 W_G(n,k)=\frac{P_J(n)}{(n)_k^\ell}.
\end{equation}
If $J_0$ is the disjoint union of the $\ell$ row cliques, then
$P_{J_0}(n)=(n)_k^\ell$.  Hence
\begin{equation}\label{eq:potts-ratio}
 W_G(n,k)=\frac{P_J(n)}{P_{J_0}(n)}.
\end{equation}
The sum of edge weights of $J$ differs from that of
$J_0$ only through the $k^2$ edges of weight $-2$ corresponding to
each edge of $G$.  Thus
\[
 \sum_{e\in E(J)}w_e-\sum_{e\in E(J_0)}w_e
 =-2k^2|E(G)|.
\]
Applying \cref{lem:graph-log} to both to $J$ and $J_0$ (both having $M=\ell k$ vertices), and subtracting,
we obtain whenever $n\geq A \ell k$, $W_G(n,k)>0$ and 
\begin{align}
 \left|\log W_G(n,k)+\frac{2k^2}{n}|E(G)|\right|
 &\le
 \left|\log\frac{P_J(n)}{n^M}-\frac1n\sum_{e\in E(J)}w(e)\right|\\
 &\quad+
 \left|\log\frac{P_{J_0}(n)}{n^M}
                       -\frac1n\sum_{e\in E(J_0)}w(e)\right|\\
 &\le2B\frac{M^3}{n^2}
 =2B\frac{\ell^3k^3}{n^2}.\label{eq:boundonWG}
\end{align}

Denote $a_\ell:=\E\tr H_{\syk}^{2\ell}$ and
    $c_\ell:=V^\ell M_\ell(q)$. Recall that
    \[M_\ell(q):=\sum_{\pi\in\cP_2(2\ell)}q^{\Cr(\pi)},\qquad \E\tr H_{\syk}^{2\ell}=V^\ell\sum_{\pi\in\cP_2(2\ell)}
          W_{G_\pi}(n,k).\]
    Now, the crossing graph $G_\pi$ of any
    $\pi\in\cP_2(2\ell)$ has $\ell$ vertices and
    $|E(G_\pi)|=\Cr(\pi)$. Moreover, $q=e^{-2k^2/n}$.
    By \cref{eq:boundonWG}, for $n\geq A\ell k$ we have $e^{-d_\ell}q^{\Cr(\pi)}
        \leq W_{G_\pi}(n,k)
        \leq e^{d_\ell}q^{\Cr(\pi)}$, where $d_\ell:=2B\ell^3k^3/n^2\leq d$
    whenever $\ell\leq L$. Multiplying by $V^\ell$
    and summing over $\pi\in\cP_2(2\ell)$, we obtain
    for every $1\leq\ell\leq L$, $ e^{-d}c_\ell\leq a_\ell\leq e^d c_\ell$. Since $c_\ell\geq0$ and $1-e^{-d}\leq e^d-1$,
    this gives $|a_\ell-c_\ell|\leq(e^d-1)c_\ell$. This inequality also holds for $\ell=0$, since
    $a_0=c_0=1$. For the remaining moments, we use the elementary
    bounds
    \[
        0\leq a_\ell,c_\ell
        \leq(2\ell-1)!!\,V^\ell
        \qquad(\ell\geq0).
    \]
    Indeed, $M_\ell(q)\leq(2\ell-1)!!$ because
    $0<q\leq1$, and $|W_G(n,k)|\leq1$ since it is
    the expectation of a sign. Also $a_\ell\geq0$
    because $H_{\syk}$ is Hermitian.

    Hence, for $|z|\leq r$,
    \begin{align*}
        |Z_n(z)-Q_n(z)|
        &\leq\sum_{\ell\leq L}|a_\ell-c_\ell|
               \frac{|z|^{2\ell}}{(2\ell)!}
             +\sum_{\ell>L}(|a_\ell|+|c_\ell|)
               \frac{|z|^{2\ell}}{(2\ell)!}\\
        &\leq(e^d-1)\sum_{\ell\leq L}c_\ell
               \frac{|z|^{2\ell}}{(2\ell)!}
             +2\sum_{\ell>L}V^\ell(2\ell-1)!!
               \frac{r^{2\ell}}{(2\ell)!}\leq(e^d-1)Q_n(|z|)
             +2\sum_{\ell>L}
               \frac{(Vr^2/2)^\ell}{\ell!},
    \end{align*}
    where we used
    $(2\ell-1)!!/(2\ell)!=1/(2^\ell\ell!)$.

    It remains to bound the tail. Write $x:=Vr^2/2$
    and choose $A_D\geq\max\{e,D/\log2\}$.
    By the definition of $L$, we have
    $L+1\geq2ex$ and $L\log2\geq D\log n$.
    In particular, $x/(L+2)\leq1/2$, so
    \[\sum_{\ell>L}\frac{x^\ell}{\ell!}
        \leq\frac{x^{L+1}}{(L+1)!}
              \sum_{j\geq0}
              \left(\frac{x}{L+2}\right)^j\leq2\frac{x^{L+1}}{(L+1)!}
        \leq2\left(\frac{ex}{L+1}\right)^{L+1}
        \leq2^{-L}\leq n^{-D},\]
    where the third inequality uses
    $(L+1)!\geq((L+1)/e)^{L+1}$.

    Substituting this bound above gives
    \[
        |Z_n(z)-Q_n(z)|
        \leq(e^d-1)Q_n(|z|)+2n^{-D},
    \]
    as desired.
\end{proof}
We now control the disorder fluctuations, showing \cref{lem:sykmomentbound} which is a key ingredient in the quenched result of ~\cref{thm:sykstrip}.
\sykmomentbound*
\begin{proof}[Proof of \cref{lem:sykmomentbound}]
    For a fixed $z = x+\iota y \in \C$ consider the function $\bm g \mapsto Z_n(z;\bm g)$ and apply~\cref{fact:gaussianpoincare}. We have that, 
    \[
    H_{\syk}(\bm g) := \sigma_{n,k}\sum_{S\in\binom{[n]}{k}} g_S \Gamma_S
    \]
    is a finite dimensional matrix with $\|H_{\syk}(\bm g)\|_{\op} \leq \sigma_{n,k}\sum_{S\in\binom{[n]}{k}} |g_S|$. As a result, the function $\bm g \mapsto Z_n(z;\bm g)$ and its first derivatives are bounded by a constant times an exponential of a finite sum of absolute Gaussians, which has finite moments of all orders. Thus, 
    \[
   \E_{\bm g} \left|Z_n(z;\bm g) - Z_n(z)\right|^2 \leq  \sum_{S\in \binom{[n]}{k}} \E_{\bm g}\left| \frac{\partial Z_n(z;\bm g)}{\partial g_S}\right|^2
    \]
    By Duhamel's formula~\cref{fact:duhamel} for a differential matrix valued function $A(t)$ 
    \[
    \frac{d}{dt}e^{A(t)} = \int_{0}^1 e^{sA(t)} A'(t) e^{(1-s)A(t)} ds.
    \]
    Applying this with $A(g_S) := z \cdot H_{\syk}(\bm g)$ we obtain: 
    \begin{align*}
            \frac{\partial e^{z H_{\syk}(\bm g)}}{\partial g_S} &= z\sigma_{n,k} \int_{0}^1 e^{s z H_{\syk}(\bm g)} \Gamma_S e^{(1-s)z H_{\syk}(\bm g)} ds \\ 
    \frac{\partial Z_n(z;\bm g)}{\partial g_S } &= z\sigma_{n,k} \int_{0}^1 \tr\left(e^{s z H_{\syk}(\bm g)} \Gamma_S e^{(1-s)z H_{\syk}(\bm g)}\right) ds= z\sigma_{n,k} \tr\left(e^{ z H_{\syk}(\bm g)} \Gamma_S\right).  
    \end{align*}
    As a result, 
    \[
     \E_{\bm g} \left|Z_n(z;\bm g) - Z_n(z)\right|^2 \leq  \sigma_{n,k}^2 |z|^2 \E_{\bm g} \sum_{S\in \binom{[n]}{k}} \left|\tr\left(e^{ z H_{\syk}(\bm g)} \Gamma_S\right)\right|^2.
    \]
    Now, note that $\{\Gamma_S\}_{|S|=k}$ is orthonormal under normalized the Hilbert-Schmidt inner product $\langle A,B\rangle = \tr(A^* B)$, thus, 
    \[
    \sum_{S\in\binom{[n]}{k}} \left|\tr\left(e^{ z H_{\syk}(\bm g)} \Gamma_S\right)\right|^2 = \sum_{S\in\binom{[n]}{k}} \left| \langle  e^{ z H_{\syk}(\bm g)}, \Gamma_S\rangle \right|^2 \leq \|e^{ z H_{\syk}(\bm g)}\|_{\operatorname{HS}}^2 \leq \tr e^{2x H_{\syk}(\bm g)}
    \]
    Now, recall that $|Z_n(z)| \leq e^{V|z|^2/2}$, resulting in $\E_{\bm g} \tr e^{2x H_{\syk}(\bm g)} = Z_n(2x) \leq e^{2x^2 V}$. Thus
    \[
     \E_{\bm g} \left|Z_n(z;\bm g) - Z_n(z)\right|^2 \leq  \sigma_{n,k}^2 |z|^2 e^{2V x^2}.\qedhere
    \]
\end{proof}

Armed with \cref{lem:partfntransfer,lem:sykmomentbound}, we can now close the proofs of \cref{thm:annealed-syk-lee-yang,thm:sykstrip} by showing \cref{cl:sykcouplingdecay,cl:sykproductbound}.

\subsection{Proof of \cref{cl:sykcouplingdecay,cl:sykproductbound}}
We now prove \cref{cl:sykproductbound} and
\cref{cl:sykcouplingdecay} in order.

\begin{proof}[Proof of \cref{cl:sykproductbound}]
Recall that $K\subset\Omega$ is compact, where
\[
    \Omega
    :=\C\setminus\{\iota y:y\in\R,\ |y|\geq\zeta_*\},
\]
and that
\[
    \cU
    :=K\cup\{z\in\C:|\operatorname{Im}z|\leq\zeta_*\}.
\]
We prove the required estimate separately on the closed strip
and on $K$.

By \cref{thm:moment-function-imaginary-rooted,thm:nearest-root},
for $0<q<1$,
\[
    \cL_q(z)
    =\prod_{j\geq1}
       \left(1+\frac{z^2}{t_j(q)^2}\right),
    \qquad
    t_j(q)\geq\frac{\zeta_*}{\sqrt{1-q}},
\]
with locally uniform convergence. Since $M_1(q)=1$, the
coefficient of $z^2$ in $\cL_q(z)$ is $1/2$. Comparing
coefficients in the finite partial products and passing to
the locally uniform limit gives
\[
    \sum_{j\geq1}\frac1{t_j(q)^2}=\frac12.
\]

Write $\tau_j:=t_j(q)/\sqrt V$. Since $q=e^{-2\alpha}$ and
$V\leq1/(2\alpha)$ by \cref{prop:sykvariancebound},
\[
    0<a:=V(1-q)
    \leq\frac{1-e^{-2\alpha}}{2\alpha}<1.
\]
Consequently,
\begin{equation}\label{eq:sykQproduct}
    \begin{gathered}
        Q_n(z)
        =\prod_{j\geq1}
           \left(1+\frac{z^2}{\tau_j^2}\right),
        \qquad
        \sum_{j\geq1}\frac1{\tau_j^2}=\frac V2,\\
        \tau_j\geq\frac{\zeta_*}{\sqrt a}>\zeta_*.
    \end{gathered}
\end{equation}

\smallskip
\noindent\emph{The closed-strip estimate.}
Let $R_*:=\zeta_*^2$. By
\cref{lem:bmrootsrelation,prop:bm-monotonicity},
\[
    (1-q)^{m-1}b_m(q)
    =\sum_{j\geq1}t_j(q)^{-2m},
    \qquad
    0<b_m(q)\leq b_m(1).
\]
Moreover, \cref{lem:endpointseries} gives
\[
    \sum_{m\geq1}b_m(1)R_*^m=2.
\]

For $0<|y|\leq\zeta_*$, we have $ay^2<R_*$ and
$|y|<\tau_1$. Thus the logarithmic expansion of the product
in \eqref{eq:sykQproduct} converges absolutely,
$Q_n(\iota y)>0$, and
\[ -\log Q_n(\iota y)
    =
    \frac1{1-q}
    \sum_{m\geq1}\frac{b_m(q)}m(ay^2)^m
    =
    Vy^2\sum_{m\geq1}
       \frac{b_m(q)}m(ay^2)^{m-1}
    \leq
    \frac{Vy^2}{R_*}
    \sum_{m\geq1}b_m(1)R_*^m
    =
    \frac{2Vy^2}{R_*}
    \leq2V.\]
Here we used $1/m\leq1$, $b_m(q)\leq b_m(1)$, and
$(ay^2)^{m-1}\leq R_*^{m-1}$. Since $Q_n(0)=1$, it follows
that
\[
    Q_n(\iota y)\geq e^{-2V}
    \qquad(|y|\leq\zeta_*).
\]
The same product also gives $Q_n(y)
    \leq
    \exp\left(y^2\sum_{j\geq1}\frac1{\tau_j^2}\right)
    =e^{Vy^2/2}$. For real $x,y$ and $t>|y|$, direct calculation yields
\[
    \frac{|1+(x+\iota y)^2/t^2|}
         {1+(x^2+y^2)/t^2}
    =
    \left(
        1-\frac{4t^2y^2}{(t^2+x^2+y^2)^2}
    \right)^{1/2}
    \geq
    \frac{1-y^2/t^2}{1+y^2/t^2}.
\]
Indeed, the expression inside the square root is minimized
at $x=0$, and $t>|y|$ makes the right-hand side positive.
Apply this inequality with $t=\tau_j$, multiply over a
finite number of factors, and pass to the limit using
\eqref{eq:sykQproduct}. For every $x\in\R$ and
$|y|\leq\zeta_*$, we obtain
\[\frac{|Q_n(x+\iota y)|}
         {Q_n\!\left(\sqrt{x^2+y^2}\right)}\geq\frac{Q_n(\iota y)}{Q_n(y)}\geq\exp\left(-2V-\frac{Vy^2}{2}\right)\geq e^{-C_*V},\]
where $ C_*:=2+\frac{R_*}{2}$. Thus
\begin{equation}\label{eq:sykQclosedstrip}
    |Q_n(z)|\geq e^{-C_*V}Q_n(|z|)
    \qquad
    (|\operatorname{Im}z|\leq\zeta_*).
\end{equation}
This bound includes both horizontal boundaries of the
strip, and $C_*$ is independent of $n,k$.

\smallskip
\noindent\emph{The compact-set estimate.}
For $z=x+\iota y\in K$ and $t\geq\zeta_*$, define
\[
    \Phi(z,t)
    :=t^2\log
       \frac{1+|z|^2/t^2}{|1+z^2/t^2|}.
\]
This is well-defined because $1+z^2/t^2=0$ would imply
$z=\pm\iota t\notin\Omega$, whereas $K\subset\Omega$.
All logarithms here are ordinary real logarithms of
positive numbers. Direct calculation gives
\begin{equation}\label{eq:sykfactoridentity}
    \Phi(x+\iota y,t)
    =
    -\frac{t^2}{2}
    \log\left(
        1-\frac{4t^2y^2}{(t^2+x^2+y^2)^2}
    \right).
\end{equation}
In particular, $\Phi(z,t)\geq0$.

The function $\Phi$ is continuous on
$K\times[\zeta_*,\infty)$. Moreover, uniformly for
$z=x+\iota y\in K$,
\[
    \frac{4t^2y^2}{(t^2+x^2+y^2)^2}
    =
    \frac{4y^2}{t^2}+O_K(t^{-4}),
\]
so \eqref{eq:sykfactoridentity} implies
\[
    \Phi(x+\iota y,t)=2y^2+O_K(t^{-2})
    \qquad(t\to\infty).
\]
Compactness on bounded intervals of $t$, together with
this uniform estimate for large $t$, therefore gives a
constant $C_K\geq0$, depending only on $K$, such that
\[
    \Phi(z,t)\leq2C_K
    \qquad(z\in K,\ t\geq\zeta_*).
\]

Apply this estimate with $t=\tau_j$, exponentiate, and
multiply over $1\leq j\leq J$. We obtain
\[
    \prod_{j=1}^J
       \left|1+\frac{z^2}{\tau_j^2}\right|
    \geq
    \exp\left(-2C_K\sum_{j=1}^J\frac1{\tau_j^2}\right)
    \prod_{j=1}^J
       \left(1+\frac{|z|^2}{\tau_j^2}\right)
    \qquad(z\in K).
\]
Passing to the limit as $J\to\infty$ and using
\eqref{eq:sykQproduct} gives
\[
    |Q_n(z)|
    \geq e^{-C_KV}Q_n(|z|)
    \qquad(z\in K).
\]

Finally, set $C:=\max\{C_*,C_K\}$. Combining the compact-set estimate with
\eqref{eq:sykQclosedstrip} gives
\[
    |Q_n(z)|\geq e^{-CV}Q_n(|z|)
    \qquad(z\in\cU).
\]
Every factor in the product for $Q_n(|z|)$ is at least
$1$, so $Q_n(|z|)\geq1$. Therefore
\[
    |Q_n(z)|
    \geq e^{-CV}Q_n(|z|)
    \geq e^{-CV}
    \qquad(z\in\cU).
\]
The constant $C=C(K)$ is independent of $n,k$.
In particular, $Q_n$ has no zeros on $\cU$, as claimed.
\end{proof}

\begin{proof}[Proof of \cref{cl:sykcouplingdecay}]
Extend the numerical definition
\[
    \sigma_{n,j}^2
    :=\frac{(j-1)!}{2j\,n^{j-1}}
\]
to every integer $1\leq j\leq n$. For $1\leq j<n$,
\[
    \frac{\sigma_{n,j+1}^2}{\sigma_{n,j}^2}
    =\frac{j^2}{n(j+1)}<1.
\]
Thus $\sigma_{n,j}^2$ is decreasing in $j$.

Choose $j:=\left\lfloor2\sqrt{\frac n{\log n}}\right\rfloor$. The assumption $k/\sqrt{n/\log n}\to\infty$ ensures that
$1\leq j\leq k<n$ for all sufficiently large $n$. Using $(j-1)!\leq j^{j-1}$, we obtain
\[
    \sigma_{n,k}^2
    \leq\sigma_{n,j}^2
    =\frac{(j-1)!}{2j\,n^{j-1}}
    \leq\left(\frac jn\right)^{j-1}
    =
    \exp\left(-(j-1)\log\frac nj\right)\leq\exp\left(-\frac12\sqrt{n\log n}\right),
\]
as desired. 
\end{proof}
\section{The Real-Rootedness of the deformed matching polynomials: Proof of \cref{thm:main-thm}}\label{sec:q-deformed-real-rooted}
We now finally present the key technical input in our proof, which is the real-rootedness of the deformed matching polynomials $\mu_{n, q}$:
\maintheorem*
Our proof strategy to prove \cref{thm:main-thm} is as follows. By \cref{fact:hermite-simple-real}, we know that $\mu_{n, 1}$ has simple, real roots. Note that being a real-rooted polynomial with simple roots is an open property,\footnote{More precisely, $\operatorname{RR}_n\cap\cS_n= \operatorname{int}(\operatorname{RR}_n)$ is an open set by \cref{fact:real-rooted-topology}} and thus there exists an open neighborhood of $q = 1$ such that $\mu_{n, q}$ is simple and real-rooted for all $q$ in the neighborhood. Thus let $Q\subset(-1, 1)$ be the \emph{maximal} open set such that $\mu_{n, q}$ is simple real-rooted for all $q\in Q$, and note that $Q$ is non-empty. We are done if we can show that $Q = (-1, 1)$.\footnote{for $q = -1$, simple real-rootedness follows from \cref{lem:mu-1-real-rooted}} For $q\in \overline{Q\cup\{-1, 1\}}$, let $\lambda_1(q) \leq \cdots \leq \lambda_n(q)$ be the roots of $\mu_{n, q}$, where $\lambda_i(q) < \lambda_{i+1}(q)$ for all $i\in[n-1]$ for $q\in Q\cup\{-1, 1\}$.

Assume for the sake of contradiction that $Q\neq(-1, 1)$. Then there exists $q^*\in(-1, 1)\setminus Q$ such that $q^*\in\partial Q$,\footnote{Note that since $Q$ is open, $Q\cap\partial Q = \emptyset$. Also, throughout this argument, $\partial Q$ shall represent the relative boundary of $Q$ in $(-1,1)$} i.e. either $\mu_{n, q^*}$ is not real-rooted, or $\mu_{n, q^*}$ is not simple. Since being real-rooted is a closed property,\footnote{More precisely, $\operatorname{RR}_n$ is a closed set under the coefficient topology by \cref{fact:real-rooted-topology}} and since $q^*\in\partial Q$, we get that $\mu_{n, q^*}$ is real-rooted, and consequently, $\mu_{n, q^*}$ is \emph{not} simple, i.e. there exists a ``cluster'' $I\seq[n]$ with $|I|\geq 2$ and some $\xi\in\R$ such that $\lambda_i(q^*) = \xi$ for all $i\in I$. Write $I:= \{i\in[n]:\lambda_i(q^*) = \xi\}$. 

We argue that such a ``cluster'' can not exist, because the roots \emph{repel} each other:
\begin{lemma}[Root Repulsion Equation]\label{lem:root-evolution}
    Suppose for some non-empty open set $Q\subset(-1, 1)$ we have that $\mu_{n, q}$ is simple and real-rooted for all $q\in Q$. Let $\lambda_1(q) < \cdots < \lambda_n(q)$ be the roots of $\mu_{n, q}$ for $q\in Q$. Then for any $q\in Q$ and any $i\in[n]$ we have 
    \[\sum_{j\neq i}\frac{1}{\lambda_i(q) - \lambda_j(q)} = \frac{b_n(q)}{2A_{i, q}}\lambda_i(q) - \frac{4q(1 - q)}{A_{i, q}}\cdot\frac{\operatorname{d}\!\lambda_i(q)}{\operatorname{d}\!q},\]
    where $A_{i, q}:= 4 + (1 - q)\lambda_i(q)^2$.
\end{lemma}
\begin{remark}
    The strategy of converting the differential equation (\cref{thm:munq-pde}) to an equation on the repulsion of the roots was inspired by the proof of~\cite[Lemma~2.4]{CsordasSV94}. Although the differential evolution considered in~\cite{CsordasSV94} is different than ours, both evolutions are of the form $\partial_t F = (-a(z)\partial_z^2 + \text{ lower order terms})F$, where $a:\R\to\R_{ > 0}$ is strictly positive everywhere. Consequently, the repulsive behavior of roots in~\cite{CsordasSV94}, as well as in our results, is similar. 
    
    See~\cite{RodgersT20} for a recent application of this technique in proving \emph{Newman's conjecture} as pertaining to the Riemann zeta function, and also see~\cite{Tao17} for an excellent general introduction into how differential evolutions of polynomials influence the ``flow'' of their roots. Also see~\cite[Section~2.4, Lemma~2.10]{Gribinski25}, where Gribinski establishes the real-rootedness of Gegenbauer polynomials by tracking their zeros through a differential evolution and using interlacing to rule out collisions.
\end{remark}
We then use \cref{lem:root-evolution} to rule out a (hypothetical) colliding cluster as $q\to0$, by studying the variance of the cluster, where for any non-empty $J\seq[n]$ we define 
\[\Var_J(q):= \sum_{j\in J}\left|\lambda_j(q) - \frac{1}{|J|}\sum_{j'\in J}\lambda_{j'}(q)\right|^2.\]

\begin{lemma}[Increase in Variance]\label{lem:root-collision}
Suppose we have $\{q_j\}_{j\in\N}\subset Q$ such that $\lim_{j\to\infty}q_j = q^*$. Then 
    \[\lim_{j\to\infty}q_j(1 - q_j)\Var'_I(q_j) = -\frac{1}{2}\binom{|I|}{2}\left(4 + (1 - q^*)\xi^2\right) < 0.\]
\end{lemma}
\begin{remark}
    A few remarks are due:
    \begin{enumerate}[(1)]
        \item  Note that by \cref{fact:differentiable-roots}, for any $I\subset[n]$, $Q\ni q\mapsto \Var_I(q)\in\R$ is a $C^1$-map and thus it makes sense to talk about the $q$-derivatives of $\Var_I$.
        \item The statement of \cref{lem:root-collision} was inspired by the \emph{virial identity} in \cite[Lemma~12(iv)]{RodgersT20}.
    \end{enumerate}
\end{remark}

Since $\Var_J(\pm 1) > 0$ for $|J|\geq2$ (because $\mu_{n, \pm 1}$ are simple polynomials), we expect that $\Var_J(q) > 0$ for all $q\in[-1, 1]$.

However, note that for the (hypothetical) cluster $I$ that we found above, we have $\Var_I(q^*) = 0$ since $\lambda_i(q^*) = \lambda_j(q^*)$ for all $i, j\in I$, and this leads to a contradiction. We now formalize this argument below to prove \cref{thm:main-thm}.

\begin{proof}[Proof of \cref{thm:main-thm} assuming \cref{lem:root-evolution,lem:root-collision}]
If $n\leq 1$, the theorem is trivial, so assume $n\geq 2$. For $q = 1$ we know by \cref{fact:hermite-simple-real} that $\mu_{n, 1}$ is a simple real-rooted polynomial. As in the discussion above, let $Q\subset(-1, 1)$ be the (non-empty) maximal open set such that $\mu_{n, q}\in\operatorname{RR}_n\cap\cS_n$ for all $q\in Q$. Assume for the sake of contradiction that $Q\neq(-1, 1)$, and thus $(-1, 1)\cap\partial Q$ is non-empty. 

\paragraph{Showing that $\partial Q\subset[-1, 0]$} Write $q^*:= \sup\{(-1, 1)\cap\partial Q\}$, and note that $q^* < 1$. Also note that $(q^*, 1)\subset Q$. Assume for the sake of contradiction $q^*\in(0, 1)$. Fix any sequence $\{q_j\}_{j\in\N}\subset(q^*, 1)$ such that $\lim_{j\to\infty}q_j = q^*$, and let $I\seq[n]$ be the ``cluster'' as in the discussion above. By \cref{lem:root-collision}, 
\[0 > \lim_{j\to\infty}q_j(1 - q_j)\Var'_I(q_j) = q^*(1 - q^*)\lim_{j\to\infty}\Var'_I(q_j)\implies \lim_{j\to\infty}\Var'_I(q_j) < 0.\]
Consequently there exists $\eps > 0$ such that $\Var'_I(q^* + \eps') < 0$ for all $0 < \eps'\leq\eps$. Since $Q\ni q\mapsto\Var'_I(q)$ is continuous, and since $[q^*, 1)\ni q\mapsto\Var_I(q)$ is also continuous, we have that
\[0 < \Var_I(q^* + \eps) = \Var_I(q^* + \eps) - \Var_I(q^*) = \int_{q^*}^{q^* + \eps}\Var'_I(q)\operatorname{d}\!q < 0,\]
leading to a contradiction.

\paragraph{Showing that $\partial Q\subset[0, 1]$} As with the $q = 1$ case, by \cref{lem:mu-1-real-rooted} we have that $\mu_{n, -1}$ is simple and real-rooted. Consequently for some $\eps_n > 0$ we have that $\mu_{n, q}$ is simple and real-rooted for all $q\in(-1, -1 + \eps_n)$, and thus $(-1, -1 + \eps_n)\subset Q$. Now write $q^*:= \inf\{(-1, 1)\cap\partial Q\}$, and note that $(-1, q^*)\subset Q$. Assume for the sake of contradiction $q^*\in(-1, 0)$. Fix some $\{q_j\}_{j\in\N}\subset(-1, q^*)$ such that $\lim_{j\to\infty}q_j = q^*$, and let $I\seq[n]$ be the ``cluster'' as in \cref{lem:root-collision}. By \cref{lem:root-collision}, 
\[0 > \lim_{j\to\infty}q_j(1 - q_j)\Var'_I(q_j) = q^*(1 - q^*)\lim_{j\to\infty}\Var'_I(q_j)\implies \lim_{j\to\infty}\Var'_I(q_j) > 0.\]
Consequently there exists $\eps > 0$ such that $\Var'_I(q^* - \eps') > 0$ for all $0 < \eps'\leq\eps$. Since $Q\ni q\mapsto\Var'_I(q)$ is continuous, and since $(-1, q^*]\ni q\mapsto\Var_I(q)$ is also continuous, we have that
\[0 > -\Var_I(q^* - \eps) = \Var_I(q^*) - \Var_I(q^* - \eps) = \int^{q^*}_{q^* - \eps}\Var'_I(q)\operatorname{d}\!q > 0,\]
leading to a contradiction.

We thus have $(-1, 1)\cap\partial Q\subset[0, 1]\cap[-1, 0] = \{0\}$. Since $(-1, 1)\cap\partial Q$ is non-empty, we must have that $(-1, 1)\cap\partial Q = \{0\}$, i.e. $Q = (-1, 1)\setminus\{0\}$. Now, since $\operatorname{RR}_n$ is closed, and since $\mu_{n,q}$ is real-rooted for all $q\in[-1, 1]\setminus\{0\}$, we obtain that $\mu_{n, 0}$ is also real-rooted. Thus it only remains to show that $\mu_{n, 0}$ is simple. Suppose not, and let $I\seq[n]$ be a cluster of size $\geq 2$ such that $\lambda_i(0) = \lambda_j(0)$ for all $i, j\in I$. We thus have $\Var_I(0) = 0$, and since $(-1, 1)\setminus\{0\}\subset Q$, we have $\Var_I(q) > 0$ for all $q\in(-1, 1)\setminus\{0\}$. By \cref{lem:root-collision}, we have that $\lim_{q\searrow 0}q(1- q)\Var_I'(q) < 0$, and since $q(1 - q) > 0$ for all $q\in(0, 1)$, we have $\Var_I'(q) < 0$ for all $q\in(0, \eps)$ for some $\eps > 0$. Consequently, since $(0, 1)\ni q\mapsto \Var_I(q)$ is differentiable, the mean value theorem yields that $0 < \Var_I(\eps) = \Var_I(\eps)-  \Var_I(0) = \eps\cdot\Var_I'(\eps') < 0$ for some $\eps'\in(0, \eps)$, leading to a contradiction. Consequently, $\mu_{n, 0}$ is simple, and thus $\mu_{n, q}$ is simple and real-rooted for all $q\in[-1, 1]$, as desired.

We now establish \cref{remark:qgeq1notrr} and show that for every real $q$ with $|q| > 1$, there exists $n = n(q)\geq 2$ such that $\mu_{n, q}$ is \textbf{not} real-rooted.

First suppose that $q>1$, and assume for the sake of contradiction that $\mu_{n,q}$ is real-rooted for every $n\geq 0$. Let $n=2m\geq 4$. Since $\mu_{n,q}$ is an even monic polynomial, its roots occur in opposite pairs. Thus there exist $b_1,\ldots,b_m\geq 0$ such that
\[
\mu_{n,q}(z)
 = \sum_{\ell=0}^{m}(-1)^\ell
   \binom{n}{2\ell}M_\ell(q)z^{n-2\ell}
 = \prod_{k=1}^{m}(z^2-b_k).
\]
Here $b_k$ is the square of the $k^{\mathrm{th}}$ nonnegative root of
$\mu_{n,q}$.

By Vi\`ete's formulas, together with $M_1(q)=1$ and $M_2(q)=2+q$,
we have
\[
\sum_{k=1}^{m} b_k
 = \binom{n}{2}M_1(q)
 = \binom{n}{2},\qquad \sum_{1\leq k<k'\leq m} b_kb_{k'}
 = \binom{n}{4}M_2(q)
 = \binom{n}{4}(2+q).
\]
Since $b_1,\ldots,b_m\geq 0$,
\[
\left(\sum_{k=1}^{m}b_k\right)^2
 = \sum_{k=1}^{m}b_k^2
   +2\sum_{1\leq k<k'\leq m}b_kb_{k'}
 \geq
 2\sum_{1\leq k<k'\leq m}b_kb_{k'}.
\]
Consequently,
\[
\binom{n}{2}^2
 \geq 2\binom{n}{4}(2+q)\implies 
q
 \leq \frac{\binom{n}{2}^2}{2\binom{n}{4}}-2
 \leq 1+\frac{15}{n-3}.
\]
Choosing any even integer $n>3+\frac{15}{q-1}$ gives a contradiction.\footnote{Alternatively, we could have also applied \cref{fact:newton} to $\mu_{n, q}$ to obtain $q\leq 1 + O(1/n)$ for $n\gg 1$, yielding a similar contradiction.} Therefore, for every $q>1$, there exists $n\geq 0$ such that $\mu_{n,q}$ is not real-rooted.

We now turn to the case $q<-1$. Once again, by Vi\`ete's formulas we obtain that if $\mu_{n, q}(z) = \prod_{k = 1}^m(z^2 - b_k)$ for some even integer $n = 2m\geq 4$, then $0 \leq \prod_{k = 1}^m b_k = M_m(q)$. On the other hand, by \cref{fact:touchard-riordan}, we have that 
\[M_m(q) = \frac{1}{(1 - q)^m}\sum_{r = 0}^m(-1)^rB_{m, r}q^{r(r + 1)/2}.\]
We claim that the term corresponding to $r = m$, which is $(-1)^mq^{m(m + 1)/2}$, is the term which dominates in magnitude as $m\to\infty$. Indeed, for any $r = m - s$, we have 
\[\frac{B_{m, r}|q|^{r(r + 1)/2}}{|q|^{m(m + 1)/2}}\leq n^s\cdot |q|^{-s(m - (s - 1)/2)} \leq \left(\frac{2m}{|q|^{(m + 1)/2}}\right)^s. \]
Thus if we choose $m$ large enough w.r.t. $|q|$, we obtain that 
\[\left|\sum_{r < m}(-1)^rB_{m, r}q^{r(r + 1)/2}\right|\leq\frac{1}{2}|q|^{m(m + 1)/2},\]
where we use the fact that $|q| > 1$. Thus
\[\operatorname{sign}\left(\frac{1}{(1 - q)^m}\sum_{r = 0}^m(-1)^rB_{m, r}q^{r(r + 1)/2}\right) = \operatorname{sign}\left((-1)^mq^{m(m + 1)/2}\right) = (-1)^{m(m + 3)/2}.\]
For $m\equiv 2, 3\bmod 4$ we have that $(-1)^{m(m + 3)/2} = -1$, i.e. $M_m(q) < 0$, leading to a contradiction.
\end{proof}

Thus it only remains to show \cref{lem:root-evolution,lem:root-collision}, which we do now.

\begin{proof}[Proof of \cref{lem:root-evolution}]
    Substituting $z = \lambda_i(q)$ in \cref{eq:munq-pde} (from \cref{thm:munq-pde}) yields 
    \begin{equation}\label{eq:eigval-pde}
        8q(1 - q)\partial_q\mu_{n, q}(\lambda_i(q)) = A_{i, q}\cdot\partial_z^2\mu_{n, q}(\lambda_i(q)) - b_n(q)\lambda_i(q)\partial_z\mu_{n, q}(\lambda_i(q)). 
    \end{equation}
    By \cref{fact:differentiable-roots} we may write 
    \[\frac{\operatorname{d}\!\lambda_i(q)}{\operatorname{d}\!q} = -\frac{\partial_q\mu_{n, q}(\lambda_i(q))}{\partial_z\mu_{n, q}(\lambda_i(q))},\]
    and thus dividing \cref{eq:eigval-pde} by $\partial_z\mu_{n, q}(\lambda_i(q))$\footnote{Note that the simplicity assumption on $\mu_{n, q}$ ensures $\partial_z\mu_{n, q}(\lambda_i(q))\neq 0$} yields 
        \[8q(1 - q)\cdot\frac{\operatorname{d}\!\lambda_i(q)}{\operatorname{d}\!q} = b_n(q)\lambda_i(q) - A_{i, q}\cdot\frac{\partial_z^2\mu_{n, q}(\lambda_i(q))}{\partial_z\mu_{n, q}(\lambda_i(q))}. \]
    Thus it suffices to show that 
    \[\frac{\partial_z^2\mu_{n, q}(\lambda_i(q))}{\partial_z\mu_{n, q}(\lambda_i(q))} = 2\sum_{j\neq i}\frac{1}{\lambda_i(q) - \lambda_j(q)}.\] 
    But this follows simply from the fact that $\mu_{n, q}(z) = \prod_{i\in[n]}(z - \lambda_i(q))$ for all $q\in Q$.
\end{proof}
 
\begin{proof}[Proof of \cref{lem:root-collision}]
    Write $\bar{\lambda}(q):= |I|^{-1}\cdot\sum_{i\in I}\lambda_i(q), y_i(q):= \lambda_i(q) - \bar{\lambda}(q)$, and note that for $q\in Q$ we have 
    \[\frac{\operatorname{d}\!\Var_I(q)}{\operatorname{d}\!q} = \frac{\operatorname{d}\!}{\operatorname{d}\!q}\sum_{i\in I}y_i(q)^2 = 2\sum_{i\in I}\left(y_i(q)\frac{\operatorname{d}\!\lambda_i(q)}{\operatorname{d}\!q} - y_i(q)\frac{\operatorname{d}\!\bar{\lambda}(q)}{\operatorname{d}\!q}\right) = 2\sum_{i\in I}y_i(q)\cdot\frac{\operatorname{d}\!\lambda_i(q)}{\operatorname{d}\!q}.\]
    Thus 
    \[8q(1 - q)\frac{\operatorname{d}\!\Var_I(q)}{\operatorname{d}\!q} = 2\sum_{i\in I}y_i(q)\cdot\left(8q(1 - q)\frac{\operatorname{d}\!\lambda_i(q)}{\operatorname{d}\!q}\right)\]
    \[ \overset{\text{\cref{lem:root-evolution}}}{=} 2\sum_{i\in I}y_i(q)\cdot\left(b_n(q)\lambda_i(q) - 2A_{i, q}\cdot\sum_{k\in[n]\setminus\{i\}}\frac{1}{\lambda_i(q) - \lambda_k(q)}\right)\]
    \[ = 2b_n(q)\Var_I(q) - 4\sum_{i\in I}y_i(q)\cdot A_{i, q}\cdot\sum_{k\in[n]\setminus\{i\}}\frac{1}{\lambda_i(q) - \lambda_k(q)}.\]
    Now by \cref{fact:real-rooted-topology} the map $\operatorname{RR}_n\ni f\mapsto (\lambda_1(f), \ldots, \lambda_n(f))\in\R^n$ is continuous, and thus for any $i\in[n]$ we have $\lim_{j\to\infty}\lambda_i(q_j) = \lambda_i(q^*)$ since $\lim_{j\to\infty}\mu_{n, q_j} = \mu_{n, q^*}$ in the coefficient topology.

    Now, since $(\lambda_1(q), \ldots, \lambda_n(q))\mapsto \Var_I(q)$ is continuous, we have $\lim_{j\to\infty}\Var_I(q_j) = \Var_I\left(\lim_{j\to\infty}q_j\right) = \Var_I(q^*) = 0$. Thus 
    \[\lim_{j\to\infty}q_j(1 - q_j)\frac{\operatorname{d}\!\Var_I}{\operatorname{d}\!q}(q_j)= -\frac{1}{2}\lim_{j\to\infty}\sum_{i\in I}A_{i, q_j}\cdot\sum_{k\in[n]\setminus\{i\}}\frac{y_i(q_j)}{\lambda_i(q_j) - \lambda_k(q_j)}\]
    \begin{align*}
&= -\frac{1}{2}
    \lim_{j\to\infty}
    \Biggl[\sum_{i\in I}
    y_i(q_j)\cdot A_{i, q_j}
    \sum_{\substack{k\in I\setminus\{i\}}}
    \frac{1}
         {\lambda_i(q_j)-\lambda_k(q_j)}
    +
    \sum_{i\in I}
    y_i(q_j)\cdot A_{i, q_j}
    \sum_{k\in[n]\setminus I}
    \frac{1}
         {\lambda_i(q_j)-\lambda_k(q_j)}
    \Biggr].
\end{align*}
   
    Since $\lambda_i(q^*) - \lambda_k(q^*)\neq 0$ for all $i\in I, k\in[n]\setminus I$, and since $\lim_{j\to\infty}y_i(q_j) = y_i(q^*) = 0$, we have
    \[\lim_{j\to\infty}
    \sum_{i\in I}
    y_i(q_j)\cdot A_{i, q_j}
    \sum_{k\in[n]\setminus I}
    \frac{1}
         {\lambda_i(q_j)-\lambda_k(q_j)} = 0.\] 
         Consequently, 
    \[\lim_{j\to\infty}q_j(1 - q_j)\frac{\operatorname{d}\!\Var_I}{\operatorname{d}\!q}(q_j) = -\frac{1}{2}
    \lim_{j\to\infty}
    \sum_{i\in I}
    y_i(q_j)A_{i, q_j}
    \sum_{\substack{k\in I\setminus\{i\}}}
    \frac{1}
         {\lambda_i(q_j)-\lambda_k(q_j)}\] 
         \[= -\frac{1}{2}
    \lim_{j\to\infty}
    \sum_{\substack{i < k\\i, k\in I}}\frac{y_i(q_j)A_{i, q_j} - y_k(q_j)A_{k, q_j}}{\lambda_i(q_j)-\lambda_k(q_j)} = -\frac{1}{2}
    \lim_{j\to\infty}
    \sum_{\substack{i < k\\i, k\in I}}\frac{(\lambda_i(q_j) - \lambda_k(q_j) + y_k(q_j))A_{i, q_j} - y_k(q_j)A_{k, q_j}}{\lambda_i(q_j)-\lambda_k(q_j)}\]
    \[ = -\frac{1}{2}
    \lim_{j\to\infty}
    \sum_{\substack{i < k\\i, k\in I}}\left(A_{i, q_j} + y_k(q_j)\cdot\frac{A_{i, q_j} - A_{k, q_j}}{\lambda_i(q_j)-\lambda_k(q_j)}\right)\] 
    \[ = -\frac{1}{2}
    \lim_{j\to\infty}
    \sum_{\substack{i < k\\i, k\in I}}\left(A_{i, q_j} + y_k(q_j)\cdot(1 - q_j)\left(\lambda_{i}(q_j) + \lambda_{k}(q_j)\right)\right)= -\frac{1}{2}\sum_{\substack{i < k\\i, k\in I}}A_{i, q^*} = -\frac{1}{2}
    \binom{|I|}{2}(4 + (1 - q^*)\xi^2),\]
    where we use the fact that $\lim_{j\to\infty}y_k(q_j) = y_k(q^*) = 0, \lim_{j\to\infty}A_{i, q_j} = A_{i, q^*} = 4 + (1 - q^*)\xi^2$ for any $i, k\in I$. 
\end{proof}

\subsection{Alternative Proof of \cref{thm:main-thm} using Real-Stability Preserving Operators}\label{subsec:alternate-proof}
We now give an alternative proof of \cref{thm:main-thm} assuming \cref{thm:munq-pde}, using the theory of real-stability preserving operators (more precisely, \cref{fact:rr-preserve}).
\begin{proof}[Proof of \cref{thm:main-thm} assuming \cref{thm:munq-pde,fact:rr-preserve}]
For any $q\neq 1$, write 
    \[\tilde{\mu}_{n,q}(z) := (1-q)^{n/2}\mu_{n,q}\left(\frac{z}{\sqrt{1-q}}\right) = \sum_{\ell = 0}^{\lfloor n/2\rfloor}(-1)^\ell\binom{n}{2\ell}F_\ell(q)z^{n-2\ell},\]
    where $F_\ell(q):= (1-q)^\ell M_\ell(q)$. Note that by \cref{fact:hermite-simple-real,lem:mu-1-real-rooted}, it suffices to prove that $\tilde{\mu}_{n,q}$ is simple and real-rooted for all $q\in(-1,1),n\geq 2$.
    
    By \cref{thm:munq-pde}, we obtain
    \[8q\partial_q\tilde{\mu}_{n,q} = \left((4+z^2)\partial_z^2-(2n+1)z\partial_z+n(n+2)\right)\cdot\tilde{\mu}_{n,q} =: D_n\tilde{\mu}_{n,q}.\]
    Note that $D_n$ is a linear map $\R[z]_{\leq n}\to\R[z]_{\leq n}$. In particular it can be abstractly viewed as a $(n+1)\times(n+1)$ matrix.
    
    Now, first assume $q\in(0,1)$. Note that for $q = e^{-8t}$, we obtain $-8q\partial_q\tilde{\mu}_{n,q} = \partial_t\tilde{\mu}_{n,q}$, and thus 
    \begin{equation}\label{eq:tildemunq-pde}
        \partial_t\tilde{\mu}_{n,e^{-8t}} = -D_n\tilde{\mu}_{n,e^{-8t}}\implies\tilde{\mu}_{n,e^{-8t}}(z) = e^{-tD_n}z^n
    \end{equation}
    where $e^{-tD_n}$ refers to the exponential of $-tD_n$, viewed as a $(n+1)\times(n+1)$ matrix. We also use the fact that $\tilde{\mu}_{n,e^{-8t}}(z)\vert_{t = 0} = \tilde{\mu}_{n,1}(z) = z^n$. Thus we can conclude that $\tilde{\mu}_{n,q}$ is real-rooted for $q\in(0,1)$ if we can show that $e^{-tD_n}z^n$ is a real-rooted polynomial for all $t\geq 0$. We claim that it suffices to show that the linear operator $\Id - hD_n$ is a real-rootedness preserving operator for all sufficiently small $h> 0$: Indeed, 
    \[e^{-tD_n}z^n = \lim_{m\to\infty}\left(\Id - \frac{tD_n}{m}\right)^mz^n,\]
    and thus if $\Id - hD_n$ is real-rootedness preserving for all sufficiently small $h>0$, then for all sufficiently large $m$, $(\Id - tD_n/m)^mz^n$ is real-rooted (since $z^n$ is real-rooted), and since the set of real-rooted polynomials of degree $\leq n$ is closed,\footnote{For monic real-rooted polynomials this follows from \cref{fact:real-rooted-topology}. For general real-rooted polynomials, this follows from \cref{item:hurwitz} of \cref{fact:complex-analysis}} we obtain that $e^{-tD_n}z^n$ is a real-rooted polynomial for all $t\geq 0$, as desired.

    Now, by \cref{fact:rr-preserve}, $(\Id-  hD_n)$ is real-rootedness preserving if 
    \[\mathcal{D}(z,w):= (\Id - hD_n)[(z + w)^n] = (z+w)^n-hD_n[(z+w)^n]\] 
    \[= (z+w)^{n-2}\left(z^2 + (2-3nh)zw+ (1-n(n+2)h)w^2-4hn(n-1)\right)\in\R[z,w]\]
    is real-stable. To see the real-stability of $\mathcal{D}(z,w)$, note that for any $h\in(0,\frac{1}{n(n+2)})$, there exist $\gamma_1,\gamma_2 > 0$ such that $\gamma_1 +  \gamma_2 = 2-3nh,\gamma_1\gamma_2=1-n(n+2)h$, and thus we may write 
    \[\mathcal{D}(z,w)= (z+w)^{n-2}\left((z+\gamma_1w)(z+\gamma_2w)-4hn(n-1)\right).\]
    Now suppose $\mathcal{D}(z,w)=0$ for some $z,w\in\C$ with $\Imm(z),\Imm(w) > 0$. If $\Imm(z),\Imm(w) > 0$, then $z+w\neq 0$, and thus $\mathcal{D}(z,w)=0\implies (z+\gamma_1w)(z+\gamma_2w)=4hn(n-1)>0$. But if we have two non-real complex numbers $z_1,z_2\in\C\setminus\R$ with $z_1z_2\in\R_{>0}$, then $\Imm(z_1),\Imm(z_2)$ must have opposite signs, which can not be the case since $\Imm(z+\gamma_iw) > 0$ for $i\in\{1,2\}$. Consequently, $\mathcal{D}$ is real-stable, and thus $\tilde{\mu}_{n,q}$ is real-rooted for all $q\in(0,1)$. Since the set of real-rooted polynomials is closed, $\tilde{\mu}_{n,0}$ is also real-rooted. Finally, for $q<0$, we have $\tilde{\mu}_{n,-e^{-8t}}(z) = e^{-tD_n}\tilde{\mu}_{n,-1}(z)$ for all $t\geq 0$.\footnote{This is obtained by evolving the differential equation in \cref{eq:tildemunq-pde} starting from $q=-1$.} By \cref{lem:mu-1-real-rooted}, $\tilde{\mu}_{n,-1}(z)$ is real-rooted, and thus $\tilde{\mu}_{n,q}$ is also real-rooted for all $q\in(-1,0)$, as desired.

    Thus it only remains to show that $\tilde{\mu}_{n,q}$ is simple for all $q\in(-1,1)$. To show that, we use \emph{Laguerre's inequality} (see \cref{fact:laguerre}). The argument is very similar to the argument in the proof of \cref{thm:moment-function-imaginary-rooted} used to show the simplicity of $\cL_q$.

    Assume for the sake of contradiction $\tilde{\mu}_{n,q_0}$ has a root $z_0\in\R$ of multiplicity $m\geq2$ for some $q_0\in(-1,1)$. 
Write $h(q,z):=\partial_z^{m-2}\tilde{\mu}_{n,q}(z), A:=\partial_z^m\tilde{\mu}_{n,q_0}(z_0)$ and observe that $h(q_0,z_0)=0, \partial_z^2h(q_0,z_0)=A\neq 0$. Also define the function $E:(-1,1)\to\R_{\geq 0}$ given as 
\[E(q):= h_z(q,z_0)^2 -h(q,z_0)h_{zz}(q,z_0),\]
where $h_z$ stands for $\partial_zh$, $h_{zz}$ stands for $\partial_z^2h$, and so on. 

By \cref{fact:laguerre}, $E(q)\geq 0$ for all $q\in(-1, 1)$. Since $E:(-1, 1)\to\R_{\geq 0}$ is differentiable, and since $E(q_0) = 0$, we have $E'(q_0) = 0$. However, 
\[E'(q_0)=\left[2h_zh_{qz}-h_qh_{zz}-h_{qzz}h\right]_{(q,z)=(q_0,z_0)}=-A\cdot h_q(q_0,z_0),
\]
Now, since $8q\partial_q\tilde{\mu}_{n,q} = D_n\tilde{\mu}_{n,q}$, applying $\partial_z^{m-2}$ on both sides yields and evaluating at $(q,z) = (q_0,z_0)$ yields
\[8q_0h_q(q_0,z_0)=(4+z_0^2)A,\]
where we observe that $\partial_z^k \tilde{\mu}_{n,q_0}(z_0) = 0$ for all $k < m$, and thus only the $(4+z^2)\partial_z^2$ operator survives being hit by $\partial_z^{m-2}$. We thus have $0= 8q_0E'(q_0)=-(4+z_0^2)A^2 < 0$, leading to a contradiction, as desired.
\end{proof}
\section{The Differential Equation Governing the deformed matching polynomials: \cref{thm:munq-pde}}
The only ingredient remaining to demonstrate \cref{thm:main-thm} is \cref{thm:munq-pde}. The way we show \cref{thm:munq-pde} is by deriving a differential recurrence (see \cref{lem:mlq-recurrence}) for $M_\ell(q)$ using \cref{fact:touchard-riordan}, and then massaging the relation to obtain a differential equation for $\mu_{n,q}$. 
\munqpde*
We also have a similar differential equation governing $\mathscr{L}_q$, where $\mathscr{L}_q(z):= \cL_q(\sqrt{1 - q}\cdot z)$.
\scrlpde*
\begin{remark}
    A few remarks are due:
    \begin{enumerate}[(1)]
        \item At $q = -1$, $\partial_q$ refers to the right derivative $q\searrow -1$.
        \item \cite[Section~6.3]{AlfinitoB26} gives a different proof of \cref{cor:scrl-pde} using the theory of modular forms. They also recover \cref{lem:mlq-recurrence} as a consequence of their proof.
    \end{enumerate}
\end{remark}
We now finally state our differential recurrence for $M_\ell(q)$.
\begin{restatable}[Differential Recurrence for $M_\ell(q)$]{lemma}{mlqrecurrence}\label{lem:mlq-recurrence}
    For any $\ell\geq 0$, we have the following polynomial identity over the variable $q$:
    \[2q F'_\ell(q) = \ell(\ell + 1)F_\ell(q) - 2\ell(2\ell - 1)F_{\ell - 1}(q),\]
    where $F_\ell(q):= (1 - q)^\ell M_\ell(q)$. In particular, 
    \begin{equation}\label{eq:mlq-recurrence}
        2q(1 - q)M'_\ell(q) = \ell(2 + (\ell - 1)(1 - q))M_\ell(q) - 2\ell(2\ell - 1)M_{\ell - 1}(q).
    \end{equation}
    Here $M'_\ell(q)$ (resp. $F'_\ell(q)$) is the derivative of $M_\ell(q)$ (resp. $F_\ell(q)$) with respect to $q$, and we set $M_{-1}(q) = F_{-1}(q)= 0$ for all $q\in\R$.
\end{restatable}

Once \cref{lem:mlq-recurrence} is in place, we can quickly conclude the proof of \cref{thm:munq-pde}:

\begin{proof}[Proof of \cref{thm:munq-pde} assuming \cref{lem:mlq-recurrence}]
    For any $\ell\geq 0$ write $n_\ell:= n - 2\ell$. Multiply both sides of \cref{eq:mlq-recurrence} by $4(-1)^\ell\binom{n}{2\ell}z^{n_\ell}$, and sum over $0\leq\ell\leq \lfloor n/2\rfloor$. The left hand side becomes 
    \[\sum_{\ell = 0}^{\lfloor n/2\rfloor}2q(1 - q)M'_\ell(q)\cdot 4(-1)^\ell\binom{n}{2\ell}z^{n_\ell} = 8q(1 - q)\sum_{\ell = 0}^{\lfloor n/2\rfloor}(-1)^\ell M'_\ell(q)\cdot \binom{n}{2\ell}z^{n_\ell}\]
    \[= 8q(1 - q)\cdot\partial_q\mu_{n, q}(z),\]
    where we recall that $\mu_{n, q}(z) = \sum_{\ell = 0}^{\lfloor n/2\rfloor}(-1)^\ell M_\ell(q)\cdot \binom{n}{2\ell}z^{n_\ell}$. The right hand side becomes 
    \begin{equation}\label{eq:munqpde-rhs}
        \sum_{\ell = 0}^{\lfloor n/2\rfloor}\left(\underbrace{4\ell(2 + (\ell - 1)(1 - q))M_\ell(q)\cdot (-1)^\ell\binom{n}{2\ell}}_{\text{(I)}} + \underbrace{8\ell(2\ell - 1)M_{\ell - 1}(q)\cdot (-1)^{\ell - 1}\binom{n}{2\ell}}_{\text{(II)}}\right)z^{n_\ell}.
    \end{equation}
    
    Now we compare the coefficients of $z^{n_\ell}$ on both sides of \cref{eq:munq-pde}. Note that the operator $(1 - q)z^2\cdot\partial^2_z - b_n(q)z\cdot\partial_z + c_n(q)$ maps any monomial $z^t$ to a scalar multiple of itself, while the operator $\partial_z^2$ reduces degree by $2$. Consequently the coefficient of $z^{n_\ell}$ looks like $C_{\ell, n, q}M_\ell(q) + D_{\ell, n, q}M_{\ell - 1}(q)$, where the $M_{\ell - 1}(q)$ term arises when $\partial_z^2$ is applied to the $M_{\ell - 1}(q)z^{n_{\ell - 1}} = M_{\ell - 1}(q)z^{n_\ell + 2}$ term in $\mu_{n, q}(z)$. 

    Now consider $C_{\ell, n, q}$, i.e. the contribution proportional to $M_\ell(q)$. Its coefficient on the right-hand side is
    \[(-1)^\ell \binom{n}{2\ell}\cdot \left((1-q)n_\ell(n_\ell-1)-b_n(q)n_\ell+c_n(q)\right) = (-1)^\ell \binom{n}{2\ell}\cdot \left(4\ell\bigl(2+(\ell-1)(1 - q)\bigr)\right),\]
    and thus (I) in \cref{eq:munqpde-rhs} is verified.

It remains to account for the contribution involving $M_{\ell - 1}(q)$, which can only come from the operator $4\partial^2_z$. Note that
\[4\partial_z^2\left((-1)^{\ell - 1}M_{\ell - 1}(q)\binom{n}{2(\ell - 1)}z^{n_{\ell - 1}}\right) = 4(-1)^{\ell-1}(n_\ell+2)(n_\ell+1)\binom{n}{2\ell-2}M_{\ell-1}(q)z^{n_\ell}.\]
Also observe that 
\[(n_\ell+2)(n_\ell+1)\binom{n}{2\ell-2} = (2\ell)(2\ell-1)\binom{n}{2\ell},\]
and thus
\[4\partial_z^2\left((-1)^{\ell - 1}M_{\ell - 1}(q)\binom{n}{2(\ell - 1)}z^{n_{\ell - 1}}\right) = 8\ell(2\ell-1)M_{\ell-1}(q)\cdot (-1)^{\ell-1}\binom{n}{2\ell}z^{n_\ell},\]
which verifies (II) in \cref{eq:munqpde-rhs}, as desired.
\end{proof}
\begin{remark}
    Although the computations in the above proof are somewhat unpleasant, the main idea is quite simple: We apply the $\partial_q$ operator to bring in $M_\ell'(q)$ from \cref{lem:mlq-recurrence}. Also, as remarked in the proof, for $M_{\ell - 1}(q)$ to make an appearance, we need to apply the $\partial_z^2$ operator so that the $z^{n_{\ell - 1}} = z^{n_\ell + 2}$ monomial associated to $M_{\ell - 1}(q)$ gets ``lowered'' to $z^{n_\ell}$. 
\end{remark}
The proof of \cref{cor:scrl-pde} also follows similarly:
\begin{proof}[Proof of \cref{cor:scrl-pde}]
    Write $F_\ell(q):= (1 - q)^\ell M_\ell(q)$, and note that 
    \[\mathscr{L}_q(z) = \sum_{\ell = 0}^\infty F_\ell(q)\frac{z^{2\ell}}{(2\ell)!}.\]
    Now, by \cref{lem:mlq-recurrence} we know that 
    \begin{equation}\label{eq:flq-recurrence}
        2q F'_\ell(q) = \ell(\ell + 1)F_\ell(q) - 2\ell(2\ell - 1)F_{\ell - 1}(q).
    \end{equation}
    Multiplying \cref{eq:flq-recurrence} by $4z^{2\ell}/(2\ell)!$ and summing for all $\ell\geq 1$ yields the desired result. 
\end{proof}

Thus it only remains to prove \cref{lem:mlq-recurrence}.
\begin{proof}[Proof of \cref{lem:mlq-recurrence} using \cref{fact:touchard-riordan}]
    Let us try to ``reverse-engineer'' \cref{eq:mlq-recurrence}. We want a recurrence relating $M_\ell(q)$ and $M_{\ell - 1}(q)$. By \cref{fact:touchard-riordan} we know that 
    \[F_\ell(q):= (1 - q)^\ell M_\ell(q) = \sum_{r = 0}^\ell(-1)^rq^{r(r + 1)/2}B_{\ell, r}.\]
    From the definition of $B_{\ell, r}$ we also know that for $0\leq r\leq\ell$,
    \[2\ell(2\ell - 1)B_{\ell - 1, r} = (\ell - r)(\ell + r + 1)B_{\ell, r} = \left(\ell(\ell + 1) - r(r + 1)\right)B_{\ell, r}, \]
    where we set $B_{\ell - 1, \ell}:= 0$. Thus 
    \[2\ell(2\ell - 1)F_{\ell - 1}(q) = \ell(\ell + 1)F_\ell(q) - \sum_{r = 0}^\ell(-1)^rq^{r(r + 1)/2}r(r + 1)B_{\ell, r}.\]
    Thus we almost have a recurrence between $M_\ell(q)$ and $M_{\ell - 1}(q)$ except for the term 
    \[\sum_{r = 0}^\ell(-1)^rq^{r(r + 1)/2}r(r + 1)B_{\ell, r}.\] 
    But this extra term also makes clear why we have to consider a derivative w.r.t. $q$, since differentiating $q^{r(r + 1)/2}$ w.r.t. $q$ yields a factor of $r(r + 1)/2$. More precisely, note that $2q\partial_q q^{r(r + 1)/2} = r(r + 1)q^{r(r + 1)/2}$, and thus 
    \[\sum_{r = 0}^\ell(-1)^rq^{r(r + 1)/2}r(r + 1)B_{\ell, r} = 2q\partial_q\sum_{r = 0}^\ell(-1)^rq^{r(r + 1)/2}B_{\ell, r} = 2q F'_\ell(q).\]
    Consequently we obtain 
    \[2\ell(2\ell - 1)F_{\ell - 1}(q) = \ell(\ell + 1)F_\ell(q) - 2q F'_\ell(q).\]
    Substituting $F_\ell(q) = (1 - q)^\ell M_\ell(q)$ and simplifying yields the desired recurrence.
\end{proof}
\section{Generalizations of \cref{thm:main-thm}}\label{sec:vertex-weighted-generalizations}
Recall \cref{thm:main-thm}:
\maintheorem*
As mentioned in \cref{sec:intro}, \cref{thm:main-thm} is a special case of the \emph{Heilmann-Lieb matching theorem}~\cite{HeilmannL72} for $q = 1$. To make this comparison more precise, rewrite $\mu_{n, q}$ as follows:
\[\mu_{n,q}(z)= \sum_{\ell = 0}^{\lfloor n/2\rfloor}(-1)^\ell\binom{n}{2\ell}M_\ell(q)z^{n-2\ell} = \sum_{\pi\in\cM_n}(-1)^{|\pi|}q^{\Cr(\pi)}z^{n-2|\pi|},\]
where recall that $\cM_n$ is the set of all matchings (not necessarily perfect) on $[n]$. Recall that for any $n\geq 0$, we say $\pi=\{e_1,\ldots,e_k\}$, with $k\geq 0$, is a \emph{matching} on $[n]$ if $e_1,\ldots,e_k\subset[n]$, $|e_i|=2$ for all $i\in[k]$, and $e_i\cap e_j= \emptyset$ for all $1\leq i < j\leq k$. Write $|\pi|= k$, and note that $\pi$ is a perfect matching if $|\pi| = n/2$. Note that the empty matching $\pi = \{\}$ also counts as a matching in $\cM_n$.

By \cref{thm:main-thm}, we know that $\mu_{n, q}$ is real-rooted. Equivalently, if we define the \emph{deformed partition function}
\[P_{n, q}(z):= \sum_{\pi\in\cM_n}q^{\Cr(\pi)}z^{n-2|\pi|},\]
then \cref{thm:main-thm} implies that all roots of $P_{n, q}$ are purely imaginary, i.e. they lie on the line $\Ree(z)= 0$. The reason we call this the deformed \emph{partition function} is because~\cite{HeilmannL72} consider the \emph{partition function} of the \emph{monomer-dimer model} on a graph $G = G([n], E)$ on the vertex set $[n]$, which is given as 
\[Z_{G, \operatorname{MD}}(z):= \sum_{\pi\in\cM_n}\1(\pi\seq E(G))\cdot z^{n-2|\pi|}.\]
As with \cref{thm:main-thm},~\cite[Lemma~4.1]{HeilmannL72}, shows that all roots of $Z_{G, \operatorname{MD}}$ are purely imaginary. Note that $Z_{K_n, \operatorname{MD}} = P_{n, 1}$, where $K_n$ is the complete graph on $n$ vertices.

To give some physical intuition for $Z_{G, \operatorname{MD}}$, consider $n$ sites and some underlying graph $G$ on the $n$ sites, which specifies a topology on the sites. Every site may participate in a \emph{dimer} with at most one of its neighbors. Consequently, the state of the system can be specified by matching $\pi\subset E(G)$, where the edges $e\in\pi$ are dimers, and some unmatched sites (a.k.a. as \emph{monomers}). Let $z$ be the \emph{activity/fugacity} of a monomer, which represents the tendency of a site to stay unmatched. Then $Z_{G, \operatorname{MD}}(z)$ represents the \emph{grand canonical partition function} for the distribution of monomers and dimers on $G$. Furthermore, the imaginary-rootedness of $Z_{G, \operatorname{MD}}$ implies that this system can not have a phase transition except when the monomer fugacity is $0$.

The above physical intuition, apart from being interesting in its own right,\footnote{It is interesting to speculate on a natural physical interpretation for $P_{n, q}$: Fix some parameter $q\in[0,1]$, and consider the monomer-dimer model, where each matching/dimer configuration $\pi$ attracts a penalty of $q^{\Cr(\pi)}$, i.e. matchings with more crossings are suppressed for energetic reasons.} also suggests potential generalizations of \cref{thm:main-thm}. A natural generalization that Heilmann-Lieb consider is to allow the $n$ sites of the system to have \emph{different} fugacities. Thus for any $\bm w = (w_1, \ldots, w_n)\in\R_{\geq0}^n$ consider the polynomial
\[Z_{G, \bm w, \operatorname{MD}}(z):= \sum_{\substack{\pi\in\cM_n\\\pi\seq E(G)}} \left(\prod_{\substack{i\\i\text{ unmatched in }\pi}}w_i\right)\cdot z^{n - 2|\pi|}.\]
\cite{HeilmannL72} then show a more general theorem, saying that for any $\bm w\in\R_{\geq 0}^n$, the (univariate) polynomial $Z_{G, \bm w, \operatorname{MD}}(z)$ is imaginary-rooted.\footnote{it is understood if $Z_{G, \bm w, \operatorname{MD}}(z)\equiv 0$, then the zero function is considered imaginary-rooted} One may ask if a similar generalization holds for our $q$-deformed matching polynomial, and the answer is yes. For any $\bm w:= (w_1,\ldots, w_n)\in\R_{\geq 0}^n$, define

\begin{align}\label{eq:weightedmudef}
    \mu_{n, q, \bm{w}}(z) &:= \sum_{\pi \in \cM_n}(-1)^{|\pi|}q^{\Cr(\pi)}\prod_{i\in[n]\setminus V(\pi)}(w_iz) = \sum_{\pi \in \cM_n}(-1)^{|\pi|}q^{\Cr(\pi)}\left(\prod_{i\in[n]\setminus V(\pi)}w_i\right)\cdot z^{n - 2|\pi|}, 
\end{align}
where $V(\pi):= \bigcup_{e\in\pi}e$ refers to vertices matched under $\pi$. Then we have

\begin{restatable}[Vertex-Weighted Generalization of \cref{thm:main-thm}]{theorem}{vertexweightedmaintheorem}\label{thm:main-thm-vertex-weighted}
    For any real $q\in[-1, 1]$, any integer $n\geq 0$, and any $\bm{w}\in\R^n_{\geq 0}$, the polynomial $\mu_{n, q, \bm{w}}$ is real-rooted.
\end{restatable}

We can also consider the extension of \cref{thm:main-thm} in another direction: Note that the underlying graph in $\mu_{n, q}$ is the complete graph $K_n$, since we consider all matchings $\pi\in\cM_n$. On the other hand,~\cite{HeilmannL72} prove imaginary-rootedness of $Z_{G, \operatorname{MD}}$ for all graphs $G$. Can we hope for something similar?

The answer turns out to be no. Let $G = G([n], E)$ be a graph, and write 
\[\mu^{(e)}_{n, q, G}(z):= \sum_{\substack{\pi\in\cM_n\\\pi\seq E(G)}}(-1)^{|\pi|}q^{\Cr(\pi)} z^{n - 2|\pi|}.\]

\begin{proposition}\label{prop:edge-weighted-failure}
Consider the graph $G$ with edges $\{\{1, 3\}, \{1, 6\}, \{2, 5\}, \{3, 4\}\}$ on $[n]$, where $n = 6$. Then $\mu:= \mu^{(e)}_{6, q, G}$ is \textbf{not} real-rooted for any $q\in[-1, 0.7]$.
\end{proposition}
\begin{proof}
    Note that $\mu(z) = z^6 - 4z^4 + (3 + q)z^2 - 1$, and the claim follows by direct computation.
\end{proof}
This proposition is perhaps not so surprising in light of the fact that the crossing number $\Cr(\pi)$ of a matching $\pi$ depends on the ordering of the vertices. 

We now turn our attention to proving \cref{thm:main-thm-vertex-weighted}. It is a fairly straightforward consequence of the Grace-Walsh-Szeg\H{o} theorem (\cref{fact:Grace-Walsh-Szego}) from the theory of real stable polynomials.
\begin{proof}
    Recalling the notion of \emph{polarization} from \cref{def:polarization}, we claim that 
    \begin{equation}\label{eq:vertexweight-polarization}
        \mu_{n, q, \bm{w}}(z) = \operatorname{Pol}_n(\mu_{n, q})\left(w_1z, w_2z, \ldots, w_nz\right).
    \end{equation}
    By \cref{fact:Grace-Walsh-Szego}, $\operatorname{Pol}_n(\mu_{n, q})$ is real-stable, since $\mu_{n, q}$ is real-rooted (by \cref{thm:main-thm}). Now suppose $w_1, \ldots, w_n > 0$, and assume for the sake of contradiction $\mu_{n, q, \bm{w}}$ was not real-rooted. Then $\mu_{n, q, \bm{w}}$ would have a non-real root, and since $\mu_{n, q, \bm{w}}$ is a polynomial with real coefficients, it would have a non-real root $z$ with strictly positive imaginary part $\Imm(z) > 0$. But then $w_1 z, \ldots, w_n z\in\C$ would satisfy $\Imm(w_iz) > 0$ for all $i\in[n]$, and also satisfy $\operatorname{Pol}_n(\mu_{n, q})\left(w_1z, w_2z, \ldots, w_nz\right) = \mu_{n, q, \bm{w}}(z) = 0$, thus contradicting the real-stability of $\operatorname{Pol}_n(\mu_{n, q})$. 

    To handle the case where we may have $w_i = 0$ for some $i\in[n]$, note that for any $\eps > 0$, $\mu_{n, q, \bm{w} + \eps}$ is real-rooted by the previous argument, where $\bm{w} + \eps := (w_1 + \eps, \ldots, w_n + \eps)$. Also note that $\mu_{n, q, \bm{w} + \eps}$ converges locally uniformly to $\mu_{n, q, \bm{w}}$ as $\eps\searrow 0$. Consequently, by \cref{item:hurwitz} of \cref{fact:complex-analysis}, if $\mu_{n, q, \bm{w}}$ had a non-real root, then for some small enough $\eps > 0$, $\mu_{n, q, \bm{w} + \eps}$ would have a non-real root too, leading to a contradiction.
    
    Thus it only remains to show \cref{eq:vertexweight-polarization}. We do this by treating $w_1, \ldots, w_n$ as formal variables, and comparing coefficients. For any $I\seq[n]$ write $w^I:= \prod_{i\in I}w_i$. Note that if $n - |I|$ is odd, then the coefficient of $w^I$ on both sides of \cref{eq:vertexweight-polarization} is $0$. Thus assume $n - |I| =: 2\ell$ is even, and note that the coefficient of $w^I$ in $\mu_{n, q, \bm{w}}(z)$ is
    \[(-1)^{\ell}z^{n - 2\ell}\cdot\sum_{\substack{\pi \in \cM_n\\ V(\pi) = [n]\setminus I}}q^{\Cr(\pi)}.\]
    Now if $[n]\setminus I = \{i_1 < \cdots < i_{2\ell}\}$, then note that the bijection $[n]\setminus I\ni i_r\mapsto r\in[2\ell]$ preserves the crossing number $\Cr(\pi)$ of the matching $\pi$. Consequently, 
    \[\sum_{\substack{\pi \in \cM_n\\ V(\pi) = [n]\setminus I}}q^{\Cr(\pi)} = \sum_{\pi\text{ is a perfect matching on }[2\ell]}q^{\Cr(\pi)} = M_\ell(q).\]
    It may now be verified from the definition of $\operatorname{Pol}_n(\mu_{n, q})\left(w_1z, w_2z, \ldots, w_nz\right)$ that the coefficient of $w^I$ is exactly $(-1)^\ell z^{n - 2\ell}M_\ell(q)$, as desired.
\end{proof}

\section{Combinatorial Consequences of Real-Rootedness: \cref{thm:log-concave,thm:polya-frequency}}
The real-rootedness of $\mu_{n, q}$ has some purely combinatorial consequences thanks to connections between real-rooted polynomials and \emph{log-concave sequences}. More precisely, write 
\[c_\ell(q):= \frac{M_\ell(q)}{|\cP_2(2\ell)|} = \frac{M_\ell(q)}{(2\ell - 1)!!},\]
and recall \cref{thm:log-concave}:
\logconcavity*

For $q = 1$, note that $c_\ell(1) = 1$ for all $\ell\geq 0$, and thus \cref{thm:log-concave} follows trivially. For $q\in[-1, 1)$ though \cref{thm:log-concave} is a consequence of \emph{Newton's inequality} (\cref{fact:newton}) applied to $\mu_{n, q}$. 

One can also get a more general class of inequalities involving $\{M_\ell(q)\}_{\ell\geq 1}$ by proving that $\bm{a}^{(n)}:= (a_0^{(n)}, a_1^{(n)}, a_2^{(n)}, \ldots, )$ given by 
    \[a^{(n)}_\ell:= \1(2\ell\leq n)\cdot\binom{n}{2\ell}M_\ell(q)\]
    is a \emph{P\'olya-frequency sequence}.

\begin{definition}[P\'olya-frequency sequences]\label{def:polyafrequency}
    Consider a sequence $\bm{a}:= (a_0, a_1, a_2, \ldots)$ of real numbers, and write $a_k := 0$ for $k < 0$. Define the \emph{Toeplitz matrix} $T(\bm{a})$ as $T(\bm{a}):= (a_{i - j})_{i, j\geq 0}$, i.e.
\[
T(\bm{a})
:=
\begin{pmatrix}
a_0 & 0 & 0 & 0 & \cdots\\
a_1 & a_0 & 0 & 0 & \cdots\\
a_2 & a_1 & a_0 & 0 & \cdots\\
a_3 & a_2 & a_1 & a_0 & \cdots\\
\vdots & \vdots & \vdots & \vdots & \ddots
\end{pmatrix}.
\]
We then say that $\bm{a}$ is a \emph{P\'olya-frequency sequence} if all finite minors of $T(\bm{a})$ are non-negative, i.e. for all finite non-empty $I, J\seq\Z_{\geq 0}$ with $|I| = |J|$, we have $\det(a_{i - j})_{i\in I,j\in J}\geq 0$. 
\end{definition}

\begin{theorem}\label{thm:polya-frequency}
    For any real $q\in[-1, 1]$, and any integer $n\geq 1$, the sequence $\bm{a}^{(n)}:= (a^{(n)}_0, a^{(n)}_1, a^{(n)}_2, \ldots )$ given by 
    \[a^{(n)}_\ell:= \1(2\ell\leq n)\cdot\binom{n}{2\ell}M_\ell(q)\]
    is a P\'olya-frequency sequence.
\end{theorem}

Just as we need \cref{fact:newton} to prove \cref{thm:log-concave}, we need the \emph{Aissen-Schoenberg-Whitney} theorem~\cite{AissenSW52}, which is a ``higher-order'' analog of Newton's inequalities, to prove \cref{thm:polya-frequency}.
\begin{fact}[Aissen-Schoenberg-Whitney theorem]\label{fact:asw}
    Let $\lambda_1, \ldots, \lambda_m\geq 0$ be real numbers, and write $e_\ell:= e_\ell(\lambda_1, \ldots, \lambda_m)$ for all $0\leq\ell\leq m$. Then the sequence 
    \[\bm{e}:= (e_0, e_1, \ldots, e_m, 0, 0, \ldots)\]
    is a P\'olya frequency sequence.
\end{fact}
\begin{proof}[Proof of \cref{thm:log-concave,thm:polya-frequency}]
We claim that 
\begin{equation}\label{eq:munq-expression}
    \mu_{n, q}(z) = z^{\1(n\text{ is odd})}\cdot\prod_{k = 1}^{\lfloor n/2\rfloor}(z^2 - \lambda_k)
\end{equation}
for non-negative real numbers $\lambda_1, \ldots, \lambda_m\geq 0$, where $m:= \lfloor n/2\rfloor$. By \cref{thm:main-thm} we know that $\mu_{n, q}$ is real-rooted. Also note that if $n$ is even, then $\mu_{n, q}$ is actually a polynomial in $z^2$. If $n$ is odd, then $\mu_{n, q}$ is $z$ times a polynomial in $z^2$. Consequently \cref{eq:munq-expression} follows. Comparing coefficients then yields $e_\ell(\lambda_1, \ldots, \lambda_m) = \binom{n}{2\ell}M_\ell(q)$ for all $0\leq\ell\leq m$.

Applying \cref{fact:newton} to $e_\ell(\lambda_1, \ldots, \lambda_m)$ yields 
\[c_\ell(q)^2\geq\frac{n-2\ell-1}{n-2\ell+1}c_{\ell-1}(q)c_{\ell+1}(q).\]
Taking $n\geq 2\ell + 2$ to be an arbitrarily large even integer yields the desired result. Similarly, \cref{thm:polya-frequency} follows by applying \cref{fact:asw} to $e_\ell(\lambda_1, \ldots, \lambda_m)$. 
\end{proof}

\begingroup
\small
\bibliographystyle{alphaurl}
\bibliography{refs}

\newcommand{\etalchar}[1]{$^{#1}$}
\begin{thebibliography}{CGAH{\etalchar{+}}17}

\bibitem[AC76]{ArikC76}
M.~Arik and D.~D. Coon.
\newblock Hilbert spaces of analytic functions and generalized coherent states.
\newblock {\em Journal of Mathematical Physics}, 17(4):524--527, 04 1976.
\newblock \href
  {https://arxiv.org/abs/https://pubs.aip.org/aip/jmp/article-pdf/17/4/524/19147370/524_1_online.pdf}
  {\path{arXiv:https://pubs.aip.org/aip/jmp/article-pdf/17/4/524/19147370/524_1_online.pdf}},
  \href {https://doi.org/10.1063/1.522937} {\path{doi:10.1063/1.522937}}.

\bibitem[ACKK25]{anschuetzStronglyInteractingFermions2025}
Eric~R. Anschuetz, Chi-Fang Chen, Bobak~T. Kiani, and Robbie King.
\newblock Strongly {{Interacting Fermions Are Nontrivial}} yet {{Nonglassy}}.
\newblock {\em Physical Review Letters}, 135(3):030602, July 2025.
\newblock \href {https://doi.org/10.1103/cbqf-d24r}
  {\path{doi:10.1103/cbqf-d24r}}.

\bibitem[Asa70]{Asano70}
Taro Asano.
\newblock Theorems on the partition functions of the heisenberg ferromagnets.
\newblock {\em Journal of the Physical Society of Japan}, 29(2):350--359, 1970.
\newblock \href {https://arxiv.org/abs/https://doi.org/10.1143/JPSJ.29.350}
  {\path{arXiv:https://doi.org/10.1143/JPSJ.29.350}}, \href
  {https://doi.org/10.1143/JPSJ.29.350} {\path{doi:10.1143/JPSJ.29.350}}.

\bibitem[ASW52]{AissenSW52}
Michael Aissen, I.~J. Schoenberg, and A.~M. Whitney.
\newblock On the generating functions of totally positive sequences i.
\newblock {\em Journal d'Analyse Math{\'e}matique}, 2(1):93--103, 1952.
\newblock \href {https://doi.org/10.1007/BF02786970}
  {\path{doi:10.1007/BF02786970}}.

\bibitem[BA26]{AlfinitoB26}
Matteo Beccaria and Eleonora Alfinito.
\newblock Modular structures in the dssyk partition function, 2026.
\newblock URL: \url{https://arxiv.org/abs/2607.11828}, \href
  {https://arxiv.org/abs/2607.11828} {\path{arXiv:2607.11828}}.

\bibitem[Bar16]{Barvinok16}
Alexander Barvinok.
\newblock {\em Combinatorics and Complexity of Partition Functions}.
\newblock Springer, 2016.
\newblock \href {https://doi.org/10.1007/978-3-319-51829-9}
  {\path{doi:10.1007/978-3-319-51829-9}}.

\bibitem[BB09a]{BorceaB09a}
Julius Borcea and Petter Br{\"a}nd{\'e}n.
\newblock The {Lee-Yang} and {P{\'o}lya-Schur} programs. i. linear operators
  preserving stability.
\newblock {\em Inventiones Mathematicae}, 177(3):541--569, 2009.
\newblock \href {https://doi.org/10.1007/s00222-009-0189-3}
  {\path{doi:10.1007/s00222-009-0189-3}}.

\bibitem[BB09b]{BorceaB09b}
Julius Borcea and Petter Brändén.
\newblock The lee-yang and pólya-schur programs. ii. theory of stable
  polynomials and applications.
\newblock {\em Communications on Pure and Applied Mathematics},
  62(12):1595--1631, 2009.
\newblock URL: \url{https://onlinelibrary.wiley.com/doi/abs/10.1002/cpa.20295},
  \href
  {https://arxiv.org/abs/https://onlinelibrary.wiley.com/doi/pdf/10.1002/cpa.20295}
  {\path{arXiv:https://onlinelibrary.wiley.com/doi/pdf/10.1002/cpa.20295}},
  \href {https://doi.org/10.1002/cpa.20295} {\path{doi:10.1002/cpa.20295}}.

\bibitem[BBN19]{Babbush_Qsim}
Ryan Babbush, Dominic~W. Berry, and Hartmut Neven.
\newblock Quantum simulation of the sachdev-ye-kitaev model by asymmetric
  qubitization.
\newblock {\em Physical Review A}, 99(4), April 2019.
\newblock URL: \url{http://dx.doi.org/10.1103/PhysRevA.99.040301}, \href
  {https://doi.org/10.1103/physreva.99.040301}
  {\path{doi:10.1103/physreva.99.040301}}.

\bibitem[BCD24]{basso2024optimizing}
Joao Basso, Chi-Fang Chen, and Alexander~M Dalzell.
\newblock Optimizing random local hamiltonians by dissipation.
\newblock {\em arXiv preprint arXiv:2411.02578}, 2024.

\bibitem[Ber26]{bergamaschi2026fastmixingalltoallquantum}
Thiago Bergamaschi.
\newblock Fast mixing of all-to-all quantum systems at high temperatures, 2026.
\newblock URL: \url{https://arxiv.org/abs/2606.26090}, \href
  {https://arxiv.org/abs/2606.26090} {\path{arXiv:2606.26090}}.

\bibitem[BINT19]{SYK_Berkooz2}
Micha Berkooz, Mikhail Isachenkov, Vladimir Narovlansky, and Genis Torrents.
\newblock Towards a full solution of the large n double-scaled syk model.
\newblock {\em Journal of High Energy Physics}, 2019(3):1--72, 2019.

\bibitem[BKM26]{BasuKM26}
Arpon Basu, Pravesh~K. Kothari, and Siddhant Midha.
\newblock Sharp bounds on ground state energy of the syk model, 2026.
\newblock URL: \url{https://arxiv.org/abs/2607.27185}, \href
  {https://arxiv.org/abs/2607.27185} {\path{arXiv:2607.27185}}.

\bibitem[BKS97]{bozejko1997q}
Marek Bo{\.z}ejko, Burkhard K{\"u}mmerer, and Roland Speicher.
\newblock q-gaussian processes: non-commutative and classical aspects.
\newblock {\em Communications in Mathematical Physics}, 185(1):129--154, 1997.

\bibitem[BLM13]{boucheron2013non}
St{\'e}phane Boucheron, G{\'a}bor Lugosi, and Pascal Massart.
\newblock {\em Concentration Inequalities: A Nonasymptotic Theory of
  Independence}.
\newblock Oxford University Press, 02 2013.
\newblock \href {https://doi.org/10.1093/acprof:oso/9780199535255.001.0001}
  {\path{doi:10.1093/acprof:oso/9780199535255.001.0001}}.

\bibitem[BM25]{SYK_Berkooz3}
Micha Berkooz and Ohad Mamroud.
\newblock A cordial introduction to double scaled syk.
\newblock {\em Reports on Progress in Physics}, 88(3):036001, 2025.

\bibitem[Br{\"a}14]{Branden14}
Petter Br{\"a}nd{\'e}n.
\newblock Unimodality, log-concavity, real-rootedness and beyond, 2014.
\newblock URL: \url{https://arxiv.org/abs/1410.6601}, \href
  {https://arxiv.org/abs/1410.6601} {\path{arXiv:1410.6601}}.

\bibitem[Bre92]{Brenti92}
Francesco Brenti.
\newblock Expansions of chromatic polynomials and log-concavity.
\newblock {\em Trans. Amer. Math. Soc.}, 332(2):729--756, 1992.
\newblock \href {https://doi.org/10.2307/2154193} {\path{doi:10.2307/2154193}}.

\bibitem[BS91]{bozejko1991example}
Marek Bo{\.z}ejko and Roland Speicher.
\newblock An example of a generalized brownian motion.
\newblock {\em Communications in Mathematical Physics}, 137(3):519--531, 1991.

\bibitem[BSS14]{BatsonSS14}
Joshua Batson, Daniel~A. Spielman, and Nikhil Srivastava.
\newblock Twice-ramanujan sparsifiers.
\newblock {\em SIAM Review}, 56(2):315--334, 2014.
\newblock \href {https://doi.org/10.1137/130949117}
  {\path{doi:10.1137/130949117}}.

\bibitem[Cay09]{Cayley_2009}
Arthur Cayley.
\newblock {\em A theorem on trees}, page 26–28.
\newblock Cambridge Library Collection - Mathematics. Cambridge University
  Press, 2009.

\bibitem[CC89]{CravenC89}
Thomas Craven and George Csordas.
\newblock {Jensen polynomials and the Tur{\'a}n and Laguerre inequalities.}
\newblock {\em Pacific Journal of Mathematics}, 136(2):241 -- 260, 1989.

\bibitem[CC94]{CravenC94}
T.~Craven and G.~Csordas.
\newblock Differential operators of infinite order and the distribution of
  zeros of entire functions.
\newblock {\em Journal of Mathematical Analysis and Applications},
  186(3):799--820, 1994.
\newblock URL:
  \url{https://www.sciencedirect.com/science/article/pii/S0022247X84713345},
  \href {https://doi.org/10.1006/jmaa.1994.1334}
  {\path{doi:10.1006/jmaa.1994.1334}}.

\bibitem[CGAH{\etalchar{+}}17]{cotler2017black}
Jordan~S Cotler, Guy Gur-Ari, Masanori Hanada, Joseph Polchinski, Phil Saad,
  Stephen~H Shenker, Douglas Stanford, Alexandre Streicher, and Masaki Tezuka.
\newblock Black holes and random matrices.
\newblock {\em Journal of High Energy Physics}, 2017(5):1--54, 2017.

\bibitem[CGPS22]{SYK_review}
Debanjan Chowdhury, Antoine Georges, Olivier Parcollet, and Subir Sachdev.
\newblock Sachdev-ye-kitaev models and beyond: Window into non-fermi liquids.
\newblock {\em Rev. Mod. Phys.}, 94:035004, Sep 2022.
\newblock URL: \url{https://link.aps.org/doi/10.1103/RevModPhys.94.035004},
  \href {https://doi.org/10.1103/RevModPhys.94.035004}
  {\path{doi:10.1103/RevModPhys.94.035004}}.

\bibitem[CKG25]{chen2025efficientexactnoncommutativequantum}
Chi-Fang Chen, Michael~J. Kastoryano, and András Gilyén.
\newblock An efficient and exact noncommutative quantum gibbs sampler, 2025.
\newblock URL: \url{https://arxiv.org/abs/2311.09207}, \href
  {https://arxiv.org/abs/2311.09207} {\path{arXiv:2311.09207}}.

\bibitem[CLV23]{chen2023optimalmixingglauberdynamics}
Zongchen Chen, Kuikui Liu, and Eric Vigoda.
\newblock Optimal mixing of glauber dynamics: Entropy factorization via
  high-dimensional expansion, 2023.
\newblock URL: \url{https://arxiv.org/abs/2011.02075}, \href
  {https://arxiv.org/abs/2011.02075} {\path{arXiv:2011.02075}}.

\bibitem[CLV24]{Chen_2024}
Zongchen Chen, Kuikui Liu, and Eric Vigoda.
\newblock Spectral independence via stability and applications to holant-type
  problems.
\newblock {\em TheoretiCS}, Volume 3, 2024.
\newblock URL: \url{http://dx.doi.org/10.46298/theoretics.24.16}, \href
  {https://doi.org/10.46298/theoretics.24.16}
  {\path{doi:10.46298/theoretics.24.16}}.

\bibitem[CSV94]{CsordasSV94}
George Csordas, Wayne Smith, and Richard~S. Varga.
\newblock Lehmer pairs of zeros, the de {B}ruijn-{N}ewman constant {$\Lambda$},
  and the {R}iemann hypothesis.
\newblock {\em Constr. Approx.}, 10(1):107--129, 1994.
\newblock \href {https://doi.org/10.1007/BF01205170}
  {\path{doi:10.1007/BF01205170}}.

\bibitem[CZZ26]{CaiZZ26}
Yiyi Cai, Yongtao Zhan, and Alexander Zlokapa.
\newblock Fast mixing of the syk model at high temperatures, 2026.
\newblock URL: \url{https://arxiv.org/abs/2610.05953}, \href
  {https://arxiv.org/abs/2610.05953} {\path{arXiv:2610.05953}}.

\bibitem[{\relax DLMF}]{NIST:DLMF}
{\it NIST Digital Library of Mathematical Functions}.
\newblock \url{https://dlmf.nist.gov/}, Release 1.2.8 of 2026-09-15.
\newblock F.~W.~J. Olver, A.~B. {Olde Daalhuis}, D.~W. Lozier, B.~I. Schneider,
  R.~F. Boisvert, C.~W. Clark, B.~R. Miller, B.~V. Saunders, H.~S. Cohl, and
  M.~A. McClain, eds.
\newblock URL: \url{https://dlmf.nist.gov/}.

\bibitem[Fis67]{fisher1967theory}
Michael~E Fisher.
\newblock The theory of equilibrium critical phenomena.
\newblock {\em Reports on progress in physics}, 30(2):615--730, 1967.

\bibitem[FK72]{fortuin1972random}
Cornelius~Marius Fortuin and Piet~W Kasteleyn.
\newblock On the random-cluster model: I. introduction and relation to other
  models.
\newblock {\em Physica}, 57(4):536--564, 1972.

\bibitem[FTW19]{feng2019spectrum}
Renjie Feng, Gang Tian, and Dongyi Wei.
\newblock Spectrum of syk model.
\newblock {\em Peking Mathematical Journal}, 2(1):41--70, 2019.

\bibitem[FTW20]{feng2020spectrum}
Renjie Feng, Gang Tian, and Dongyi Wei.
\newblock Spectrum of syk model iii: Large deviations and concentration of
  measures.
\newblock {\em Random Matrices: Theory and Applications}, 9(02):2050001, 2020.

\bibitem[FTW21]{feng2021spectrum}
Renjie Feng, Gang Tian, and Dongyi Wei.
\newblock Spectrum of syk model ii: Central limit theorem.
\newblock {\em Random Matrices: Theory and Applications}, 10(04):2150037, 2021.

\bibitem[G{\'A}EL{\etalchar{+}}17]{Garcia_Qsim}
L.~Garc{\'{\i}}a-{\'A}lvarez, I.~L. Egusquiza, L.~Lamata, A.~del Campo,
  J.~Sonner, and E.~Solano.
\newblock Digital quantum simulation of minimal $\mathrm{AdS}/\mathrm{CFT}$.
\newblock {\em Phys. Rev. Lett.}, 119:040501, Jul 2017.
\newblock URL: \url{https://link.aps.org/doi/10.1103/PhysRevLett.119.040501},
  \href {https://doi.org/10.1103/PhysRevLett.119.040501}
  {\path{doi:10.1103/PhysRevLett.119.040501}}.

\bibitem[Gam21]{Gamarnik_2021}
David Gamarnik.
\newblock The overlap gap property: A topological barrier to optimizing over
  random structures.
\newblock {\em Proceedings of the National Academy of Sciences}, 118(41),
  October 2021.
\newblock URL: \url{http://dx.doi.org/10.1073/pnas.2108492118}, \href
  {https://doi.org/10.1073/pnas.2108492118}
  {\path{doi:10.1073/pnas.2108492118}}.

\bibitem[GGV16]{GarciaVerbaaschotPRD}
Antonio~M. Garc{\'{\i}}a-Garc{\'{\i}}a and Jacobus J.~M. Verbaarschot.
\newblock Spectral and thermodynamic properties of the sachdev-ye-kitaev model.
\newblock {\em Phys. Rev. D}, 94:126010, Dec 2016.
\newblock URL: \url{https://link.aps.org/doi/10.1103/PhysRevD.94.126010}, \href
  {https://doi.org/10.1103/PhysRevD.94.126010}
  {\path{doi:10.1103/PhysRevD.94.126010}}.

\bibitem[GNV23]{goel2023semiclassicalgeometrydoublescaledsyk}
Akash Goel, Vladimir Narovlansky, and Herman Verlinde.
\newblock Semiclassical geometry in double-scaled syk, 2023.
\newblock URL: \url{https://arxiv.org/abs/2301.05732}, \href
  {https://arxiv.org/abs/2301.05732} {\path{arXiv:2301.05732}}.

\bibitem[GPSZ26]{gamarnik2026free}
David Gamarnik, Francisco Pernice, Alexander Schmidhuber, and Alexander
  Zlokapa.
\newblock The free energy limit of the syk model at high temperature.
\newblock {\em arXiv preprint arXiv:2605.02768}, 2026.

\bibitem[Gri25]{Gribinski25}
Aurelien Gribinski.
\newblock A bivariate approach to real-rootedness of special polynomials.
\newblock {\em Integral Transforms and Special Functions}, 36(10):824--845,
  2025.
\newblock \href
  {https://arxiv.org/abs/https://doi.org/10.1080/10652469.2025.2460013}
  {\path{arXiv:https://doi.org/10.1080/10652469.2025.2460013}}, \href
  {https://doi.org/10.1080/10652469.2025.2460013}
  {\path{doi:10.1080/10652469.2025.2460013}}.

\bibitem[He26]{he2026spectraledgequarticsyk}
Yukun He.
\newblock Free energy and spectral edge of the syk model, 2026.
\newblock URL: \url{https://arxiv.org/abs/2607.18998}, \href
  {https://arxiv.org/abs/2607.18998} {\path{arXiv:2607.18998}}.

\bibitem[Hig08]{higham2008functions}
Nicholas~J. Higham.
\newblock {\em Functions of Matrices}.
\newblock Society for Industrial and Applied Mathematics, 2008.
\newblock URL: \url{https://epubs.siam.org/doi/abs/10.1137/1.9780898717778},
  \href
  {https://arxiv.org/abs/https://epubs.siam.org/doi/pdf/10.1137/1.9780898717778}
  {\path{arXiv:https://epubs.siam.org/doi/pdf/10.1137/1.9780898717778}}, \href
  {https://doi.org/10.1137/1.9780898717778}
  {\path{doi:10.1137/1.9780898717778}}.

\bibitem[HL72]{HeilmannL72}
Ole~J. Heilmann and Elliott~H. Lieb.
\newblock {Theory of monomer-dimer systems}.
\newblock {\em Communications in Mathematical Physics}, 25(3):190 -- 232, 1972.

\bibitem[HMS20]{HarrowMS20}
Aram Harrow, Saeed Mehraban, and Mehdi Soleimanifar.
\newblock Classical algorithms, correlation decay, and complex zeros of
  partition functions of quantum many-body systems.
\newblock In {\em Proceedings of the 52nd Annual ACM SIGACT Symposium on Theory
  of Computing}, pages 378--386, 2020.
\newblock \href {https://arxiv.org/abs/1910.09071} {\path{arXiv:1910.09071}},
  \href {https://doi.org/10.1145/3357713.3384322}
  {\path{doi:10.1145/3357713.3384322}}.

\bibitem[HO22]{hastings2022optimizing}
Matthew~B Hastings and Ryan O'Donnell.
\newblock Optimizing strongly interacting fermionic hamiltonians.
\newblock In {\em Proceedings of the 54th annual ACM SIGACT symposium on theory
  of computing}, pages 776--789, 2022.

\bibitem[H{\"o}r90]{Hormander90}
Lars H{\"o}rmander.
\newblock {\em An Introduction to Complex Analysis in Several Variables},
  volume~7 of {\em North-Holland Mathematical Library}.
\newblock North-Holland, 3 edition, 1990.

\bibitem[Huh23]{Huh23}
June Huh.
\newblock Combinatorics and {H}odge theory.
\newblock In {\em I{CM}---{I}nternational {C}ongress of {M}athematicians.
  {V}ol. 1. {P}rize lectures}, pages 212--239. EMS Press, Berlin, [2023]
  \copyright 2023.
\newblock \href {https://doi.org/10.4171/ICM2022/205}
  {\path{doi:10.4171/ICM2022/205}}.

\bibitem[Iss18]{Isserlis18}
L.~Isserlis.
\newblock On a formula for the product-moment coefficient of any order of a
  normal frequency distribution in any number of variables.
\newblock {\em Biometrika}, 12(1-2):134--139, 11 1918.
\newblock \href
  {https://arxiv.org/abs/https://academic.oup.com/biomet/article-pdf/12/1-2/134/481266/12-1-2-134.pdf}
  {\path{arXiv:https://academic.oup.com/biomet/article-pdf/12/1-2/134/481266/12-1-2-134.pdf}},
  \href {https://doi.org/10.1093/biomet/12.1-2.134}
  {\path{doi:10.1093/biomet/12.1-2.134}}.

\bibitem[ISV87]{IsmailSV87}
Mourad~E.H. Ismail, Dennis Stanton, and G{\'e}rard Viennot.
\newblock The combinatorics of q-hermite polynomials and the askey—wilson
  integral.
\newblock {\em European Journal of Combinatorics}, 8(4):379--392, 1987.
\newblock URL:
  \url{https://www.sciencedirect.com/science/article/pii/S019566988780046X},
  \href {https://doi.org/10.1016/S0195-6698(87)80046-X}
  {\path{doi:10.1016/S0195-6698(87)80046-X}}.

\bibitem[JV13]{VergesCumulants}
Matthieu Josuat-Verg{\'e}s.
\newblock Cumulants of the q-semicircular law, tutte polynomials, and heaps.
\newblock {\em Canadian Journal of Mathematics}, 65(4):863–878, August 2013.
\newblock URL: \url{http://dx.doi.org/10.4153/CJM-2012-042-9}, \href
  {https://doi.org/10.4153/cjm-2012-042-9} {\path{doi:10.4153/cjm-2012-042-9}}.

\bibitem[JV18]{SYK_Jia}
Yiyang Jia and Jacobus~JM Verbaarschot.
\newblock Large n expansion of the moments and free energy of sachdev-ye-kitaev
  model, and the enumeration of intersection graphs.
\newblock {\em Journal of High Energy Physics}, 2018(11):1--34, 2018.

\bibitem[JZL{\etalchar{+}}22]{jafferis_qsim}
Daniel Jafferis, Alexander Zlokapa, Joseph~D Lykken, David~K Kolchmeyer,
  Samantha~I Davis, Nikolai Lauk, Hartmut Neven, and Maria Spiropulu.
\newblock Traversable wormhole dynamics on a quantum processor.
\newblock {\em Nature}, 612(7938):51--55, 2022.

\bibitem[Kit15]{Kitaev15}
Alexei Kitaev.
\newblock A simple model of quantum holography.
\newblock {\em Entanglement in strongly-correlated quantum matter}, page~38,
  2015.

\bibitem[Kru56]{kruskal1956shortest}
Joseph~B Kruskal.
\newblock On the shortest spanning subtree of a graph and the traveling
  salesman problem.
\newblock {\em Proceedings of the American Mathematical society}, 7(1):48--50,
  1956.

\bibitem[Lev80]{Levin80}
B.~Ja. Levin.
\newblock {\em Distribution of Zeros of Entire Functions}, volume~5 of {\em
  Translations of Mathematical Monographs}.
\newblock American Mathematical Society, Providence, RI, revised edition, 1980.
\newblock \href {https://doi.org/10.1090/mmono/005}
  {\path{doi:10.1090/mmono/005}}.

\bibitem[LS81]{LiebS81}
Elliott~H. Lieb and Alan~D. Sokal.
\newblock {A general Lee-Yang theorem for one-component and multicomponent
  ferromagnets}.
\newblock {\em Communications in Mathematical Physics}, 80(2):153 -- 179, 1981.

\bibitem[LSS19]{LiuSS19}
Jingcheng Liu, Alistair Sinclair, and Piyush Srivastava.
\newblock The {Ising} partition function: Zeros and deterministic
  approximation.
\newblock {\em Journal of Statistical Physics}, 174:287--315, 2019.
\newblock \href {https://doi.org/10.1007/s10955-018-2199-2}
  {\path{doi:10.1007/s10955-018-2199-2}}.

\bibitem[LY52]{LY52}
T.~D. Lee and C.~N. Yang.
\newblock Statistical theory of equations of state and phase transitions. ii.
  lattice gas and ising model.
\newblock {\em Phys. Rev.}, 87:410--419, Aug 1952.
\newblock URL: \url{https://link.aps.org/doi/10.1103/PhysRev.87.410}, \href
  {https://doi.org/10.1103/PhysRev.87.410} {\path{doi:10.1103/PhysRev.87.410}}.

\bibitem[LYL{\etalchar{+}}19]{luo_qsim}
Zhihuang Luo, Yi-Zhuang You, Jun Li, Chao-Ming Jian, Dawei Lu, Cenke Xu, Bei
  Zeng, and Raymond Laflamme.
\newblock Quantum simulation of the non-fermi-liquid state of sachdev-ye-kitaev
  model.
\newblock {\em npj Quantum Information}, 5(1):53, 2019.

\bibitem[MH21]{Mann_2021}
Ryan~L. Mann and Tyler Helmuth.
\newblock Efficient algorithms for approximating quantum partition functions.
\newblock {\em Journal of Mathematical Physics}, 62(2), February 2021.
\newblock URL: \url{http://dx.doi.org/10.1063/5.0013689}, \href
  {https://doi.org/10.1063/5.0013689} {\path{doi:10.1063/5.0013689}}.

\bibitem[MS16]{MaldacenaS16}
Juan Maldacena and Douglas Stanford.
\newblock Remarks on the sachdev-ye-kitaev model.
\newblock {\em Phys. Rev. D}, 94:106002, Nov 2016.
\newblock URL: \url{https://link.aps.org/doi/10.1103/PhysRevD.94.106002}, \href
  {https://doi.org/10.1103/PhysRevD.94.106002}
  {\path{doi:10.1103/PhysRevD.94.106002}}.

\bibitem[MSS15a]{MarcusSS15a}
{Adam W.} Marcus, {Daniel A.} Spielman, and Nikhil Srivastava.
\newblock Interlacing families i: Bipartite ramanujan graphs of all degrees.
\newblock {\em Annals of Mathematics}, 182(1):307--325, 2015.
\newblock \href {https://doi.org/10.4007/annals.2015.182.1.7}
  {\path{doi:10.4007/annals.2015.182.1.7}}.

\bibitem[MSS15b]{MarcusSS15}
Adam~W. Marcus, Daniel~A. Spielman, and Nikhil Srivastava.
\newblock Interlacing families {II}: {M}ixed characteristic polynomials and the
  {K}adison-{S}inger problem.
\newblock {\em Ann. of Math. (2)}, 182(1):327--350, 2015.

\bibitem[MSS16]{maldacena2016bound}
Juan Maldacena, Stephen~H Shenker, and Douglas Stanford.
\newblock A bound on chaos.
\newblock {\em Journal of High Energy Physics}, 2016(8):106, 2016.

\bibitem[MSS22]{MarcusSS22}
Adam~W. Marcus, Daniel~A. Spielman, and Nikhil Srivastava.
\newblock Finite free convolutions of polynomials.
\newblock {\em Probability Theory and Related Fields}, 182(3--4):807--848,
  2022.
\newblock \href {https://arxiv.org/abs/1504.00350} {\path{arXiv:1504.00350}},
  \href {https://doi.org/10.1007/s00440-021-01105-w}
  {\path{doi:10.1007/s00440-021-01105-w}}.

\bibitem[New07]{Newton1707}
Isaac Newton.
\newblock {\em Arithmetica universalis: sive de compositione et resolutione
  arithmetica liber}.
\newblock Typis Academicis, Cambridge, 1707.
\newblock URL: \url{https://rmda.kulib.kyoto-u.ac.jp/en/item/rb00028902}.

\bibitem[New74]{Newman74}
Charles~M. Newman.
\newblock Zeros of the partition function for generalized ising systems.
\newblock {\em Communications on Pure and Applied Mathematics}, 27(2):143--159,
  1974.
\newblock URL:
  \url{https://onlinelibrary.wiley.com/doi/abs/10.1002/cpa.3160270203}, \href
  {https://arxiv.org/abs/https://onlinelibrary.wiley.com/doi/pdf/10.1002/cpa.3160270203}
  {\path{arXiv:https://onlinelibrary.wiley.com/doi/pdf/10.1002/cpa.3160270203}},
  \href {https://doi.org/10.1002/cpa.3160270203}
  {\path{doi:10.1002/cpa.3160270203}}.

\bibitem[NS06]{NicaS06}
A.~Nica and R.~Speicher.
\newblock {\em Lectures on the Combinatorics of Free Probability}.
\newblock Number v. 13 in Lectures on the combinatorics of free probability.
  Cambridge University Press, 2006.
\newblock URL: \url{https://books.google.com/books?id=zsw_-E-QrOkC}.

\bibitem[Oku23]{Okuyama23}
Kazumi Okuyama.
\newblock High temperature expansion of double scaled syk.
\newblock {\em Physics Letters B}, 843:138036, 2023.
\newblock URL:
  \url{https://www.sciencedirect.com/science/article/pii/S0370269323003702},
  \href {https://doi.org/10.1016/j.physletb.2023.138036}
  {\path{doi:10.1016/j.physletb.2023.138036}}.

\bibitem[PR16]{polchinski2016spectrum}
Joseph Polchinski and Vladimir Rosenhaus.
\newblock The spectrum in the sachdev-ye-kitaev model.
\newblock {\em Journal of High Energy Physics}, 2016(4):1--25, 2016.

\bibitem[PR22]{PatelR22}
Viresh Patel and Guus Regts.
\newblock Approximate counting using {Taylor}'s theorem: A survey.
\newblock {\em Bulletin of the EATCS}, 138, 2022.
\newblock URL: \url{https://arxiv.org/abs/2212.08143}, \href
  {https://arxiv.org/abs/2212.08143} {\path{arXiv:2212.08143}}.

\bibitem[Rio75]{riordan1975distribution}
John Riordan.
\newblock The distribution of crossings of chords joining pairs of 2n points on
  a circle.
\newblock {\em Mathematics of Computation}, 29(129):215--222, 1975.

\bibitem[Ros19]{rosenhaus2019introduction}
Vladimir Rosenhaus.
\newblock An introduction to the syk model.
\newblock {\em Journal of Physics A: Mathematical and Theoretical},
  52(32):323001, 2019.

\bibitem[RS02]{RahmanS02}
Q~I Rahman and G~Schmeisser.
\newblock {\em Analytic Theory of Polynomials}.
\newblock Oxford University Press, 09 2002.
\newblock \href {https://doi.org/10.1093/oso/9780198534938.001.0001}
  {\path{doi:10.1093/oso/9780198534938.001.0001}}.

\bibitem[RSFA26]{Rouz__2026}
Cambyse Rouzé, Daniel Stilck~França, and Álvaro~M. Alhambra.
\newblock Optimal quantum algorithm for gibbs state preparation.
\newblock {\em Physical Review Letters}, 136(6), February 2026.
\newblock URL: \url{http://dx.doi.org/10.1103/lhht-svmn}, \href
  {https://doi.org/10.1103/lhht-svmn} {\path{doi:10.1103/lhht-svmn}}.

\bibitem[RT20]{RodgersT20}
Brad Rodgers and Terence Tao.
\newblock The de bruijn–newman constant is non-negative.
\newblock {\em Forum of Mathematics, Pi}, 8:e6, 2020.
\newblock \href {https://doi.org/10.1017/fmp.2020.6}
  {\path{doi:10.1017/fmp.2020.6}}.

\bibitem[Rud76]{Rudin76}
Walter Rudin.
\newblock {\em Principles of Mathematical Analysis}.
\newblock McGraw-Hill, New York, 3 edition, 1976.
\newblock URL:
  \url{https://www.lehman.edu/faculty/rbettiol/lehman_teaching/2020mat320/baby_Rudin.pdf}.

\bibitem[Sac24]{Sachdev24}
Subir Sachdev.
\newblock {Quantum spin glasses and Sachdev-Ye-Kitaev models}, 2 2024.
\newblock \href {https://arxiv.org/abs/2402.17824} {\path{arXiv:2402.17824}}.

\bibitem[SF71]{SuzukiF71}
Masuo Suzuki and Michael~E. Fisher.
\newblock Zeros of the partition function for the heisenberg, ferroelectric,
  and general ising models.
\newblock {\em Journal of Mathematical Physics}, 12(2):235--246, 02 1971.
\newblock \href
  {https://arxiv.org/abs/https://pubs.aip.org/aip/jmp/article-pdf/12/2/235/19190307/235_1_online.pdf}
  {\path{arXiv:https://pubs.aip.org/aip/jmp/article-pdf/12/2/235/19190307/235_1_online.pdf}},
  \href {https://doi.org/10.1063/1.1665583} {\path{doi:10.1063/1.1665583}}.

\bibitem[Sok05]{Sokal_2005}
Alan~D. Sokal.
\newblock {\em The multivariate Tutte polynomial (alias Potts model) for graphs
  and matroids}, page 173–226.
\newblock Cambridge University Press, July 2005.
\newblock URL: \url{http://dx.doi.org/10.1017/CBO9780511734885.009}, \href
  {https://doi.org/10.1017/cbo9780511734885.009}
  {\path{doi:10.1017/cbo9780511734885.009}}.

\bibitem[SS03]{SteinS03}
Elias~M. Stein and Rami Shakarchi.
\newblock {\em Complex analysis}, volume~2 of {\em Princeton Lectures in
  Analysis}.
\newblock Princeton University Press, Princeton, NJ, 2003.

\bibitem[Sta89]{Stanley89}
Richard~P. Stanley.
\newblock Log-concave and unimodal sequences in algebra, combinatorics, and
  geometry.
\newblock {\em Annals of the New York Academy of Sciences}, 576(1):500--535,
  1989.
\newblock URL:
  \url{https://nyaspubs.onlinelibrary.wiley.com/doi/abs/10.1111/j.1749-6632.1989.tb16434.x},
  \href
  {https://arxiv.org/abs/https://nyaspubs.onlinelibrary.wiley.com/doi/pdf/10.1111/j.1749-6632.1989.tb16434.x}
  {\path{arXiv:https://nyaspubs.onlinelibrary.wiley.com/doi/pdf/10.1111/j.1749-6632.1989.tb16434.x}},
  \href {https://doi.org/10.1111/j.1749-6632.1989.tb16434.x}
  {\path{doi:10.1111/j.1749-6632.1989.tb16434.x}}.

\bibitem[SY93]{SachdevY93}
Subir Sachdev and Jinwu Ye.
\newblock Gapless spin-fluid ground state in a random quantum heisenberg
  magnet.
\newblock {\em Phys. Rev. Lett.}, 70:3339--3342, May 1993.
\newblock URL: \url{https://link.aps.org/doi/10.1103/PhysRevLett.70.3339},
  \href {https://doi.org/10.1103/PhysRevLett.70.3339}
  {\path{doi:10.1103/PhysRevLett.70.3339}}.

\bibitem[SZ26]{SchmidhuberZ26}
Alexander Schmidhuber and Alexander Zlokapa.
\newblock Polynomial-time classical algorithms for mean-field models up to the
  glass transition, 2026.
\newblock URL: \url{https://arxiv.org/abs/2610.06807}, \href
  {https://arxiv.org/abs/2610.06807} {\path{arXiv:2610.06807}}.

\bibitem[Sza09]{Szabłowski09}
Paweł~J. Szabłowski.
\newblock q-gaussian distributions: Simplifications and simulations.
\newblock {\em Journal of Probability and Statistics}, 2009(1):752430, 2009.
\newblock URL:
  \url{https://onlinelibrary.wiley.com/doi/abs/10.1155/2009/752430}, \href
  {https://arxiv.org/abs/https://onlinelibrary.wiley.com/doi/pdf/10.1155/2009/752430}
  {\path{arXiv:https://onlinelibrary.wiley.com/doi/pdf/10.1155/2009/752430}},
  \href {https://doi.org/10.1155/2009/752430} {\path{doi:10.1155/2009/752430}}.

\bibitem[Sze39]{szeg1939orthogonal}
Gabor Szeg\H{o}.
\newblock {\em Orthogonal polynomials}, volume~23.
\newblock American Mathematical Soc., 1939.

\bibitem[Tao17]{Tao17}
Terence Tao.
\newblock Heat flow and zeroes of polynomials, 2017.
\newblock What's New (blog).
\newblock URL:
  \url{https://terrytao.wordpress.com/2017/10/17/heat-flow-and-zeroes-of-polynomials/}.

\bibitem[Tou52]{touchard1952probleme}
Jacques Touchard.
\newblock Sur un probleme de configurations et sur les fractions continues.
\newblock {\em Canadian Journal of Mathematics}, 4:2--25, 1952.

\bibitem[Tut54]{tutte1954contribution}
William~Thomas Tutte.
\newblock A contribution to the theory of chromatic polynomials.
\newblock {\em Canadian journal of mathematics}, 6:80--91, 1954.

\bibitem[Vin26]{Vinzant26}
Cynthia Vinzant.
\newblock Log-concave polynomials for matroids and expansion, 2026.
\newblock URL: \url{https://epubs.siam.org/doi/abs/10.1137/25M1807344}, \href
  {https://arxiv.org/abs/https://epubs.siam.org/doi/pdf/10.1137/25M1807344}
  {\path{arXiv:https://epubs.siam.org/doi/pdf/10.1137/25M1807344}}, \href
  {https://doi.org/10.1137/25M1807344} {\path{doi:10.1137/25M1807344}}.

\bibitem[Wag09]{Wagner09}
David~G. Wagner.
\newblock Multivariate stable polynomials: theory and applications, 2009.
\newblock URL: \url{https://arxiv.org/abs/0911.3569}, \href
  {https://arxiv.org/abs/0911.3569} {\path{arXiv:0911.3569}}.

\bibitem[Wat62]{Watson62}
G.N. Watson.
\newblock {\em A Treatise on the Theory of Bessel Functions}.
\newblock University Press, 1962.
\newblock URL: \url{https://books.google.com/books?id=PaTOzwEACAAJ}.

\bibitem[Wic50]{Wick50}
G.~C. Wick.
\newblock The evaluation of the collision matrix.
\newblock {\em Phys. Rev.}, 80:268--272, Oct 1950.
\newblock URL: \url{https://link.aps.org/doi/10.1103/PhysRev.80.268}, \href
  {https://doi.org/10.1103/PhysRev.80.268} {\path{doi:10.1103/PhysRev.80.268}}.

\bibitem[YL52]{LY52_0}
C.~N. Yang and T.~D. Lee.
\newblock Statistical theory of equations of state and phase transitions. i.
  theory of condensation.
\newblock {\em Phys. Rev.}, 87:404--409, Aug 1952.
\newblock URL: \url{https://link.aps.org/doi/10.1103/PhysRev.87.404}, \href
  {https://doi.org/10.1103/PhysRev.87.404} {\path{doi:10.1103/PhysRev.87.404}}.

\bibitem[ZK26]{zlokapaSYKThermalExpectations2026}
Alexander Zlokapa and Bobak~T. Kiani.
\newblock {{SYK}} thermal expectations are classically easy at any temperature,
  February 2026.
\newblock \href {https://arxiv.org/abs/2602.22619} {\path{arXiv:2602.22619}},
  \href {https://doi.org/10.48550/arXiv.2602.22619}
  {\path{doi:10.48550/arXiv.2602.22619}}.

\bibitem[Zlo26]{zlokapaRigorousQuasipolynomialtimeClassical2026}
Alexander Zlokapa.
\newblock A rigorous quasipolynomial-time classical algorithm for {{SYK}}
  thermal expectations, April 2026.
\newblock \href {https://arxiv.org/abs/2604.21089} {\path{arXiv:2604.21089}},
  \href {https://doi.org/10.48550/arXiv.2604.21089}
  {\path{doi:10.48550/arXiv.2604.21089}}.

\end{thebibliography}
\endgroup

\end{document}